\documentclass[aps,prb,twocolumn,shortbibliography,superscriptaddress]{revtex4-1}
\usepackage{epsfig}
\usepackage{epstopdf}
\usepackage{amsmath}
\usepackage{amsfonts}
\usepackage{amssymb}
\usepackage{hyperref}
\usepackage{bm}
\usepackage{makecell}
\usepackage{rotating}
\usepackage{hyperref}
\usepackage{multirow}

\usepackage{chemformula} 
\usepackage[T1]{fontenc} 
\usepackage{subcaption}
\usepackage[percent]{overpic}
\usepackage{academicons} 
\usepackage{xcolor}
\usepackage{ragged2e}

\usepackage{graphicx}
\usepackage{dcolumn}
\usepackage{bm}
\usepackage{color}

\usepackage[english]{babel}
\usepackage{soul}
\usepackage{orcidlink}
\usepackage{siunitx}
\usepackage{nicefrac}
\usepackage{xspace}
\usepackage{float}
\usepackage{dsfont}
\usepackage{textcomp}
\usepackage{gensymb}
\usepackage{braket}
\usepackage{times}
\usepackage{standalone}
\newcommand{\tc}[2]{\textcolor{#1}{#2}}

\usepackage{pifont}       
\usepackage[dvipsnames]{xcolor} 
\newcommand{\cmark}{\textcolor{ForestGreen}{\textbf{\ding{51}}}} 
\newcommand{\xmark}{\textcolor{Red}{\textbf{\ding{55}}}}         
\usepackage{array} 
\usepackage[table, dvipsnames]{xcolor}
\definecolor{PastelBlue}{RGB}{235, 245, 255} 
\definecolor{HeaderGray}{RGB}{220, 220, 220} 

\DeclareMathAlphabet\mathbfcal{OMS}{cmsy}{b}{n}

\usepackage{physics}
\usepackage{comment}

\usepackage{etoolbox} 
\usepackage{graphicx}

\let\oldincludegraphics\includegraphics
\renewcommand{\includegraphics}[2][]{%
  \oldincludegraphics[width=0.7\linewidth, #1]{#2}%
}

\usepackage{ifthen}
\usepackage{filecontents}

\newcommand{\bigfigures}{%
  "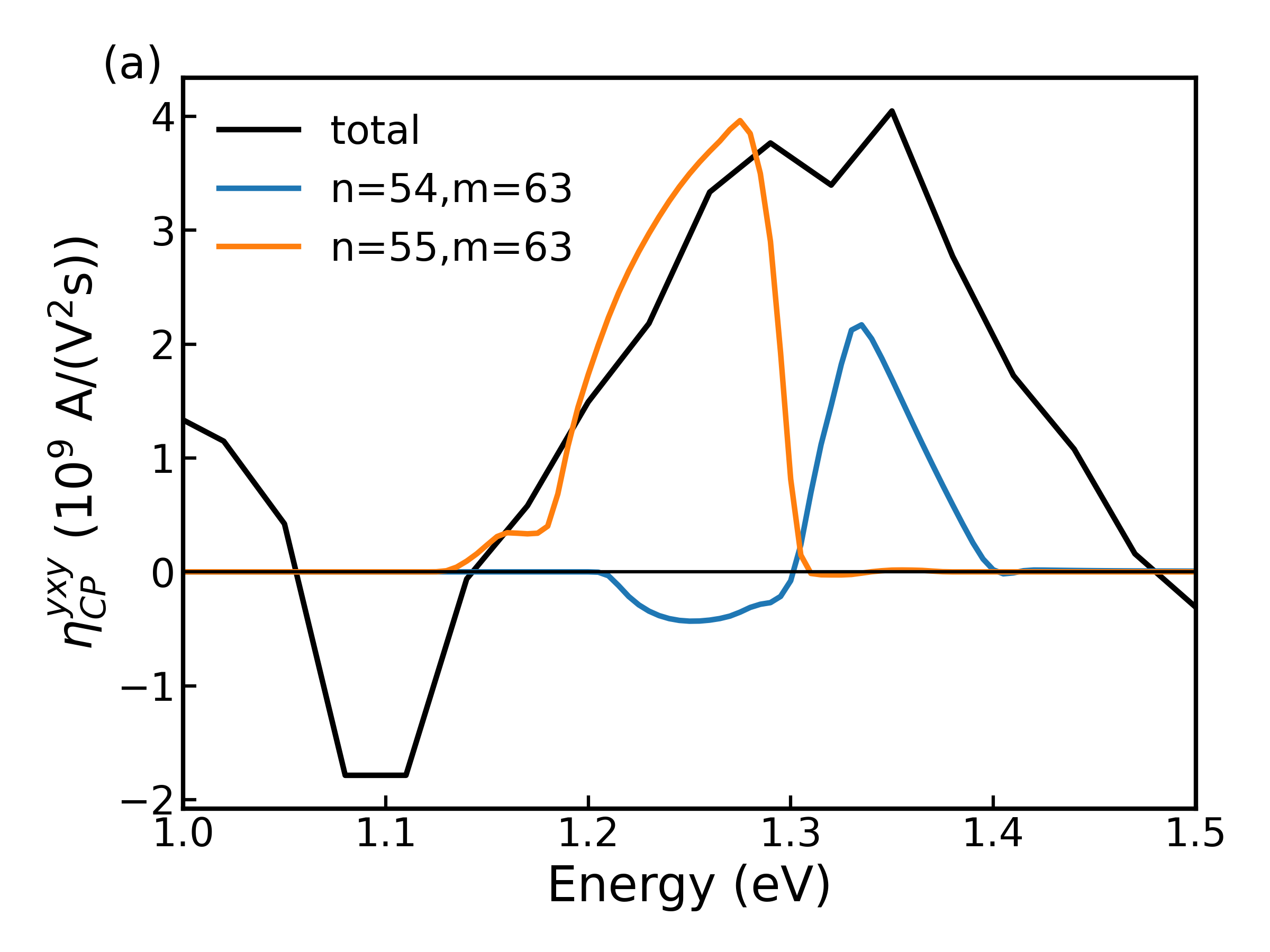",%
  "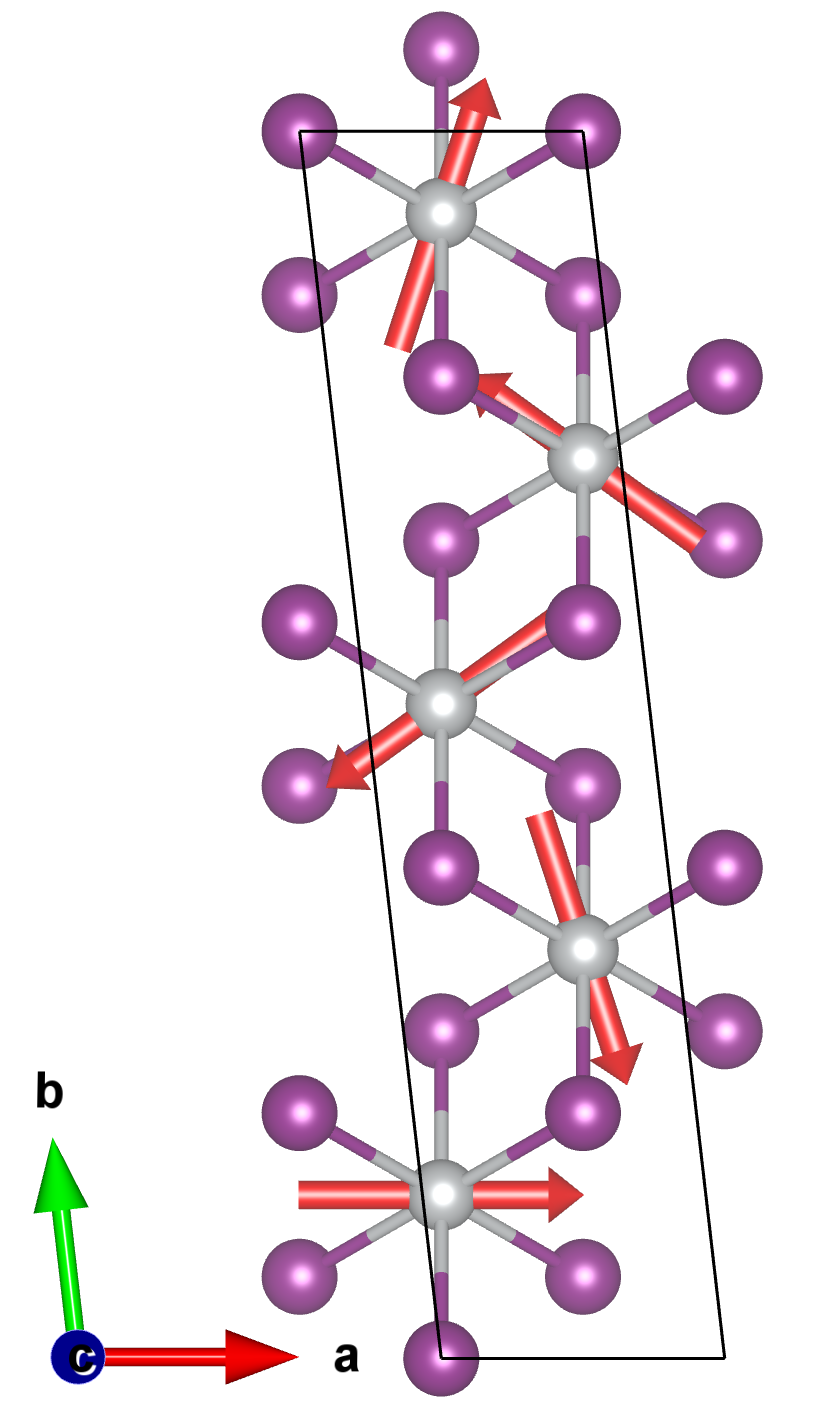",%
  "Charge_linear_shift_1x5_cycloid.eps",%
  "Charge_linear_injection_1x5_cycloid.eps",%
  "Charge_circular_injection_1x5_cycloid.eps"%
}

\makeatletter
\pretocmd{\includegraphics}{%
  \def\@tempa{#2}%
  \def\@found{0}%
  \@for\@f:=\bigfigures\do{%
    \ifx\@f\@tempa
      \def\@found{1}%
    \fi
  }%
  \ifnum\@found=1
    \oldincludegraphics[draft,width=0.7\linewidth]{#2}%
    \expandafter\@gobble
  \fi
}{}{}
\makeatother

\usepackage{tikz,xcolor,hyperref}

\definecolor{lime}{HTML}{A6CE39}
\DeclareRobustCommand{\orcidicon}{%
	\begin{tikzpicture}
	\draw[lime, fill=lime] (0,0)
	circle [radius=0.16]
	node[white] {{\fontfamily{qag}\selectfont \tiny ID}};
	\draw[white, fill=white] (-0.0625,0.095)
	circle [radius=0.007];
	\end{tikzpicture}
	\hspace{-2mm}
}

\foreach \x in {A, ..., Z}{%
	\expandafter\xdef\csname orcid\x\endcsname{\noexpand\href{https://orcid.org/\csname orcidauthor\x\endcsname}{\noexpand\orcidicon}}
}

\begin{document}

\title{Charge and spin photogalvanic effects in the $p$-wave magnet NiI$_2$}

\author{Giuseppe Cuono\orcidA}
\email[]{giuseppe.cuono@unimib.it}
\affiliation{Department of Materials Science, University of Milan-Bicocca, Via Roberto Cozzi 55, 20125 Milan, Italy}
\affiliation{Consiglio Nazionale delle Ricerche CNR-SPIN, c/o Universit\'{a} degli Studi "G. D’Annunzio", 66100 Chieti, Italy}

\author{Srdjan Stavri\'c\orcidB}
\affiliation{Vin\v{c}a Institute of Nuclear Sciences - 
National Institute of the Republic of Serbia, University of Belgrade, P. O. Box 522, RS-11001 Belgrade, Serbia}

\author{Javier Sivianes Casta\~no\orcidC}
\affiliation{Centro de F\'{i}sica de Materiales (CSIC-UPV/EHU), 20018, Donostia-San Sebasti\'{a}n, Spain}

\author{Julen Iba\~{n}ez-Azpiroz\orcidD}
\affiliation{Centro de F\'{i}sica de Materiales (CSIC-UPV/EHU), 20018, Donostia-San Sebasti\'{a}n, Spain}
\affiliation{IKERBASQUE, Basque Foundation for Science, 48009 Bilbao, Spain}
\affiliation{Donostia International Physics Center (DIPC), 20018
Donostia-San Sebasti\'{a}n, Spain}

\author{Paolo Barone\orcidE}
\affiliation{Consiglio Nazionale delle Ricerche CNR-SPIN, Area della Ricerca di Tor Vergata, Via del Fosso del Cavaliere, 100, I-00133 Rome, Italy}

\author{Andrea Droghetti\orcidF}
\email{andrea.droghetti@unive.it}
\affiliation{Department of Molecular Sciences and Nanosystems, Ca’ Foscari University of Venice, via Torino 155, 30170, Mestre, Venice, Italy}

\author{Silvia Picozzi\orcidG}
\affiliation{Department of Materials Science, University of Milan-Bicocca, Via Roberto Cozzi 55, 20125 Milan, Italy}
\affiliation{Consiglio Nazionale delle Ricerche CNR-SPIN, c/o Universit\'{a} degli Studi "G. D’Annunzio", 66100 Chieti, Italy}

\date{\today}


\begin{abstract}
NiI$_2$ is an exotic van der Waals material in which a noncollinear spin spiral breaks spatial inversion symmetry without sizeable structural distortion, generating improper ferroelectric polarization, and stabilizing $p$-wave magnetic states with electron-volt-scale odd-parity spin splitting. Using first-principles calculations, here we establish that nonlinear optical transport can directly probe and separate these effects.
Magnetically--induced inversion breaking associated with the spin spiral produces a photogalvanic shift current under linearly polarized light, with conductivities exceeding those of conventional ferroelectrics. In contrast, a large photogalvanic injection current under circularly polarized light originates from helicity-selective transitions between spin-split states at opposite crystal momenta, directly exposing the nonrelativistic $p$-wave spin texture. We further predict pure spin photocurrents whose flow direction exchanges with that of the charge current under linear and circular excitation. The ability to generate and control pure spin currents without accompanying charge currents makes NiI$_2$ a promising material platform for all-optical spin injection in van der Waals heterostructures.
\end{abstract}

\maketitle

\begin{figure*}[t!]
\centering
\includegraphics[width=0.9\textwidth]{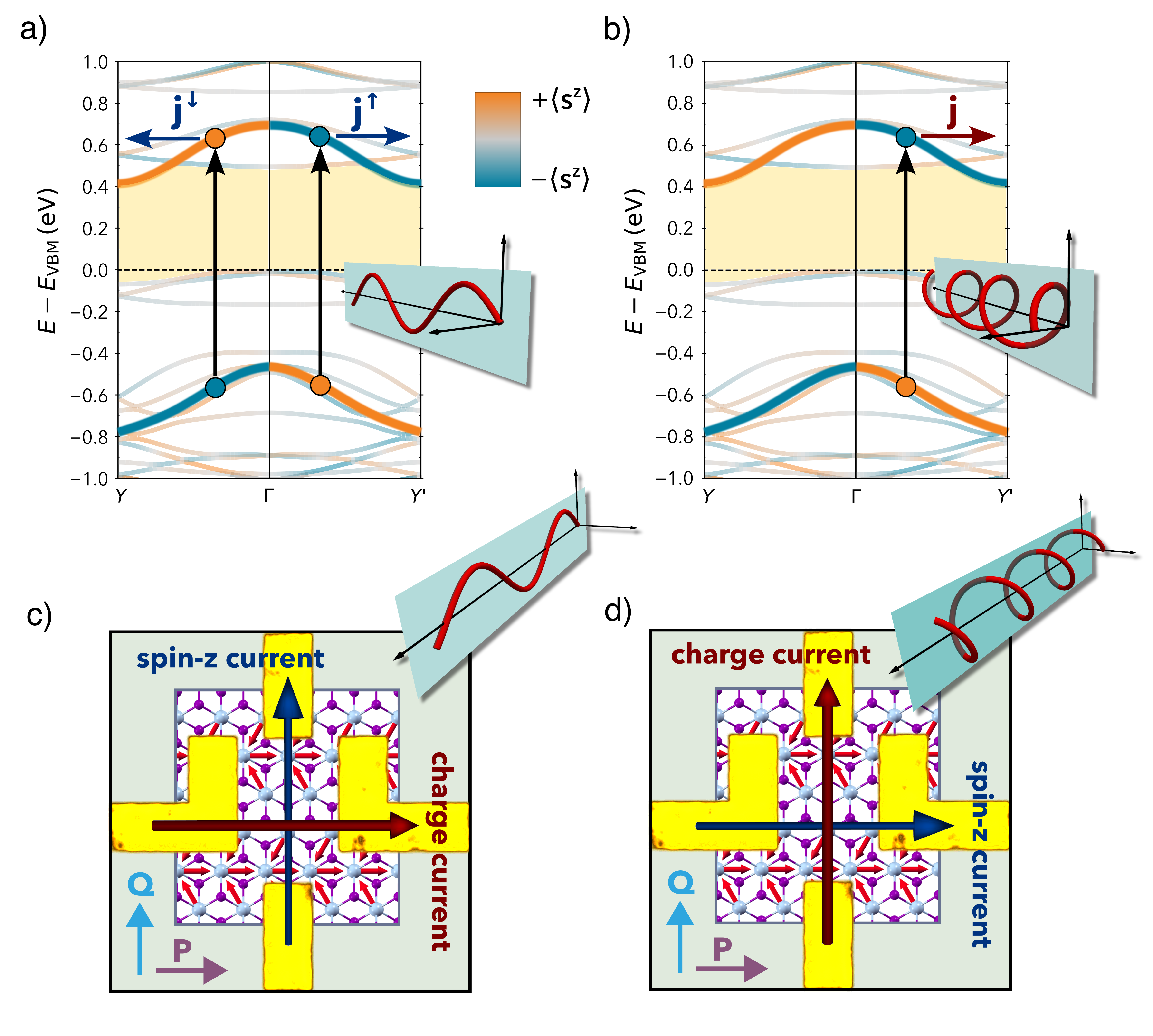}
\caption{\justifying Schematic illustrating (a) Spin$-z$ injection current, $j^z=j^\uparrow -j^\downarrow$, under linearly polarized light. $j^{\uparrow(\downarrow)}$ is the current associated to spin up (down) electrons, assuming $z$ spin quantization axis. (b) Charge injection current under circularly polarized light. (c) Device illuminated by linearly polarized light driving charge (shift) and spin-$z$ (injection) current along the electrical polarization $\mathbf{P}$ and the spiral propagation vector $\mathbf{Q}$, respectively. (d) Device illuminated by circularly polarized light driving charge (injection) and spin-$z$ (shift) current along $\mathbf{Q}$ and $\mathbf{P}$, respectively. 
}\label{cartoon}
\end{figure*}

\section{Introduction}

Broken symmetries underpin the emergent electronic, magnetic, and optical properties of quantum materials. When multiple symmetries are lifted simultaneously, nontrivial couplings can arise between distinct degrees of freedom, and new phenomena may appear. A paradigmatic example is provided by noncollinear spin-spiral magnetic structures, which break spatial inversion symmetry via a helical spin arrangement giving rise to an improper electric polarization and magnetoelectric coupling 
\cite{Cheong2007,Hur2004,Katsura2005, Lawes2005,Kenzelmann2005,Taniguchi2006,Mostovoy2006,Xiang.prl.2011,kaplan.prb.2011,edstrom.2025}. Furthermore, similar noncollinear magnetic structures, under certain symmetry conditions, can host odd-parity ($p$-wave) magnetic states \cite{hellenes2024,yamada2025,Song25}. These are characterized by zero net magnetization, but present an electron-volt-scale spin splitting, with opposite spin polarization at inversion-related momenta in reciprocal space ---even in the nonrelativistic limit without spin-orbit coupling (SOC). $p$-wave magnets are promising candidates for next-generation spintronics, potentially enabling spin-polarized transport with ultrafast spin dynamics and minimal stray fields.

The coexistence of spin-spiral type-II multiferroicity \cite{Kurumaji.prb.2013, Song22, Ju21, Amini24}
and $p$-wave magnetism \cite{Song25} has recently been reported in the van der Waals material NiI$_2$, where it was shown to enable voltage control of spin polarization \cite{Song25}. Experimentally, these phenomena have been probed via photogalvanic measurements, in which a dc charge current arises as a second-order bulk response to optical excitation \cite{Sipe00, Dai23}.

In photogalvanic measurements, two major effects appear. The linear photogalvanic effect (LPGE) \cite{Belinicher1980} occurs in noncentrosymmetric materials and, among these, has been most extensively studied in polar systems, where it directly reflects polar order: a symmetry-allowed macroscopic polarization generates a finite dc photocurrent under linearly polarized illumination \cite{Dai23, Young12, Young12b, Butler15, Tan16}. Thus, in most materials, LPGE originates from intrinsic crystal inversion asymmetry. In NiI$_2$, however, the lattice is centrosymmetric in the absence of magnetism, and inversion symmetry is broken primarily by the spin-spiral order. How such magnetically-induced inversion breaking modifies nonlinear optical responses -- compared to conventional crystal-structure-driven mechanisms -- has only recently attracted attention \cite{PhysRevB.110.L041104, Sivianes25} and remains largely unexplored.

In contrast, circular photogalvanic effect (CPGE) \cite{Belinicher1980, PismaZhETF.27.640}, induced by circularly polarized light, is sensitive to spin-split electronic structures \cite{Ganichev2000, Ganichev2003,BELKOV2003,GOLUB2003,Ganichev14} as first demonstrated in tellurium \cite{Asnin1978, Asnin1979} and later in quantum wells \cite{ganichev2003spinphotocurrentsquantumwells1, ganichev2003spinphotocurrentsquantumwells2}, in a variety of Rashba systems \cite{Ganichev_2003,Weber2005,Giglberger2007, Lechner2011, Taniuchi25, Hirose2018, Wang22, Liu2020, Niesner18}, transition metal dichalcogenides \cite{Yuan2014, Quereda2018}, topological insulators \cite{Hosur11,  McIver2012, Yu20,Sun2021},  Weyl semimetals \cite{deJuan2017,Ji2019}, and recently also predicted in collinear altermagnets \cite{Jiang2025}. By analogy, CPGE has been employed as a probe of $p$-wave magnetism in NiI$_2$ \cite{Song22}; however, a direct microscopic connection between the CPGE response and the underlying $p$-wave spin-split band structure has yet to be established.

Beyond charge currents, optical excitation can also generate spin currents in systems with spin-split electronic bands through selective coupling to states of opposite spin \cite{Bhat2005,Young2013, Xu2021,Lihm2022}, as predicted for surfaces with Rashba states \cite{Sivianes25_2}, transition metal dichalcogenides \cite{Xu2021}, and topological insulators\cite{Xu2021, Kim1017}. This mechanism is known as the spin photogalvanic effect. The resulting spin currents may correspond to spin-polarized charge currents or to pure spin currents, in which carriers with opposite spin propagate in opposite directions with vanishing net charge flow. To date, however, optically induced spin currents have not been studied in $p$-wave magnets. Their demonstration would provide a route toward optical spin injection in spintronic devices based on these materials.

In this work, we employ first-principles calculations based on density functional theory (DFT) to study photogalvanic effects in NiI$_2$. We confirm that LPGE reported in experiments originates from spin-spiral-induced inversion symmetry breaking and that the resulting photoconductivity can reach remarkably large values, exceeding those of common bulk oxide ferroelectrics. We further show that the CPGE is strongly enhanced by optical transitions between nonrelativistic spin-split bands, thereby establishing a direct microscopic connection between the CPGE response and the underlying $p$-wave magnetic state. Finally, we predict that linearly polarized light can drive pure spin currents, with potential for optically controlled spin transport. Taken together, our results establish spin-spiral $p$-wave magnets as an intriguing material platform in which broken symmetries intertwine electronic, magnetic and optical properties, enabling fundamentally new nonlinear and spin-selective phenomena.

\begin{table*}[t]
\centering
\renewcommand{\arraystretch}{1.5}
{\large \sffamily 
\begin{tabular}{|l|c|c|}
\hline\hline
\rowcolor{HeaderGray}  
 & \textbf{\enspace Linearly Polarized Light \enspace} & \textbf{\enspace Circularly Polarized Light \enspace} \\ \hline
 \rowcolor{PastelBlue} 
\enspace Charge shift current   & \quad even \cmark \quad odd \cmark \quad \quad & \quad even \xmark \quad odd \cmark \quad \quad \\ \hline
\rowcolor{PastelBlue} 
\enspace Charge injection current \enspace & even \xmark \quad odd \cmark & even \cmark \quad odd \cmark \\ \hline
\rowcolor{PastelBlue}
\enspace Spin shift current      & even \xmark \quad odd \cmark & even \cmark \quad odd \cmark \\ \hline
\rowcolor{PastelBlue}
\enspace Spin injection current   & even \cmark \quad odd \cmark & even \xmark \quad odd \cmark \\ \hline\hline
\end{tabular}
}
\caption{\justifying Symmetry-allowed second-order photocurrent contributions for even and odd magnetic period $n$. The label $\cmark$ denotes allowed, whereas $\xmark$ denotes forbidden contribution.}
\label{tab:allowed_currents}
\end{table*}

\section{Results}

\subsection{Crystal structure, magnetic order, and symmetry analysis}
Bulk NiI$_2$ crystallizes in the rhombohedral $R\overline{3}m$ structure, which consists of I–Ni–I layers with a triangular Ni$^{2+}$ lattice sandwiched between two iodine planes to form edge-sharing NiI$_6$ octahedra. The layers are stacked along the $c$ axis and weakly coupled by van der Waals interactions. Each Ni ion carries a nominal spin $S=1$, and magnetic frustration leads to competing phases and a sequence of two magnetic transitions at $T_{N1}\simeq 75$~K and $T_{N2}\simeq 59.5$~K \cite{KUINDERSMA1981, Friedt76, BILLEREY1977,Tseng25}. Earlier experimental characterizations identified a collinear antiferromagnetic phase, with propagation vector $\mathbf{Q}=(0,0,3/2)$ (in hexagonal reciprocal lattice units, r.l.u.) below $T_{N1}$ \cite{KUINDERSMA1981}, whereas recent resonant X-ray scattering experiments suggest an amplitude-modulated collinear spin density wave with $\mathbf{Q}=(0.087,0.087,1.5)$ (r.l.u.)\cite{Tseng25}. Below $T_{N2}$, an incommensurate helimagnetic phase sets in, accompanied by a rhombohedral-to-monoclinic structural transition \cite{KUINDERSMA1981,Tseng25}. Given the negligible monoclinic distortion, the helimagnetic phase can be described in the hexagonal cell as a spin spiral tilted by $\sim 35^\circ$ from the $c$ axis, with propagation vector $\mathbf{Q}\simeq(0.14,0,1.47)$ (r.l.u.) and spins rotating in a plane orthogonal to $\mathbf{Q}$ \cite{KUINDERSMA1981,Tseng25}.

SOC locks spins to the lattice, such that spin and spatial symmetries transform jointly. As a result, the helimagnetic order breaks nearly all rotational and mirror symmetries of the crystal, retaining only a single twofold ($C_2$) axis perpendicular to the in-plane projection of $\mathbf{Q}$ and allowing the appearance of a ferroelectric polarization along this unique $C_2$ axis\cite{Kurumaji.prb.2013, Song22}. The resulting type-II multiferroic order persists down to the few-layer limit and even in the monolayer \cite{Song22,Ju21,Amini24}, where it forms a proper-screw spin helix with a reduced critical temperature $T\simeq 21$~K. Notably, because reversal of the spin chirality necessarily reverses the polarization, the magnetoelectric coupling is symmetry protected.

The helimagnetic order of bulk NiI$_2$, propagating perpendicular to the spin-spiral plane, can be approximated by a commensurate magnetic wavevector $\mathbf{Q}=(1/7,0,3/2)$ (r.l.u.), whose magnetic symmetries belong to the Type-IV magnetic space group (MSG) $C_c2$. This group comprises the aforementioned twofold axis $C_2$, as well as time-reversal (TR) symmetry combined with a fractional translation along the $c$ lattice vector. As a consequence, TR is an element of the magnetic point group (MPG), which in turn forces all macroscopic TR-odd physical quantities to vanish. We note that this MSG is preserved for any spin-rotation plane containing the $C_2$ axis. Accordingly, spin spirals can be decomposed into in-plane (cycloidal) and out-of-plane (proper-screw) components. The symmetry of the triangular lattice further allows for six helimagnetic domains, characterized by the in-plane component of the magnetic wavevector being parallel to $(100)$, $(010)$, or $(1\bar{1}0)$, each with two spin-chiral states of opposite handedness.

Since a fully realistic treatment of the bulk spin helix would require prohibitively large supercells, following Ref.~[\onlinecite{Song25}] we adopt 2D commensurate models that preserve the essential symmetry properties of the real system described above. Specifically, we consider a single NiI$_2$ layer with a helical spin configuration, either a cycloid or a proper-screw, characterized by the wavevector $\mathbf{Q}=(0,1/n,0)$ (r.l.u.), where $n$ defines the (commensurate) magnetic period (see Fig. S1 in the Supporting Information (SI). In real space, this corresponds to a modulation along the $y \parallel [120]$ direction in the hexagonal cell, with $x \parallel [100]$.

As in the bulk, both a cycloid, in which the magnetic moments rotate within the $xy$ plane, and a proper-screw spiral, with moments rotating in the $xz$ plane, preserve the $C_2$ axis and exhibit identical MSG symmetries. 
For odd values of $n$, the MSG is Type-III $C2$ (No.~5.13), whereas for even values it is Type-IV $P_b2$ (No.~3.4). 
A key difference emerges, however, for odd-$n$ structures: their MPG no longer contains TR symmetry, making TR-odd quantities symmetry-allowed. This may represent an emergent feature of NiI$_2$ in the monolayer limit.
Without loss of generality, we focus here on the cycloidal configuration, as shown in Fig. ~\ref{bands}a for $n=3$, while the proper-screw case is discussed in Secs. S4 and S9 of the SI.

\subsection{Electronic structure and band spin-polarization}\label{sec.band_structure}
Our DFT calculations are performed using the VASP package \cite{Kresse93,Kresse96,Kresse96b} within the generalized-gradient approximation, employing the Perdew--Burke--Ernzerhof exchange--correlation functional \cite{Perdew96} (further computational details are provided in Sec. S1 of the SI). NiI$_2$ is correctly described as an insulator, although the band gap is expected to be underestimated relative to the true value, which is however not known experimentally.

\begin{figure*}[t!]
\centering
\includegraphics[width=\textwidth]{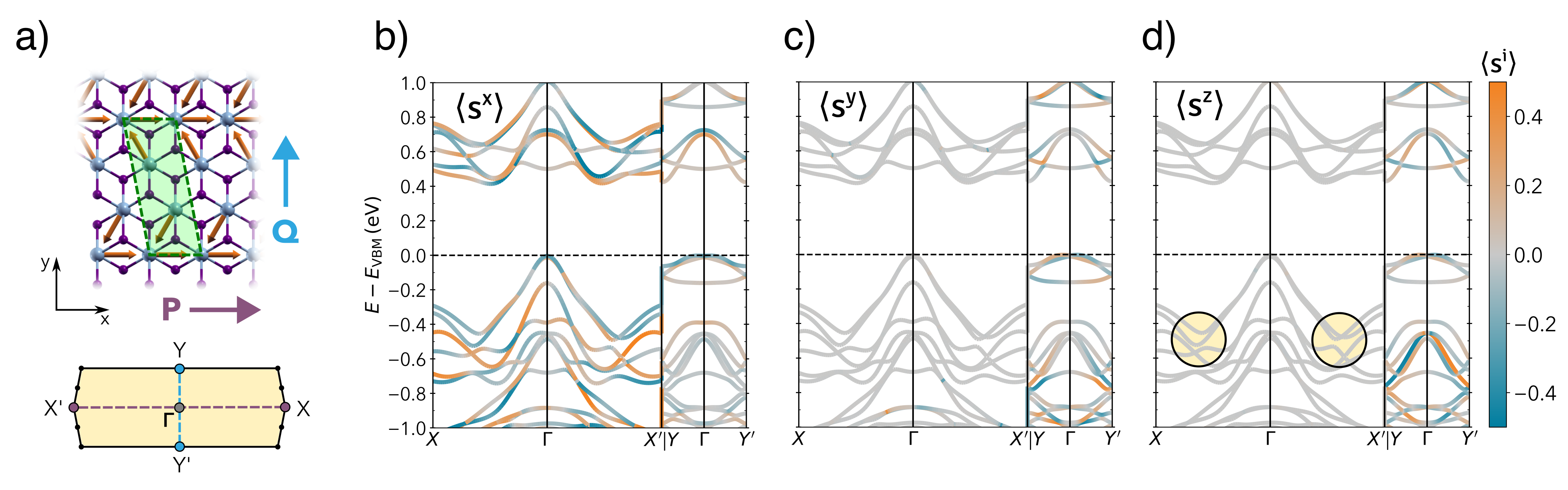}
\caption{\justifying (a) Wigner–Seitz cell (green parallelogram) of the $n=3$ cycloidal structure and the corresponding Brillouin zone (BZ).
(b–d) Band structures with the color scale showing the expectation values of the spin $x$-, $y$-, and $z$-components. The yellow circles in (d) indicate pronounced $k_x\rightarrow -k_x$ band asymmetries.}\label{bands}
\end{figure*}

Figure~\ref{bands}b-d shows the band structure for magnetic period $n=3$, along two mutually perpendicular directions in the Brillouin zone (BZ), depicted in the inset of panel~a. The color scale represents the band spin polarization, i.e.,
the expectation value of the spin components $\langle s^i (\mathbf{k})\rangle = \langle \psi_\alpha(\mathbf{k}) | s^i | \psi_\alpha(\mathbf{k}) \rangle \quad (i = x,y,z)$
of the Bloch state $\psi_\alpha(\mathbf{k})$ for each band $\alpha$, with energy $\epsilon_\alpha(\mathbf{k})$, and crystal momentum $\mathbf{k}$.
The in-plane spin components $\langle s^x(\mathbf{k}) \rangle$ and $\langle s^y(\mathbf{k}) \rangle$ give rise to a spin texture of relativistic origin, as confirmed by their complete suppression when SOC is switched off in the calculations (while still constraining the cycloidal order).

In contrast, the out-of-plane component $\langle s^z(\mathbf{k}) \rangle$ vanishes along the X–$\Gamma$–X$^{\prime}$ path but, for some bands, is large and antisymmetric along the Y–$\Gamma$–Y$^{\prime}$ path  (i.e. odd under $k_y \rightarrow -k_y$), which is parallel to the spin-helix propagation direction $\mathbf{Q}$. These bands therefore display $p$-wave texture. In this case, turning off SOC removes some band splittings at the $\Gamma$ point; however, the magnitude and antisymmetry of $\langle s^z(\mathbf{k}) \rangle$ remain unchanged. This interpretation is further supported by comparing the symmetry constraints imposed by the MSG with those of spin groups \cite{brinkman66, litvin77}, whose elements act independently on spin and real space. Spin groups thus describe the approximate symmetries of magnetic structures in the absence of SOC \cite{Etxebarria.stensor.25}. Within the tensorial framework provided by the theory of invariants \cite{radaelli.25}, one finds that a nonrelativistic spin polarization is allowed only along the direction perpendicular to the spin-spiral plane and for crystal momenta parallel to the magnetic wavevector (see Sec.~S2 of the SI). The spin splitting can therefore be effectively described by an odd-parity Zeeman field $h_\perp(\mathbf{k}) \propto \mathbf{k}\cdot\mathbf{Q}$ \cite{Song25}.

This picture remains valid for the proper-screw spiral, which shares the same $C_2$ symmetry as the cycloid. The only notable difference is that the nonrelativistic spin polarization is oriented along $y$ (i.e., $\langle s^y(\mathbf{k}) \rangle$) instead of $z$ (see Sec. S4 of the SI). This is consistent with the spin-group symmetry analysis (Sec.~S2 of the SI).

In addition to the $p$-wave magnetism, the band structure exhibits another notable feature. The absence of TR symmetry in the MPG of the cycloid with odd periodicity ($n=3$) allows for a finite, albeit very small, net in-plane magnetization perpendicular to $\mathbf{Q}$ ($M_x \sim 0.02~\mu_\mathrm{B}$ per supercell), as well as for a nonreciprocal band dispersion along the X--$\Gamma$--X$^{\prime}$ direction, i.e., $\varepsilon_\alpha(k_x) \neq \varepsilon_\alpha(-k_x)$. A similar effect has recently been described in detail for the multiferroic phase of EuO \cite{Stavric25}, where the band asymmetry appears along a direction perpendicular to both the magnetization and the polarization. In contrast, in the NiI$_2$ monolayer the MPG constrains the nonreciprocal band dispersion to occur along a direction that, in real space, is parallel to both the magnetization and the ferroelectric polarization. While nonreciprocity is, in principle, present across all energy bands, its magnitude varies; for clarity, the most pronounced $k_x \rightarrow -k_x$ asymmetries are highlighted by yellow circles in Fig.~\ref{bands}d.

Both the weak ferromagnetic moment and the band nonreciprocity, being TR-odd properties, are symmetry-forbidden for even-$n$ helices and in the bulk. Moreover, they are relativistic effects that disappear in the absence of SOC and therefore constitute an additional manifestation ---alongside polar order and spin texture--- of symmetry breaking \cite{Stavric25}. Each gives rise to distinct contributions to the second-order photoconductivity, as discussed in the following.

\begin{figure}[htbp]
    \centering
    \includegraphics[width=\columnwidth]{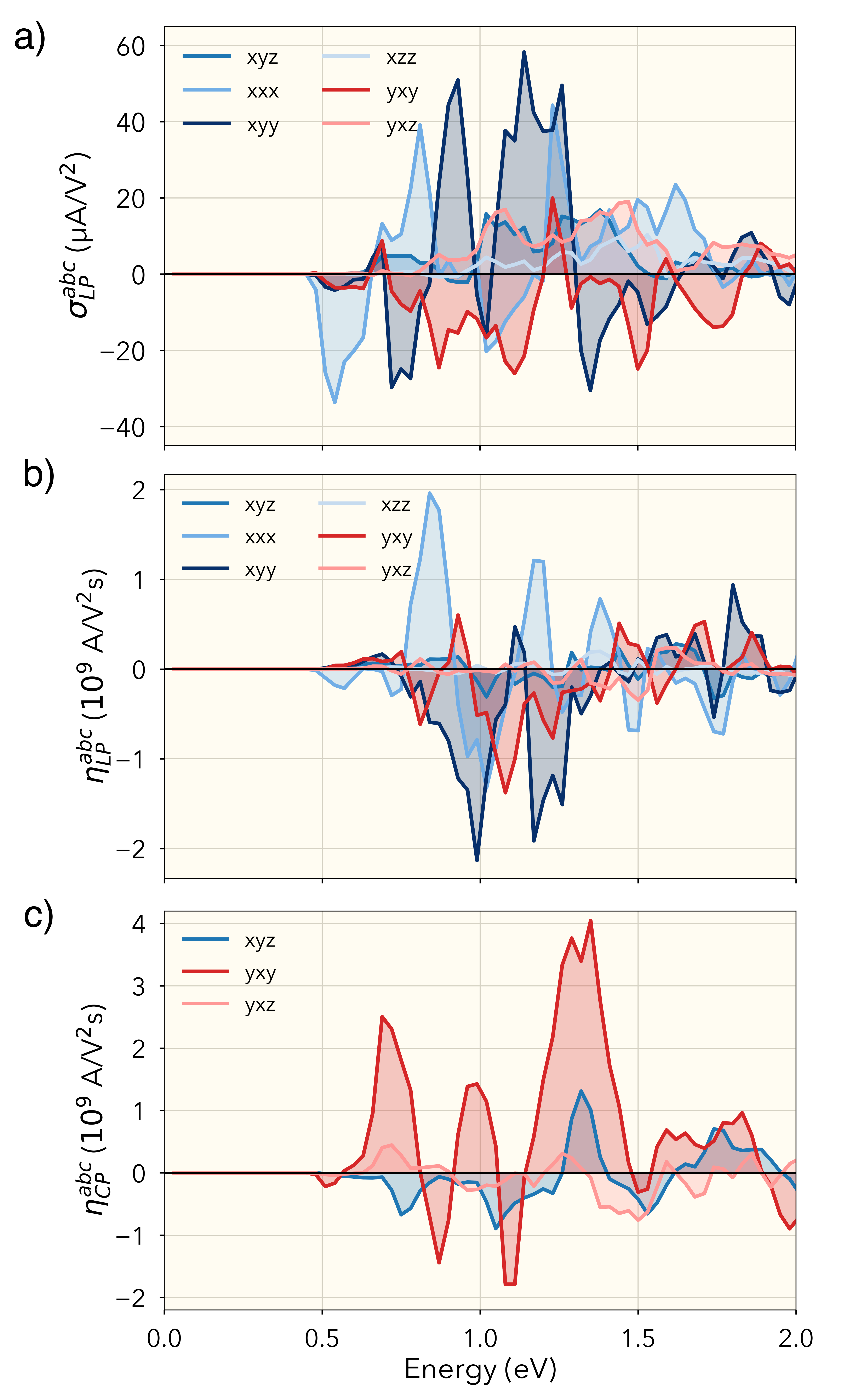}
    \caption{\justifying  Non-zero components of the (a) charge LP shift photoconductivity, (b) LP injection photoconductivity, and (c) CP injection photocurrent tensors for the $n=3$ cycloid structure.}
    \label{photocurrents}
\end{figure}

\subsection{Photoconductivity}
The second-order dc charge or spin photocurrent densities driven by an electric field $\mathbfcal{E}(\omega)$ of frequency $\omega$  is
\begin{equation}\begin{split}
j^{(s),a}= 2[ \sigma^{(s),abc}_\mathrm{LP}(\omega)+ \tau \eta^{(s),abc}_\mathrm{LP}(\omega)]\Re[\mathcal{E}^b(\omega)\mathcal{E}^c(-\omega)]\\
+2i[ \sigma^{(s),abc}_\mathrm{CP}(\omega)+ \tau \eta^{(s),abc}_\mathrm{CP}(\omega)]\Im[\mathcal{E}^b(\omega)\mathcal{E}^c(-\omega)]
\end{split}\label{eq.j}
\end{equation}
where $a$ labels the current direction, and $b,c$ the polarization directions of the electric field.
The index $s=x,y,z$ denotes the spin polarization of the spin current and
$\tau$ is the charge carrier relaxation time or the spin lifetime. The real and imaginary terms in Eq. (\ref{eq.j}) describe the LPGE and the CPGE, respectively. The photoconductivity -- a rank-three tensor -- is decomposed into two contributions, the so-called shift and injection photoconductivities, $\sigma_\mathrm{LP(CP)}$ and $\eta_\mathrm{LP(CP)}$, for linearly (circularly) polarized light. These are computed using the expressions from Refs. [\onlinecite{Puente23}] and~[\onlinecite{Lihm2022}] for charge and spin, respectively. 

The shift current is an interband coherence effect that can be interpreted, in the charge channel, as a real-space displacement of carriers during an optical transition \cite{Sipe00, Young12, Ibanez2018}. In the spin channel, the corresponding shift becomes spin dependent and produce a dc flow of spin angular momentum even in the absence of net charge transport \cite{Young2013}. In contrast, the injection current originates from the asymmetric population of Bloch states in momentum space \cite{Aversa1995,Sipe00,Nastos2012}. 
For charge currents this reflects an imbalance of carrier velocities \cite{Zhang2019,Stavric25, Wang2020}, whereas for spin currents it corresponds to an asymmetric population of spin-polarized states at opposite momenta. Both charge and spin currents may coexist, but a pure spin current can arise only when the charge contributions cancel while opposite momenta carry different spin projections.


The symmetry-adapted form of photoconductivity tensors is dictated by the MPG of the target magnetic material. Notably, LP charge injection (spin shift) under LP light and CP charge shift (spin injection) photoconductivity tensors are odd under time reversal, implying that they are forbidden for MPGs that include TR, such as for the commensurate helimagnetic phase of bulk NiI$_2$ or for the single-layer cycloid with even-$n$ period. In Table ~\ref{tab:allowed_currents} we summarize the allowed photogalvanic responses, while the full symmetry-adapted tensors are given in Sec. S2 of the SI.
The spectra of the various components of $\sigma_\mathrm{LP(CP)}$ and $\eta_\mathrm{LP(CP)}$ are computed using a Wannier-interpolation scheme \cite{Marzari1997,Mostofi2008} applied to the DFT band structure via the {\sc Wannier90} package \cite{Pizzi2020}. 

{\it Charge LPGE.} The shift current under LP light is predicted to produce the LPGE for both odd and even $n$. The corresponding photoconductivity tensor exhibits six independent nonzero components: $\sigma^{xyz}_\mathrm{LP}$, $\sigma^{xxx}_\mathrm{LP}$, $\sigma^{xyy}_\mathrm{LP}$, $\sigma^{xzz}_\mathrm{LP}$, $\sigma^{yxy}_\mathrm{LP}$ and $\sigma^{yxz}_\mathrm{LP}$. Owing to the intrinsic symmetry of this tensor under exchange of its last two indices, all remaining nonvanishing components follow directly from these by permutation of the polarization indices. Spin-group symmetries do not introduce additional constraints (see Sec. S2 of the SI), implying that SOC is not required for LP charge photocurrents to appear.

The resulting spectra of the photoconductivity components for $n=3$ are shown in Fig.~\ref{photocurrents}a, while those for $n=4$ are reported in the Sec. S7 of the SI. The onset of the response coincides with the optical band gap, confirming its interband origin of the shift current. The spectra exhibit pronounced, sharp features, which arise from the reduced dimensionality of the system and the associated Van Hove singularities in the electronic structure, similar to those reported in other low-dimensional materials \cite{Rangel17,Ibanez2018}. In the visible range, the $\sigma^{xxx}_\mathrm{LP}$ and $\sigma^{xyy}_\mathrm{LP}$ components dominate, corresponding to a photocurrent along $x$ (i.e., parallel to the polarization direction) driven by light with electric field polarized along the $x$ and $y$ direction, respectively, with peak values reaching $\sim 50$~$\mu$A/V$^2$.
Similar results were obtained in previous calculations of shift photocondutivities in NiI$_2$, albeit performed for a different model system \cite{PhysRevB.110.L041104}. This peak value is an order of magnitude smaller than those predicted for other ferroelectric monolayers, such as GeSe, GeS \cite{Rangel17}, or GeTe \cite{Tiwari22}. Nevertheless, it substantially exceeds the response of prototypical bulk ferroelectrics in the visible range ---approximately $0.05 \,\mu$A/V$^2$ and $5 \,\mu$A/V$^2$ for BiFeO$_3$ and BaTiO$_3$ \cite{Young12,Young12b}--- and is comparable to theoretical predictions for high-performance bulk ferroelectrics such as SbSI \cite{Cuono25} and~ $\alpha$-GeTe  \cite{Tiwari22}. Remarkably, this large response occurs despite the improper ferroelectric character of the system, in which inversion symmetry is broken by electronic and magnetic degrees of freedom, resulting in a polarization that is roughly four orders of magnitude smaller than in conventional ferroelectrics. Thus, our results establish spin-spiral order as an efficient mechanism to activate large LPGE responses even in structurally centrosymmetric crystals.

Notably, odd-period spin spirals, such as the $n=3$ system studied here, exhibit a sizeable LP injection photoconductivity in addition to the shift contribution. This is enabled by the absence of TR symmetry in the corresponding MPG (see Table~\ref{tab:allowed_currents}). Some components reach peak values on the order of $10^9$~A/(V$^2$s). These are comparable to those predicted for EuO in Ref. [\onlinecite{Stavric25}], and can be similarly related to the nonreciprocity of the band structure activated by the TR symmetry breaking\cite{Stavric25}, as described in Sec. \ref{sec.band_structure}. 

The LP injection current adds to the LP shift current discussed above, but it is activated by SOC and vanishes identically under spin-group symmetries (Sec.~S2 of the SI). At the same time, in contrast to shift currents, injection currents scale with the relaxation time; their characteristic temperature dependence therefore provides, in principle, a means to experimentally disentangle the two contributions.
We emphasize, however, that this injection contribution disappears in systems where TR-odd properties -- such as the net magnetization and the nonreciprocal band-structure effect -- are symmetry forbidden.

{\it Charge CPGE.} A shift current under circularly polarized light emerges only for odd $n$ (see Table \ref{tab:allowed_currents}) and requires SOC (being forbidden under spin-group symmetries, see SI), while it is symmetry-forbidden for even $n$ and in the real magnetic structure of bulk NiI$_2$. The dominant and symmetry-allowed contribution to the CPGE in all cases is instead the CP injection photoconductivity, which displays the same symmetry-adapted tensorial form under magnetic or spin-group symmetries (Sec. S2 of the SI) and directly reflects the spin texture of the electronic bands. It originates from the transfer of angular momentum from circularly polarized photons to the electronic system, resulting in spin-selective optical excitations, as schematically shown in Fig. \ref{cartoon}(b). When the electron spin is locked to the crystal momentum, this produces an asymmetric distribution of photoexcited carriers in momentum space and consequently a net charge current.

In particular, in our $p$-wave system with $\mathbf{Q}\parallel y$, states at equal energy and opposite momenta $k_y$ and $-k_y$ carry opposite spin projections. For a suitable light helicity, the corresponding optical transition probabilities differ, leading to an imbalance of excited carriers and hence to a net current flowing along the $y$ direction (i.e., parallel to the cycloid propagation direction). This microscopic picture is fully consistent with the symmetry analysis and is confirmed by the first-principles calculations.

For both even and odd values of $n$, the CP injection photoconductivity exhibits three independent nonzero components: $\eta_\mathrm{CP}^{xyz}$, $\eta_\mathrm{CP}^{yxy}$, and $\eta_\mathrm{CP}^{yzx}$. All other nonvanishing components follow from antisymmetry under exchange of the last two indices, $\eta^{abc} = -\eta^{acb}$. As shown in Fig.~\ref{photocurrents}(c), the component $\eta_\mathrm{CP}^{yxy}$, dominates the response, reaching peak values of approximately $4 \times 10^{9}$ A/V$^{2}$s. By resolving the contributions of individual interband transitions (see  Sec. S6 of the SI), we find that the two largest peaks, at 1.27 and 1.35 eV, originate primarily from transitions between valence bands around $-0.6$ eV and conduction bands near $0.7$ eV, which show a very pronounced spin-$z$ splitting. This provides direct microscopic evidence that this component of the CP injection response is related to the nonrelativistic $p$-wave magnetism. 

The remaining components, $\eta_\mathrm{CP}^{xyz}$ and $\eta_\mathrm{CP}^{yzx}$, are associated with the in-plane spin texture induced by SOC. Interestingly, as discussed above, these components are still symmetry-allowed by spin-group symmetries in the nonrelativistic limit; however, SOC is required to generate the spin-$x$ and spin-$y$ polarizations and thus activate these contributions. In the system considered here, this SOC-induced spin polarization is much smaller than the $p$-wave spin-$z$ polarization, which is directly reflected in the significantly smaller values of $\eta_\mathrm{CP}^{xyz}$ and $\eta_\mathrm{CP}^{yzx}$ compared to $\eta_\mathrm{CP}^{yxy}$.

In systems with a proper-screw spiral instead of a cycloid, the roles of the tensor components associated with SOC and $p$-wave magnetism are interchanged, and the $p$-wave spin polarization is slightly reduced; however, the same qualitative considerations still apply (see Sec.~S9 of the SI).

Notably, the rank-three CP injection photoconductivity tensor can be mapped onto an effective rank-two pseudotensor through the relation\cite{Tsirkin2018}
\begin{equation}
\gamma_{ab} = -i\, \epsilon_{bcd}\,\eta_\mathrm{CP}^{acd},
\end{equation}
where $\epsilon_{bcd}$ is the Levi--Civita symbol. 
Thus, the CP injection current can be rewritten in the compact form \cite{Ganichev_2003}
\begin{equation}
j^a_{\mathrm{CP}} = \tau\, \gamma_{ab}\,
\big\{i[(\boldsymbol{\mathcal{E}}(\omega)\times\boldsymbol{\mathcal{E}}^*(\omega)]_b\big\}.
\end{equation}

In this representation, the driving field is the optical helicity vector
$i\,\boldsymbol{\mathcal{E}}\times\boldsymbol{\mathcal{E}}^*$,
which is proportional to the photon angular momentum and reverses sign upon switching the light helicity. The tensor $\gamma_{ab}$ therefore directly quantifies the conversion of photon angular momentum into dc charge current. Because the helicity vector is an axial vector whereas the current is a polar vector, $\gamma_{ab}$ transforms as a pseudotensor, reflecting the chiral nature of the CPGE response.

Within this framework\cite{Ganichev_2003}, for the cycloid system considered here, the nonzero components $\gamma_{xx}$ and $\gamma_{yy}$ (stemming from $\eta_\mathrm{CP}^{xyz}$ and $\eta_\mathrm{CP}^{yzx}$) directly mirror the spin-momentum couplings $k_xs^x$, $k_ys^y$, that characterize the SOC-induced spin texture, while $\gamma_{yz}$ (stemming from $\eta_\mathrm{CP}^{yxy}$) reflects the spin-momentum coupling $k_ys^z$ of the $p$-wave spin textures. This makes explicit that the CP injection current probes the underlying spin--momentum locking: photon angular momentum selectively excites spin-polarized states, and the resulting momentum imbalance produces a dc charge current. In this sense, the CPGE provides a direct nonlinear optical fingerprint of the spin texture of the electronic structure.

{\it Spin LPGE.} The LP spin injection current is the only symmetry-allowed contribution to the spin LPGE that is present across all systems considered. It can be viewed as the spin counterpart of the linear charge injection current; however, unlike its charge counterpart, it does not require band nonreciprocity but instead relies on a nontrivial spin texture. Even when the photoexcited carrier distribution in momentum space remains symmetric, the excited populations can be spin imbalanced because states at $k_y$ and $-k_y$ carry different spin-$z$ projections, giving rise to a finite pure spin-$z$ current along $y$, as schematically shown in Fig. \ref{cartoon}a and c. 

\begin{figure}[htbp]
    \includegraphics[width=\columnwidth]{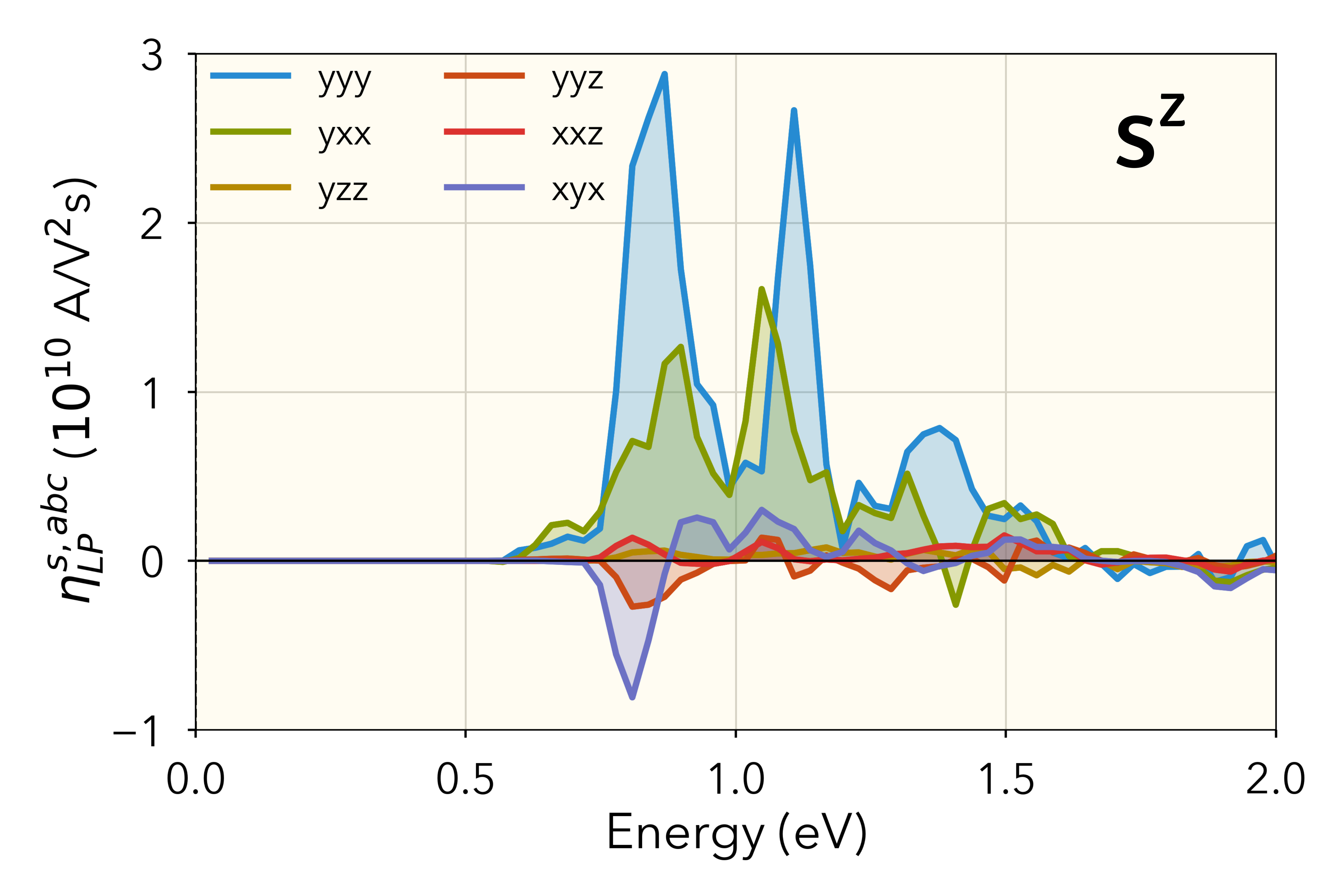}
    \caption{\justifying LP spin injection photoconductivity components for the $n=4$ cycloid structure and for the spin projection $s_z$.}
    \label{spincurrent}
\end{figure}

Consistent with this microscopic picture, the calculations give a large spin-$z$ injection conductivity components $\eta^{z,yyy}_\mathrm{LP}$ and $\eta^{z,yxx}_\mathrm{LP}$, which are shown in Fig. \ref{spincurrent}. These exceed by approximately an order of magnitude the next largest spin conductivities, $\eta^{x,xxx}_\mathrm{LP}$ and $\eta^{y,yyy}_\mathrm{LP}$ (reported in Fig. S14 of the SI), which correspond to spin-$x$ and spin-$y$ currents, respectively, and originate from the symmetry-allowed SOC-related spin texture. The nonrelativistic origin of the dominant spin-$z$ response is further confirmed by symmetry analysis: its components remain allowed under spin-group symmetries, whereas the spin-$x$ and spin-$y$ photoconductivities are permitted only by the MSG (see Sec. S2 of the SI).
Overall, these results demonstrate that the nonrelativistic $p$-wave magnetic state enables highly efficient optical generation of pure spin currents, exceeding those achievable from conventional relativistic spin splitting.

{\it Spin CPGE.} The CP spin shift current is the only contribution symmetry-allowed for all values of the magnetic period $n$. The photoconductivity tensor components are reported in the Sec. S10 of the SI. A clear hierarchy among the spin channels emerges: the spin-$z$ response, the only one allowed by spin-group symmetries, dominates, with $\sigma_\mathrm{CP}^{z,xyx}$ reaching approximately 40~$\mu$A/V$^2$, whereas the spin-$x$ and spin-$y$ conductivities are slightly smaller. This again reflects the underlying $p$-wave magnetic state.

Notably, the direction of the resulting spin current is determined not by the momentum direction associated with the $p$-wave spin splitting. The odd-parity spin selectivity associated with $k_y\!\leftrightarrow\!-k_y$ can generate a dominant spin-$z$ shift current flowing along the polar axis $x$. Thus, under CP light, the nonrelativistic $p$-wave magnetism is related to pure spin-$z$ currents whose flow direction is orthogonal to that induced under LP excitation (compare Fig. \ref{cartoon}c and d) .

\section{Discussion and conclusion}

In this work, we have presented a comprehensive first-principles investigation of charge and spin photogalvanic effects in the $p$-wave magnet NiI$_2$. We have shown that inversion symmetry breaking induced by the spin-spiral order activates a large LP shift current, despite the improper nature of the ferroelectricity, thereby establishing an efficient route to bulk photovoltaic responses in otherwise centrosymmetric crystals. Moreover, we demonstrated that the CP injection current provides a direct microscopic probe of the nonrelativistic $p$-wave spin splitting.

Our calculations reproduce the experimentally observed directional selectivity of the photocurrent response experimentally observed in Ref. [\onlinecite{Song22}]. In the device geometry with two orthogonal electrode pairs — one along the ferroelectric polarization and the other along the $\mathbf{Q}$ direction — the dominant LPGE signal is detected along the polarization axis, whereas the current from the CPGE component is measured along the spin-spiral propagation direction. 
This directional contrast directly reflects the distinct microscopic mechanisms identified in our model systems: the LPGE is dominated by the shift response permitted along the polar $x$ axis, while the CPGE is dominated by injection processes arising from the $p$-wave spin texture that generate current along the $y$ direction.

This analysis, while carried out for cycloidal configurations, remains valid more generally for magnetic structures that preserve the essential symmetry properties of the bulk. For instance, for a proper-screw spiral—where the $p$-wave polarization develops in the spin-$y$ component—the dominant LPGE contributions remain unchanged. For the CPGE, the underlying mechanism, governed by the $p$-wave spin texture, also remains the same, yielding a finite current along $y$; however, this requires an electric-field component perpendicular to the layers, reflecting the change in the spin-polarization direction.

Experimentally, the CPGE reaches its maximum at a photon energy of $\sim 1.8$ eV (680 nm). In our calculations, the corresponding maximum appears at approximately 1.3 eV. The quantitative offset is consistent with the known band-gap underestimation of standard DFT and does not affect the qualitative agreement.

Beyond the charge photocurrent measured experimentally, our calculations further predict sizable spin photocurrents with a polarization dependence complementary to that of the charge response. For LP light with in-plane electric-field components, the charge current flows along the ferroelectric polarization axis, while the dominant pure spin-$z$ current is generated along the spin-spiral propagation direction $\mathbf{Q}$ [Fig. \ref{cartoon}(a)]. For CP light, the roles are reversed: the charge current flows along $\mathbf{Q}$, whereas a pure spin-$z$ current is generated parallel to the polarization axis [Fig. \ref{cartoon}(b)]. Thus, the two intrinsic symmetry directions of the system are exchanged between charge and spin channels under linear versus circular excitation. These predicted spin currents could be detected via spin-to-charge conversion in an adjacent heavy metal or transition-metal dichalcogenide layer through the inverse spin Hall effect \cite{RevModPhys.87.1213}.

Overall, our findings establish spin-spiral $p$-wave magnets as a promising platform for optically driven charge and spin transport. The coexistence of relativistic and nonrelativistic spin-splitting mechanisms within the same material enables a controlled disentanglement of distinct symmetry-driven contributions to nonlinear transport. The ability to generate helicity-controlled pure spin currents without accompanying charge flow may open new avenues for ultrafast, contact-free spin injection and for the optical manipulation of unconventional magnetic order.

\section{Acknowledgments}
The authors are grateful to Qian Song and Riccardo Comin for useful discussions. S.S. acknowledges financial support from the Vin\v{c}a Institute, provided by the Ministry of Science, Technological Development and Innovation of the Republic of Serbia through the contract No. 451-03-33/2026-03/200017.
G.C. , S.S., P.B., A.D. and S.P. acknowledge support from the Ministry of Foreign Affairs of Italy and the Ministry of Science, Technological Development, and Innovation of Serbia through the bilateral project ``Van der Waals Heterostructures for Altermagnetic Spintronics'', realized under the executive programme for scientific and technological cooperation between the two countries.
JS and JI-A acknowledge financial support of the European Union's Horizon 2020 research and innovation programme under the European Research Council (ERC) grant agreement No.~946629, and the Spanish Ministry of Science, Innovation and Universities (MICIU) under the ``Proyectos de Generación de Conocimiento'' grant No.~PID2023-147324NA-I00.
Computational resources and support were
provided by CINECA under the ISCRA IsB28 HEXTIM, IsCc2 SFERA, and IsCc9 BRIMS projects.

\clearpage
\onecolumngrid

\section*{Supporting Information}
\addcontentsline{toc}{section}{Supporting Information}

\setcounter{section}{0}
\setcounter{figure}{0}
\setcounter{table}{0}

\renewcommand{\thesection}{S\arabic{section}}
\renewcommand{\thefigure}{S\arabic{figure}}
\renewcommand{\thetable}{S\arabic{table}}

\section{Computational details}

Density functional theory (DFT) calculations are performed using the Vienna Ab initio Simulation Package (VASP) \cite{Kresse93,Kresse96,Kresse96b}. The exchange–correlation functional is treated within the generalized gradient approximation (GGA) using the Perdew–Burke–Ernzerhof (PBE) formulation \cite{Perdew96}.

The NiI$_2$ monolayer is modeled with an in-plane lattice constant of $a = 3.97$ {\AA} and a Ni–I bond length of 2.746 {\AA}. These parameters are the same already used in a previous work \cite{Song25}.  A vacuum spacing of 25 {\AA} is introduced along the $c$-axis (out-of-plane direction) to eliminate spurious interactions between periodic images \cite{Song25}.

A plane-wave energy cutoff of 500 eV is employed. Brillouin-zone integrations for the self-consistent calculations are carried out using a $\Gamma$-centered $17 \times 10 \times 1$ $k$-point mesh. Spin–orbit coupling (SOC) is included in all calculations.

Photocurrents are computed following the Wannier-based approach introduced in Refs.~\cite{Azpiroz18,Puente23,Lihm22}. The DFT band structure is interpolated using the Wannier90 code \cite{Marzari97,Mostofi08,Pizzi2020}. The Wannier basis is constructed from Ni $s$ and $d$ orbitals and I $p$ orbitals, resulting in a total of 72 spinor bands in the presence of SOC.

The photocurrent calculations are performed on a dense $100 \times 100 \times 1$ $k$-point grid. A constant broadening of 0.02 eV is applied to account for van Hove singularities characteristic of low-dimensional systems, following the approach of Ref.~\cite{Azpiroz18}. Convergence with respect to the $k$-point sampling is carefully verified.

The photoconductivity tensor of the two-dimensional system is obtained by rescaling the calculated three-dimensional response to a single active layer, as commonly done in first-principles studies of low-dimensional materials \cite{Azpiroz18,Rangel17,Tiwari22,Cuono25}. Specifically, the monolayer photoconductivity is given by $\sigma^{abc}_{ML} = \frac{c}{w_{z}} \sigma^{abc}$, where $w_{z} = 3.03$ {\AA} represents the effective thickness of the monolayer along the out-of-plane direction.

The spin current has the dimensions of energy. Therefore, to facilitate comparison with charge photoconductivity, the spin photoconductivity is rescaled by a factor $2e/\hbar$, thereby expressing it in the same units.

\newpage

\section{Symmetry analysis}\label{sec.symmetry}
We review here the symmetry constraints on spin polarization and the symmetry-adapted form of the photoconductivity tensors. Following Ref.~\cite{Etxebarria.stensor.25}, we distinguish between relativistic and nonrelativistic effects by comparing the symmetry constraints imposed by magnetic point groups (MPGs) with those imposed by spin point groups (SpPGs). The latter comprise symmetry operations that act independently on spin and real space, and thus describe the symmetry properties of systems in the absence of SOC, which otherwise entangles spin, orbital, and lattice degrees of freedom. 

Even if vanishingly small, SOC is always present in real materials, and SpPGs can be regarded as approximate symmetries that, under certain conditions, act as supergroups of the MSG of the magnetic structure. As a result, SpPGs impose additional constraints on tensor properties, reducing the number of nonzero components in their symmetry-adapted form and isolating those of nonrelativistic origin \cite{Sivianes25}. Heuristically, contributions arising solely from SOC are therefore expected to be comparatively weak.

To exploit the group–subgroup relation between the MPG and the SpPG, one must fix the global orientation of the spin system with respect to the lattice \cite{Etxebarria.stensor.25}. All spin-spiral phases considered here are coplanar structures, whose global orientation can be specified by a unit vector $\hat{n}$ perpendicular to the spin-spiral plane. Using the Cartesian reference frame introduced in the main text, with the magnetic wavevector parallel to $y$, the in-plane cycloid has $\hat{n} \parallel z$, whereas a proper-screw spiral has $\hat{n} \parallel y$. These configurations share the same MPG symmetries but differ in their SpPG. In contrast, a helimagnetic modulation propagating along $y$ direction with magnetic moments rotating in the $yz$ plane ($\hat{n}_x\parallel x$) would belong to a different Type-IV MSG, namely $C_cc$, with MPG $m1^\prime$.

\subsection{Symmetry Constraints on the Spin Polarization of Electronic Bands} 
We adopt the same tensorial framework recently discussed in Ref.~\cite{radaelli.25}, and express a general spin texture as
\begin{equation}
s_i(\mathbf{k}) = T^{(N)}_{i,\alpha\beta\gamma\ldots}, k_\alpha k_\beta k_\gamma \ldots ,
\end{equation}
where summation over repeated indices is implied, and $T^{(n)}_{i,\alpha\beta\gamma\ldots}$ is an axial tensor of rank $N+1$, symmetric in the Greek indices and time-reversal (TR) odd for even $N$.

For the commensurate magnetic structure of bulk NiI$_2$, whose MPG includes TR, the spin polarization depends only on odd powers of $\mathbf{k}$. The allowed linear-in-$k$ terms are listed in Table~\ref{tab:spinpol}. Since symmetry-adapted tensors depend solely on the point-group operations, the same tensorial form applies to both bulk and single-layer spin spirals.

The key difference in the single-layer case compared to the bulk is the absence of TR symmetry in the MPG of helimagnetic phases with odd-$n$ magnetic periodicity. Because of that, terms involving even powers of $\mathbf{k}$, being TR-odd, become symmetry-allowed, consistently with the emergence of a finite net magnetic moment, as discussed in the main text for the $n=3$ cycloid system.

\begin{table}[ht]
\begin{tabular}{|c|c|c|c|c|}
\hline
spin-spiral & unitary PG &\multicolumn{3}{c|}{{Spin components $S_i(\bm k)$}}\\
\cline{3-5}
plane & element 		 & $s_x$			& $s_y$				& $s_z$\\
\hline
$\hat{n}_z$ 	&\multirow{2}*{$C_2$}& $T_{x,x}$		& $T_{y,y}$\; $(T_{y,z})^*$   	& \tc{red}{$T_{z,y}$}\; \tc{red}{$(T_{z,z})^*$}\\
\cline{1-1}\cline{3-5}
$\hat{n}_y$ 	& &$T_{x,x}$		& \tc{red}{$T_{y,y}$}\; \tc{red}{$(T_{y,z})^*$}   	& $T_{z,y}$\; $(T_{z,z})^*$\\
\hline
$\hat{n}_x$ 	& $m$ &\tc{red}{$T_{x,y}$}\; \tc{red}{$(T_{x,z})^*$} & $T_{y,x}$ & $T_{z,x}$ \\
\hline
\end{tabular}
\caption{\justifying Allowed linear-in-$\mathbf{k}$ terms of the band spin polarization for (commensurate) bulk and single-layer NiI$_2$. We distinguish between the symmetry-adapted forms derived from point groups with a $C_2$ axis (first two rows, relevant for the bulk helix and single-layer in-plane cycloid and proper-screw spiral) and those from point groups with a vertical mirror plane $m$ (relevant for a cycloid with a spin-spiral plane perpendicular to the NiI$_2$ layer and containing the magnetic wavevector). Red-colored terms indicate contributions allowed by coplanar spin groups, while all listed terms are present when SOC is included and MPG symmetries apply. Terms relevant only to the bulk case are indicated in parentheses $()^*$.}\label{tab:spinpol}
\end{table}

From inspection of Table~\ref{tab:spinpol}, one immediately observes that a nonrelativistic spin texture is allowed for all orientations of the spin-spiral plane (red-colored terms). Moreover, the nonrelativistic spin polarization develops exclusively along the direction perpendicular to the spin-spiral plane and only for crystal momenta parallel to the magnetic wavevector. 
Specifically, for the systems considered in this work, this implies that the spin polarization is oriented along $z$ for the cycloidal structure and along $y$ for the proper-screw structure, consistent with the DFT results presented in the main text and in Sec.~\ref{sec.screw}, respectively. 
This spin polarization is odd under $k_y \mapsto -k_y$, thereby defining a unique nodal line at $k_y = 0$ and identifying a $p$-wave magnetic phase.

\subsection{Symmetry-Adapted Forms of the Charge and Spin Photoconductivity Tensors}
As discussed in the main text, each photoconductivity can be decomposed into four tensors of the same rank [see Eq. (1)] but with distinct intrinsic symmetry properties. Among these, two are TR-even and generally symmetry-allowed, while the other two are TR-odd and therefore vanish identically in the (commensurate) helimagnetic phase of bulk NiI$_2$, as well as in single-layer spirals with even periodicity.

{\it TR-even charge photoconductivity tensors }describe charge linear shift/circular injection currents and spin circular shift/linear injection currents. Their symmetry-allowed form is the same for all considered systems, since the corresponding MPGs differ only by the presence or absence of TR symmetry, which is absent in single-layer spirals with odd periodicity. Using the same Cartesian reference frame introduced in the main text, the independent component of the charge photoconductivity tensor are
\begin{eqnarray}
\mbox{LP charge shift} &:&\quad \sigma^{xyz}_\mathrm{LP},\; \sigma^{xxx}_\mathrm{LP},\; \sigma^{xyy}_\mathrm{LP},\; \sigma^{xzz}_\mathrm{LP},\; \sigma^{yxy}_\mathrm{LP},\; \sigma^{yxz}_\mathrm{LP}\,\quad (\sigma^{zxz}_\mathrm{LP}, \sigma^{zxy}_\mathrm{LP})^*\\
\mbox{CP charge injection} &:&\quad \eta^{yxy}_\mathrm{CP},\;\eta^{yzx}_\mathrm{CP},\;\eta^{xyz}_\mathrm{CP}\quad (\eta^{zxy}_\mathrm{CP},\;\eta^{zxz}_\mathrm{CP})^*.
\end{eqnarray}
All other nonvanishing components follow from intrinsic symmetry or antisymmetry, namely $\sigma^{abc}_\mathrm{LP} = \sigma^{acb}_\mathrm{LP}$ and $\eta^{abc}_\mathrm{CP} = -\eta^{acb}_\mathrm{CP}$. The coefficients listed in parentheses $()^*$ are relevant only for bulk NiI$_2$ as they imply an out-of-plane current.

We note that SpPGs do not impose additional constraints, implying that TR-even charge photoconductivity tensors can exist even in the absence of SOC. However, as discussed in the main text, the CP injection photoconductivity is sensitive to the underlying spin texture and reflects the spin–momentum couplings $k_xs^x$, $k_ys^y$, and $k_ys^z$, consistent with the symmetry-adapted form of the spin polarization given in Table~\ref{tab:spinpol}. Among these, only $k_xs^x$ has a purely relativistic origin for both orientations of the spin-spiral plane compatible with $C_2$

For the in-plane cycloid, the nonrelativistic $p$-wave spin texture corresponds to $k_ys^z$, implying that $\eta^{yxy}_\mathrm{CP}$ is predominantly nonrelativistic in origin, consistent with the results of the DFT calculations in Sec \ref{sec.interband}. In contrast, when the spin-spiral plane lies in the $xz$ plane (as in the single-layer proper-screw spiral), the nonrelativistic contribution is expected for the $\eta^{yzx}_\mathrm{CP}$ component, associated with the spin–momentum coupling $k_ys^y$ and a nonrelativistic spin polarization along the $y$ direction, as further discussed in Sec. \ref{sec.screw_conductivity}.

{\it TR-even spin photoconductivity tensors } have independent components
\begin{eqnarray}
\mbox{CP spin shift} &:&\quad \sigma^{x,yxy}_\mathrm{CP},\; \sigma^{x,yxz}_\mathrm{CP},\; \sigma^{x,xyz}_\mathrm{CP}\quad (\sigma^{x,zxy}_\mathrm{CP},\;\sigma^{x,zxz}_\mathrm{CP})^*\nonumber\\
&&\quad \sigma^{y,yyz}_\mathrm{CP},\; \sigma^{y,xxy}_\mathrm{CP},\; \sigma^{y,xxz}_\mathrm{CP}\quad (\sigma^{y,zyz}_\mathrm{CP})^*\nonumber\\
&&\quad \sigma^{z,yyz}_\mathrm{CP},\; \sigma^{z,xxy}_\mathrm{CP},\; \sigma^{z,xxz}_\mathrm{CP}\quad (\sigma^{z,zyz}_\mathrm{CP})^*\\
\mbox{LP spin injection} &:&\quad \eta^{x,yxz}_\mathrm{LP},\; \eta^{x,xyy}_\mathrm{LP},\; \eta^{x,xxx}_\mathrm{LP},\; \eta^{x,xzz}_\mathrm{LP},\; \eta^{x,yyz}_\mathrm{LP},\; \eta^{x,xyz}_\mathrm{LP}\quad (\eta^{x,zxz}_\mathrm{LP},\;\eta^{x,zxy}_\mathrm{LP})^*\nonumber\\
&&\quad \eta^{y,yyy}_\mathrm{LP},\; \eta^{y,yxx}_\mathrm{LP},\; \eta^{y,yzz}_\mathrm{LP},\; \eta^{y,yyz}_\mathrm{LP},\; \eta^{y,xxz}_\mathrm{LP},\; \eta^{y,xyx}_\mathrm{LP}\quad (\eta^{y,zyy}_\mathrm{LP},\;\eta^{y,zxx}_\mathrm{LP},\;\eta^{y,zzz}_\mathrm{LP},\;\eta^{y,zyz}_\mathrm{LP})^*\nonumber\\
&&\quad \eta^{z,yyy}_\mathrm{LP},\; \eta^{z,yxx}_\mathrm{LP},\; \eta^{z,yzz}_\mathrm{LP},\; \eta^{z,yyz}_\mathrm{LP},\; \eta^{z,xxz}_\mathrm{LP},\; \eta^{z,xyx}_\mathrm{LP}\quad (\eta^{z,zyy}_\mathrm{LP},\;\eta^{z,zxx}_\mathrm{LP},\;\eta^{z,zzz}_\mathrm{LP},\;\eta^{z,zyz}_\mathrm{LP})^*\nonumber.\\
\end{eqnarray}
As before, all other nonvanishing components follow from intrinsic symmetry or antiysmmetry, being $\sigma^{s,abc}_\mathrm{CP}=-\sigma^{s,acb}_\mathrm{CP}$ and $\eta^{s,abc}_\mathrm{LP}=\eta^{s,acb}_\mathrm{LP}$. We further note that the coefficients associated with the spin components $s^y$ and $s^z$ exhibit the same spatial dependence, i.e., they share the same index combinations $abc$.

We emphasize that the same symmetry-adapted form applies to all considered systems, as well as to bulk NiI$_2$, for which the coefficients listed in parentheses $()^*$ must also be included. In this case, SpPGs impose an additional constraint, forbidding all components that do not correspond to nonrelativistic spin polarization. Specifically, only $s^z$ components are allowed for in-plane cycloids, while $s^y$ components are the only symmetry-allowed ones for the proper-screw spiral, when the magnetic moments rotate in the $xz$ plane.

{\it TR-odd photoconductivity tensors }are exactly zero according to SpPG. vanish identically under the constraints imposed by SpPGs. Therefore, both CP charge shift currents and LP charge injection currents are purely relativistic effects, activated by SOC in single-layer spirals with odd periodicity. On the other hand, TR-odd spin photoconductivities, within the SpPG framework, exhibit nonzero components only for spin polarizations lying in the spin-spiral plane.

\newpage
\section{Structures}

\begin{figure}[h!]
    \centering
    
    \begin{overpic}[width=3.7cm]{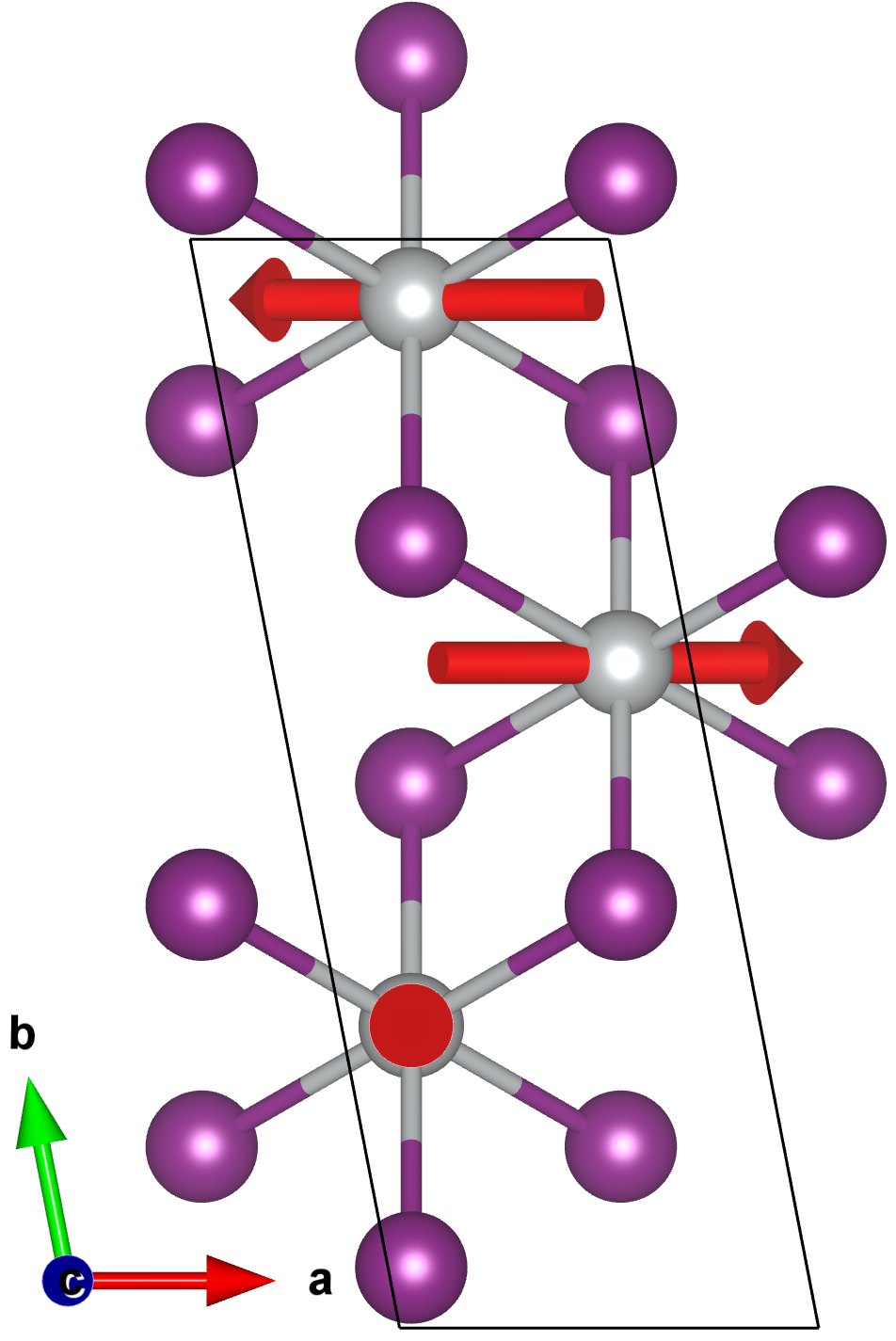}
        \put(1,105){\LARGE \textbf{(a)}}  
    \end{overpic}%
    \hspace{0.3cm} 
    \begin{overpic}[width=3.7cm]{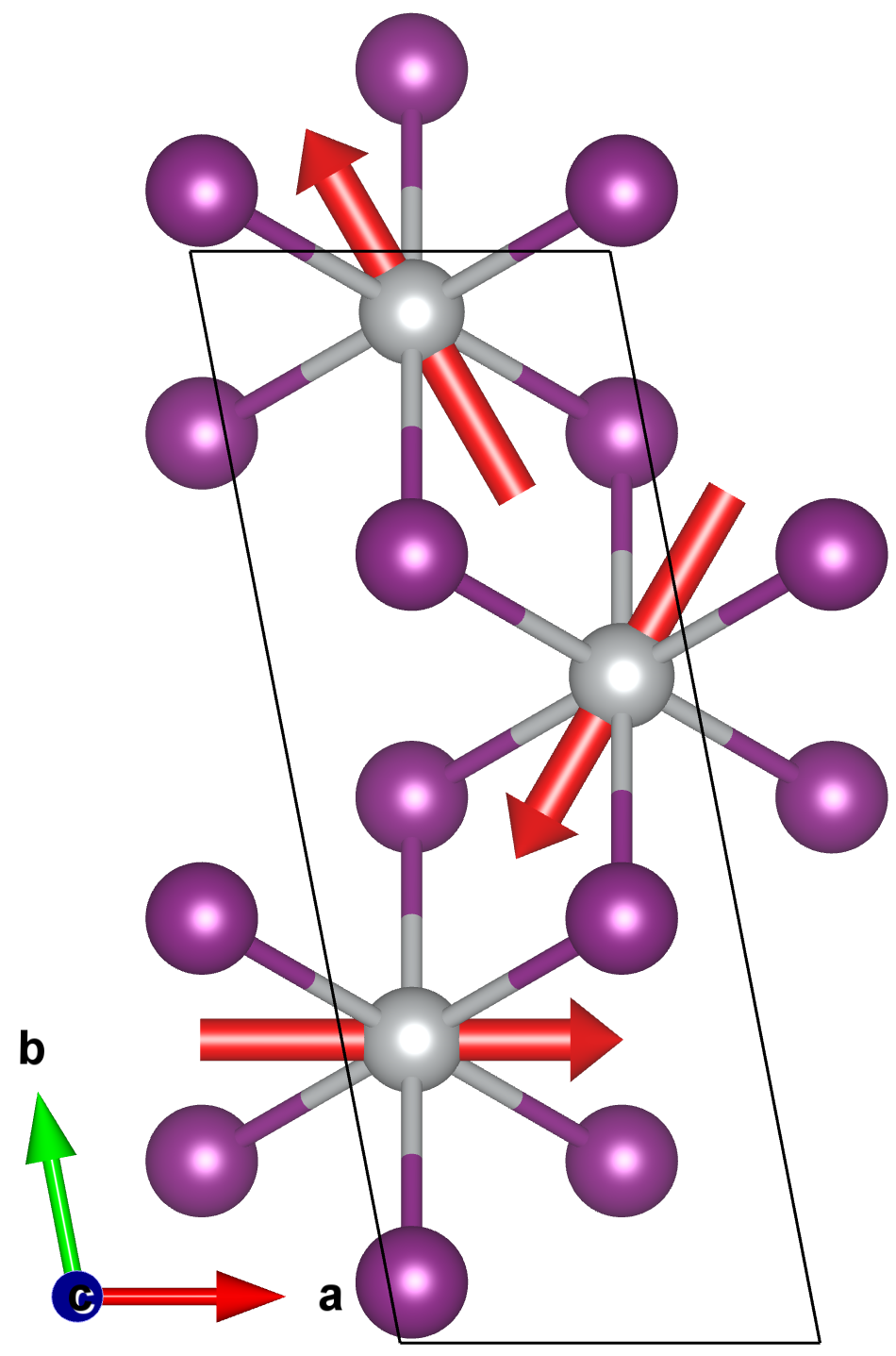}
        \put(1,105){\LARGE \textbf{(b)}}
    \end{overpic}%
    \hspace{0.3cm}
    \begin{overpic}[width=3.7cm]{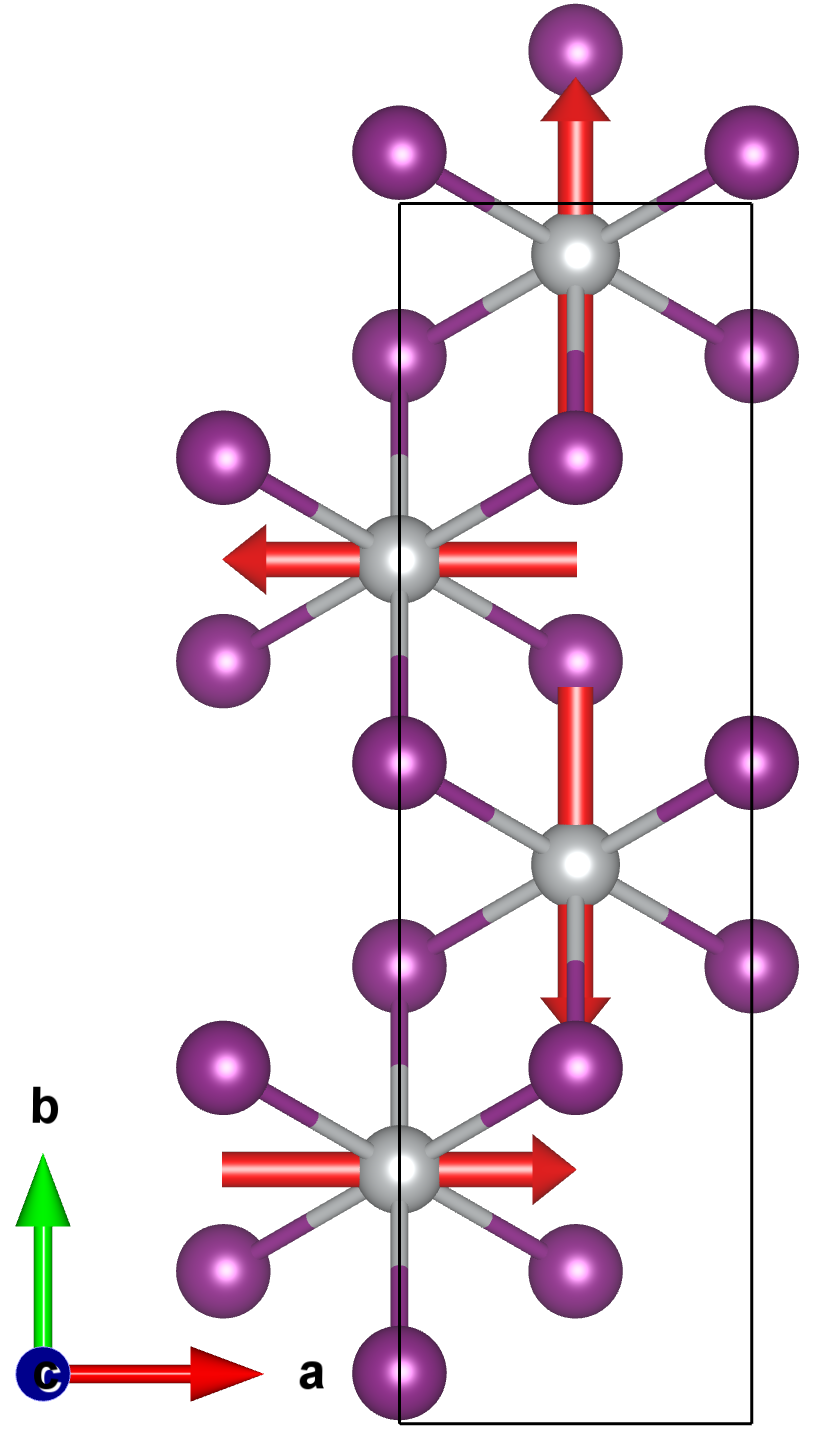}
        \put(1,90){\LARGE \textbf{(c)}}
    \end{overpic}%
    \hspace{0.3cm}
    \begin{overpic}[width=3.7cm]{1x5_cycloid.png}
        \put(1,90){\LARGE \textbf{(d)}}
    \end{overpic}

    \caption{(a) Proper screw spiral structure for $n=3$ and (b) cycloidal structures for $n=3$, (c) $n=4$ and (d) $n=5$.}
    \label{Structures}
\end{figure}

Figure~\ref{Structures} shows the structures considered in the simulations, with spin-helix propagation wavevector $\mathbf{Q}=(0,1/n,0)$. In our models, as in Ref.~\cite{Song25}, the spin helix is decomposed into a cycloidal component, with spins rotating within the layer and a proper-screw component, with spins rotating in a plane perpendicular to the propagation vector.

We consider the proper-screw spiral configuration for $n=3$, shown in Fig.~\ref{Structures}(a), and the cycloidal structures with $n=3$, $4$, and $5$, shown in Figs.~\ref{Structures}(b), (c), and (d), respectively.

\newpage

\section{Band Structure for the proper-screw spin spiral configuration}\label{sec.screw}

\begin{figure}[h!]
    \centering
    \includegraphics[width=7.5cm,angle=0]{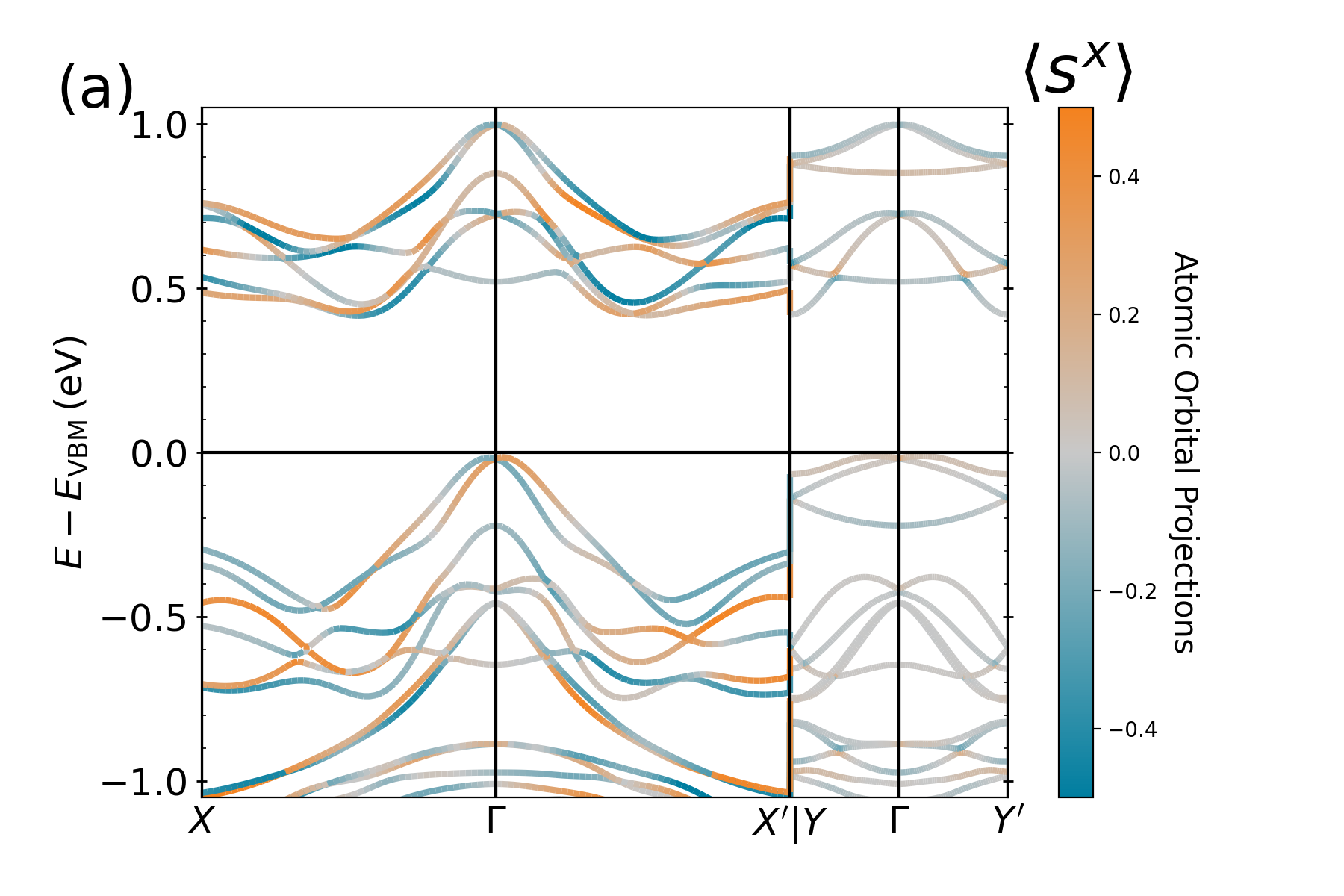}
    \includegraphics[width=7.5cm,angle=0]{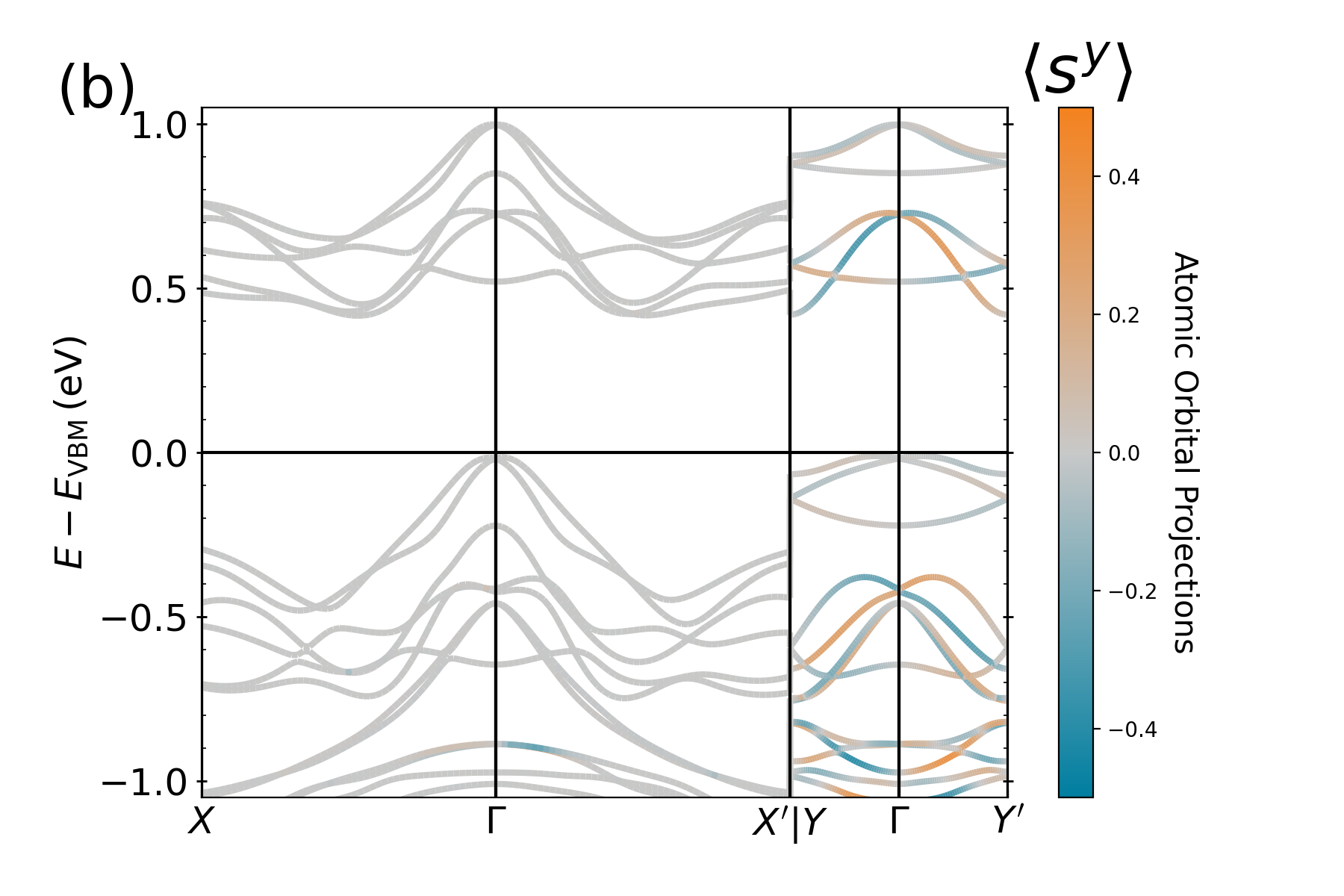}
    \includegraphics[width=7.5cm,angle=0]{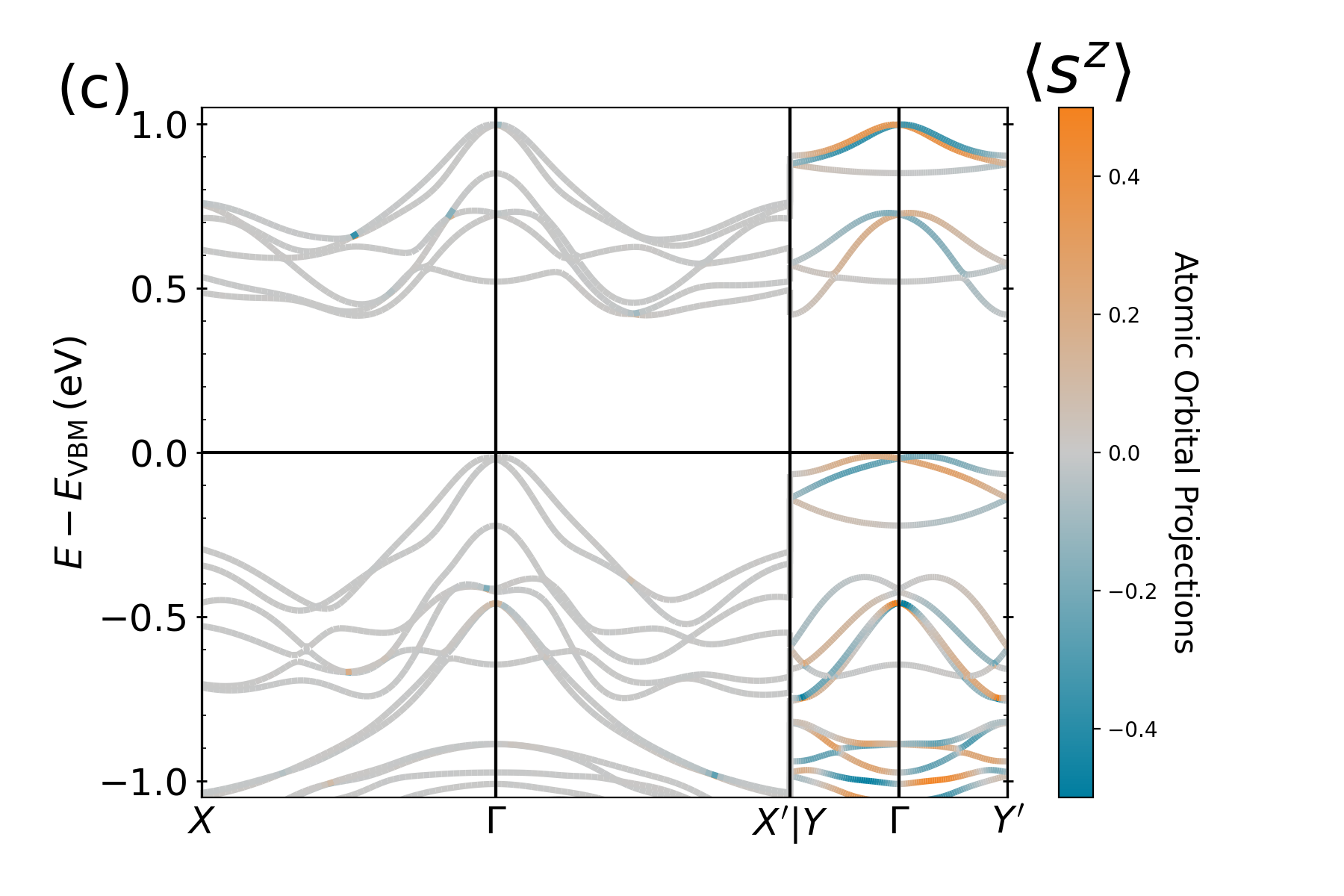}
    \caption{\justifying Band structures for the proper-screw spin spiral structure with $n=3$ magnetic period. The color scale show the expectation values of the spin $x$-, $y$-, and $z$-components.}
    \label{Bandshelical}
\end{figure}

Figure~\ref{Bandshelical} shows the band structure of the proper-screw spin-spiral configuration in Fig.~\ref{Structures}(a), with magnetic period $n=3$. The color scale represents the band spin polarization, $\langle s^i(\mathbf{k}) \rangle$ ($i = x,y,z$). As discussed in Sec.~\ref{sec.symmetry}, in this system the $p$-wave magnetism is associated with the spin-$y$ component, i.e., $\langle s^y(\mathbf{k}) \rangle$ represents the spin polarization of the bands linked to the $p$-wave state.

The spin components $\langle s^x(\mathbf{k}) \rangle$ and $\langle s^z(\mathbf{k}) \rangle$ reflect the spin texture of relativistic origin. Notably, in this system the $p$-wave spin polarization is substantially reduced compared to the cycloidal case, becoming comparable in magnitude to the SOC-induced spin polarizations. This is directly reflected in the CP injection photoconductivity, as discussed in Sec.~\ref{sec.screw_conductivity}.

As in the cycloidal band structure discussed in the main text, the absence of TR symmetry in the MPG for odd periodicity ($n=3$) allows for a nonreciprocal band dispersion along the X--$\Gamma$--X$^{\prime}$ direction, i.e., $\varepsilon_\alpha(k_x) \neq \varepsilon_\alpha(-k_x)$.

\newpage

\section{Wannierized Band Structure}

\begin{figure}[t!]
\centering
\includegraphics[width=6.5cm,angle=270]{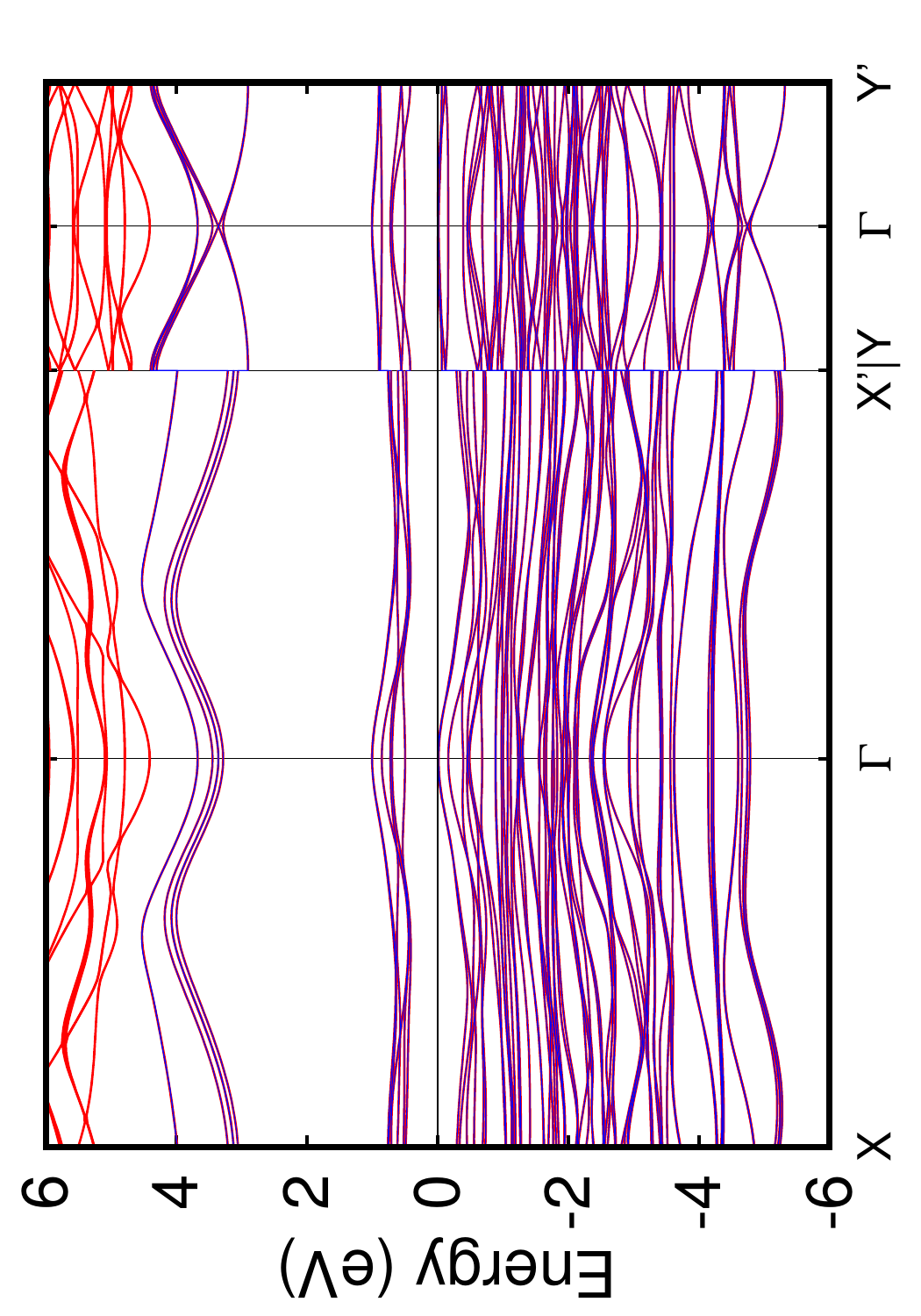}
\caption{\justifying DFT band structure (red) and Wannier-interpolated bands (blue) for the $n=3$ cycloidal structure along the high-symmetry paths X-$\Gamma$-X' and Y-$\Gamma$-Y' in the BZ. The Fermi energy is set to 0 eV.} \label{Wannierization_1x3_cycloid}
\end{figure}

Figure~\ref{Wannierization_1x3_cycloid} shows the DFT band structure of the system with a cycloidal spin configuration of magnetic period $n=3$, together with the Wannier-interpolated bands along the X–$\Gamma$–X$'$ and Y–$\Gamma$–Y$'$ paths in the BZ. The Wannier basis includes Ni $s$ and $d$ orbitals, as well as I $p$ orbitals.

The tight-binding Hamiltonian reproduces the DFT band structure with excellent accuracy over a wide energy window around the Fermi level, extending from $-6$ eV to $4$ eV.

\newpage

\section{Charge photoconductivities for the $n=3$ cycloidal magnetic structure}

This section complements the analysis of the charge photoconductivity for the cycloidal magnetic structure with period $n=3$ by providing additional results that were not included in Fig.~3 of the manuscript.

\subsection{Contributions of Individual Interband Transitions to the CP Charge Injection Photoconductivity}\label{sec.interband}

\begin{figure}[h!]
    \centering
    \includegraphics[width=7.5cm,angle=0]{yxy_total_and_selected_transitions.png}
    \includegraphics[width=7.5cm,angle=0]{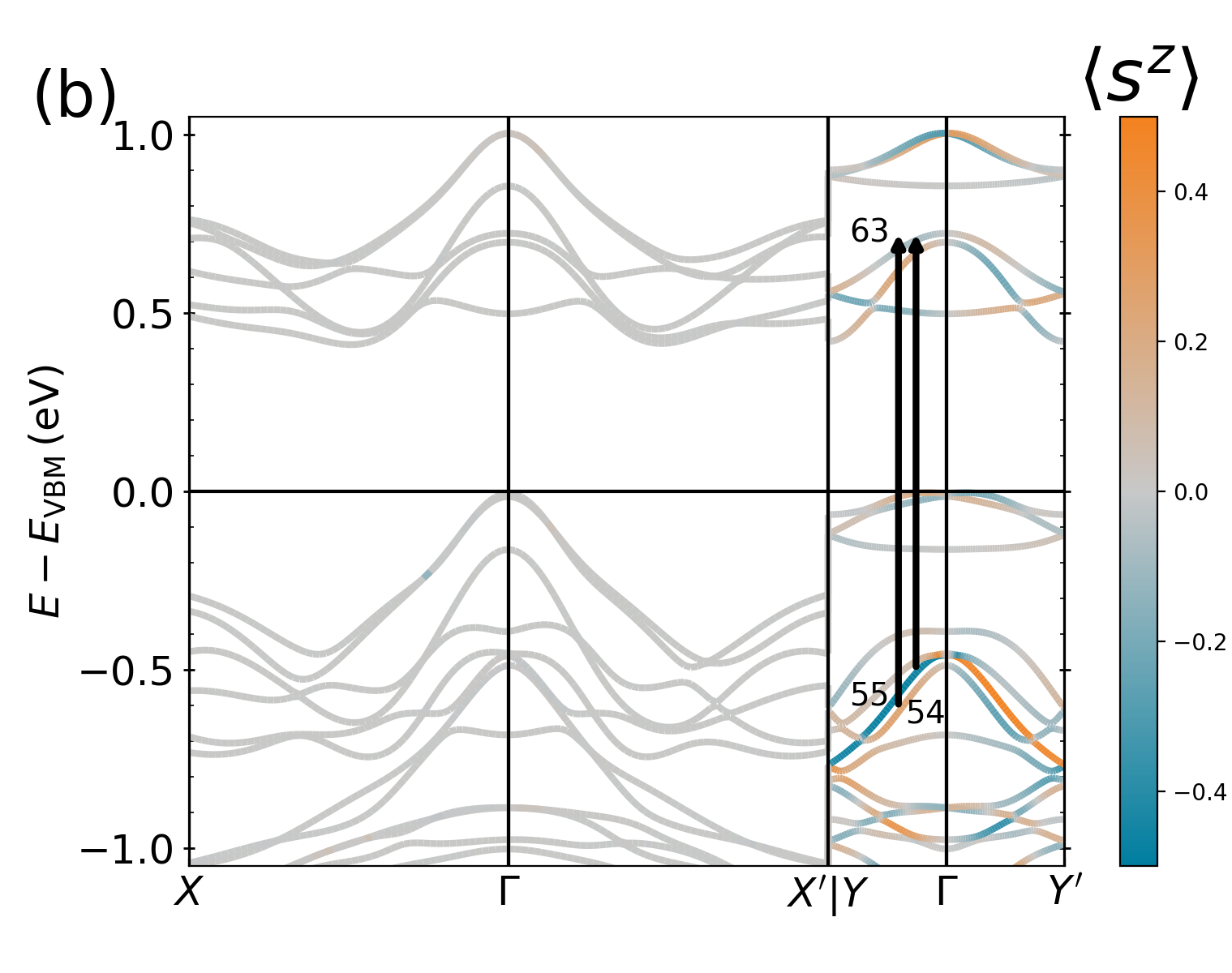}
    \caption{\justifying (a) The total $\eta_\mathrm{CP}^{yxy}$ (black line) is predominantly contributed by transitions from the occupied bands $n=54$ and $55$ to the unoccupied band $m=63$ (orange and blue lines). (b) Band structure and dominant transitions, indicated by black arrows. The color scale represents the band spin polarization $\langle s^z \rangle$.}
    \label{Circinj1x3dominanttransitions}
\end{figure}

Figure~\ref{Circinj1x3dominanttransitions} shows the contributions of different interband electronic transitions to the largest CP charge injection photoconductivity component $\eta_\mathrm{CP}^{yxy}$ for the cycloidal magnetic structure with period $n=3$.

Panel (a) demonstrates that the total $\eta_\mathrm{CP}^{yxy}$ (black line) is predominantly contributed by transitions from the occupied bands $54$ and $55$ to the unoccupied band $63$ (orange and blue lines). These bands are indicated in the panel (b) and exhibit a pronounced $p$-wave magnetic character. 

This provides direct microscopic evidence that the dominant injection response under CP light is intrinsically linked to the underlying nonrelativistic $p$-wave magnetism.

\subsection{CP Charge Shift Photoconductivity}

\begin{figure}[t!]
    \centering
    \includegraphics[width=7.5cm,angle=0]{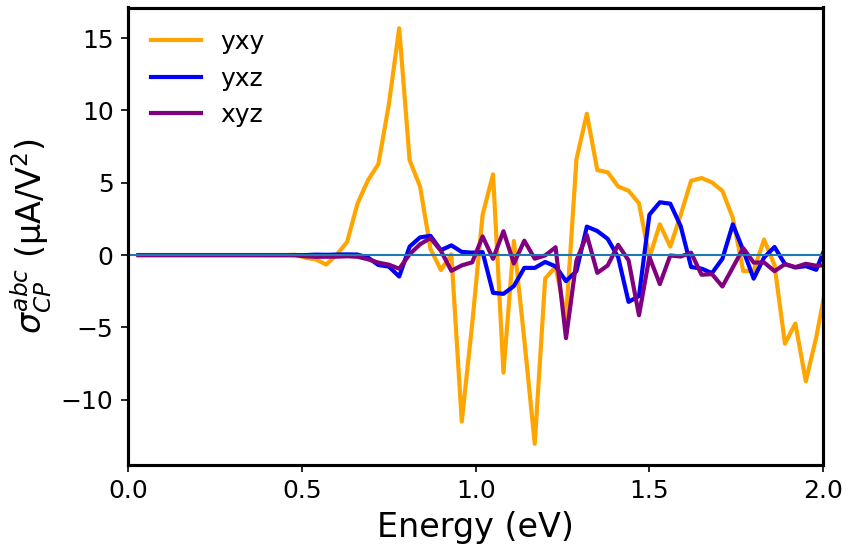}
    \caption{\justifying Non-zero components of the charge CP shift photoconductivity for the $n=3$ cycloid structure.}
    \label{Circshift1x3}
\end{figure}

Figure~\ref{Circshift1x3} displays the charge shift photoconductivity under circularly polarized (CP) light for the cycloidal magnetic structure with period $n=3$. 
The independent components allowed by symmetry are 
\[
\sigma_\mathrm{CP}^{yxy},\;  
\sigma_\mathrm{CP}^{yxz},\;  
\sigma_\mathrm{CP}^{xyz}.
\]
All other non-vanishing components follow from permutation symmetry of the last two indices, 
$\sigma^{abc} = -\sigma^{acb}$. Quantitatively, we observe that even the largest component $\sigma_\mathrm{CP}^{yxy}$ (orange line) reaches a peak value lower than that of the shift photoconductivity under LP light, suggesting that the CP charge shift current is relatively small.

\newpage

\section{Charge Photoconductivities for the $n=4$ Cycloidal Magnetic Structure}

This section analyzes the charge photoconductivities of the cycloidal magnetic structure with period $n=4$. 
The MSG is $P_{b}2$ (No.~3.4). In our supercell construction, the $x$ axis is chosen parallel to the $C_2$ rotational axis and therefore coincides with the direction of the spontaneous electric polarization.

In contrast to the $n=3$ case, for $n=4$ only the LP shift current and the CP injection current are symmetry allowed, as discussed in Sec. \ref{sec.symmetry}.

\begin{figure}[h!]
    \centering
    \includegraphics[width=7.5cm,angle=0]{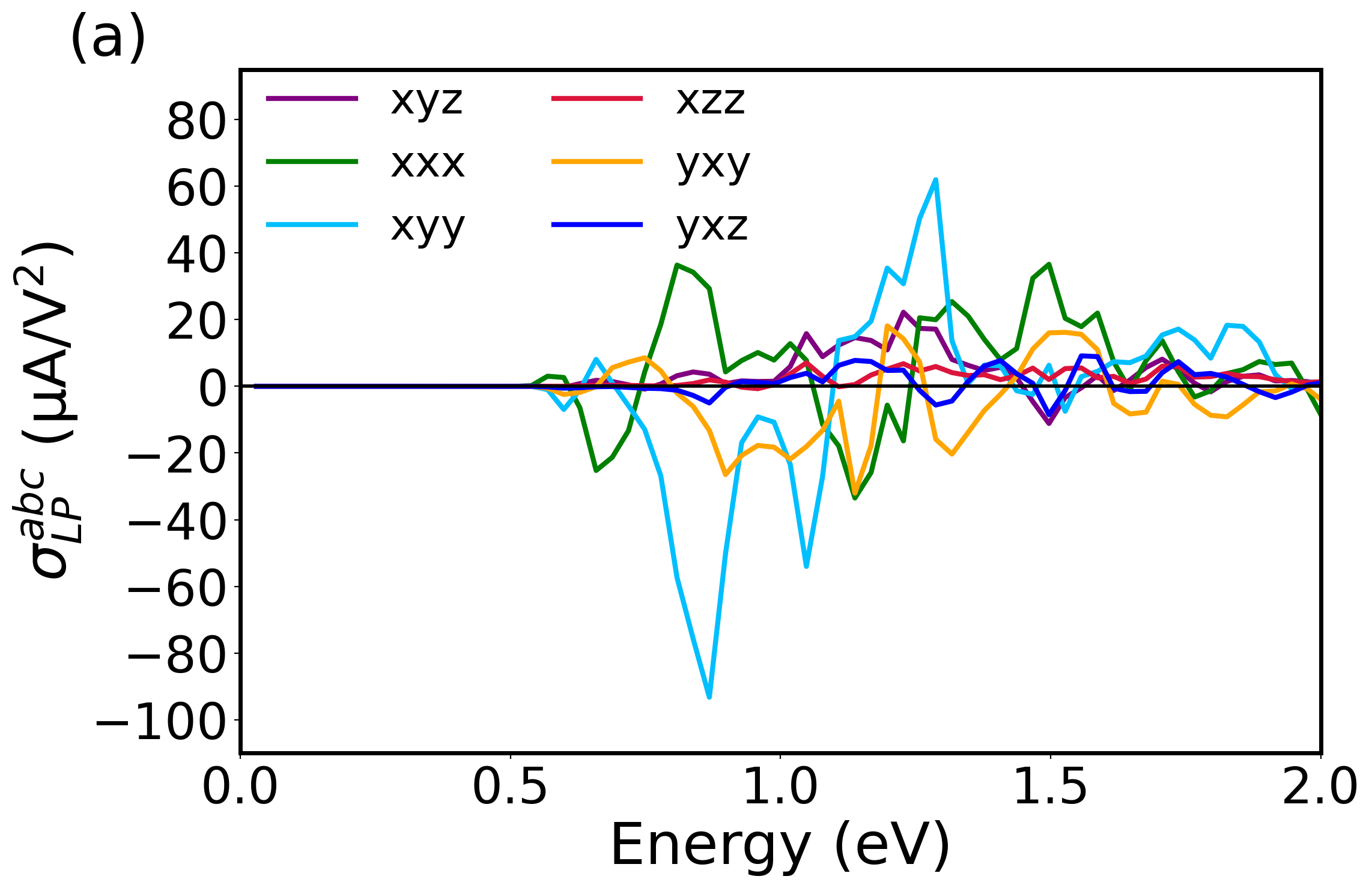}
    \includegraphics[width=7.5cm,angle=0]{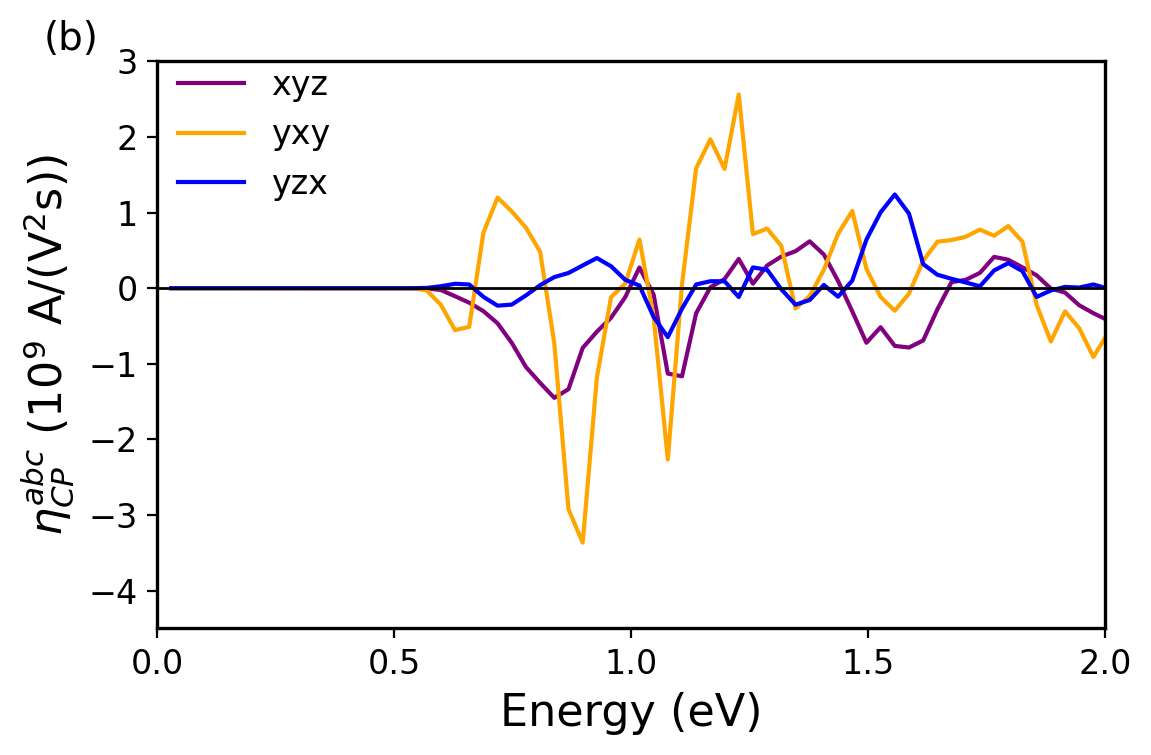}
    \caption{Non-zero components of the (a) charge LP shift photoconductivity and (b) CP injection photoconductivity tensors for the $n=4$ cycloid structure.}
    \label{Chargephotocurrents1x4}
\end{figure}

\subsection{LP Charge Shift Photoconductivity}

The shift photoconductivity tensor of the $n=4$ system under LP light exhibits the same six independent nonzero components as in the $n=3$ case:
\[
\sigma^{xyz}_\mathrm{LP},\;
\sigma^{xxx}_\mathrm{LP},\;
\sigma^{xyy}_\mathrm{LP},\;
\sigma^{xzz}_\mathrm{LP},\;
\sigma^{yxy}_\mathrm{LP},\;
\sigma^{yxz}_\mathrm{LP},
\]
in agreement with the symmetry analysis of Sec. \ref{sec.symmetry}. Owing to the intrinsic symmetry of the shift tensor under exchange of its last two indices, $\sigma^{abc}_\mathrm{LP} = \sigma^{acb}_\mathrm{LP}$, all remaining nonvanishing components follow from permutation of the polarization indices. 

The spectra are shown in Fig.~\ref{Chargephotocurrents1x4}(a). As in the $n=3$ case, the components $\sigma^{xxx}_\mathrm{LP}$ and $\sigma^{xyy}_\mathrm{LP}$ dominate the response. These correspond to a photocurrent flowing along $x$ (i.e., parallel to the polarization direction) driven by linearly polarized light with electric field components along $x$ and $y$, respectively. The component $\sigma^{yxy}_\mathrm{LP}$ also provides a sizable contribution.

\subsection{CP Charge Injection Photoconductivity}

The CP injection photoconductivity exhibits three independent nonzero components:
\[
\eta_\mathrm{CP}^{xyz},\;
\eta_\mathrm{CP}^{yxy},\;
\eta_\mathrm{CP}^{yzx},
\]
in agreement with the symmetry analysis of Sec. \ref{sec.symmetry}. All other nonvanishing components follow from antisymmetry under exchange of the last two indices, $\eta^{abc}_\mathrm{CP} = -\eta^{acb}_\mathrm{CP}$.

The spectra are shown in Fig.~\ref{Chargephotocurrents1x4}(b). As in the $n=3$ case, the component $\eta_\mathrm{CP}^{yxy}$ dominates the response, reaching peak absolute values of approximately $3 \times 10^{9}$~A/V$^{2}$s. An analysis analogous to that presented in Sec.~\ref{sec.interband} indicates that the main spectral peaks originate from interband transitions between bands exhibiting a pronounced $p$-wave magnetic texture.

\newpage

\section{Charge Photoconductivities for the $n=5$ Cycloidal Magnetic Structure}

The cycloidal magnetic structure with period $n=5$ exhibits shift and injection currents under both LP and CP light, as explained in Sec. \ref{sec.symmetry}. 
The allowed components of the corresponding photoconductivity tensors are the same as in the $n=3$ case, since the MSG is identical.

\begin{figure}[h!]
    \centering
    \includegraphics[width=5.5cm,angle=270]{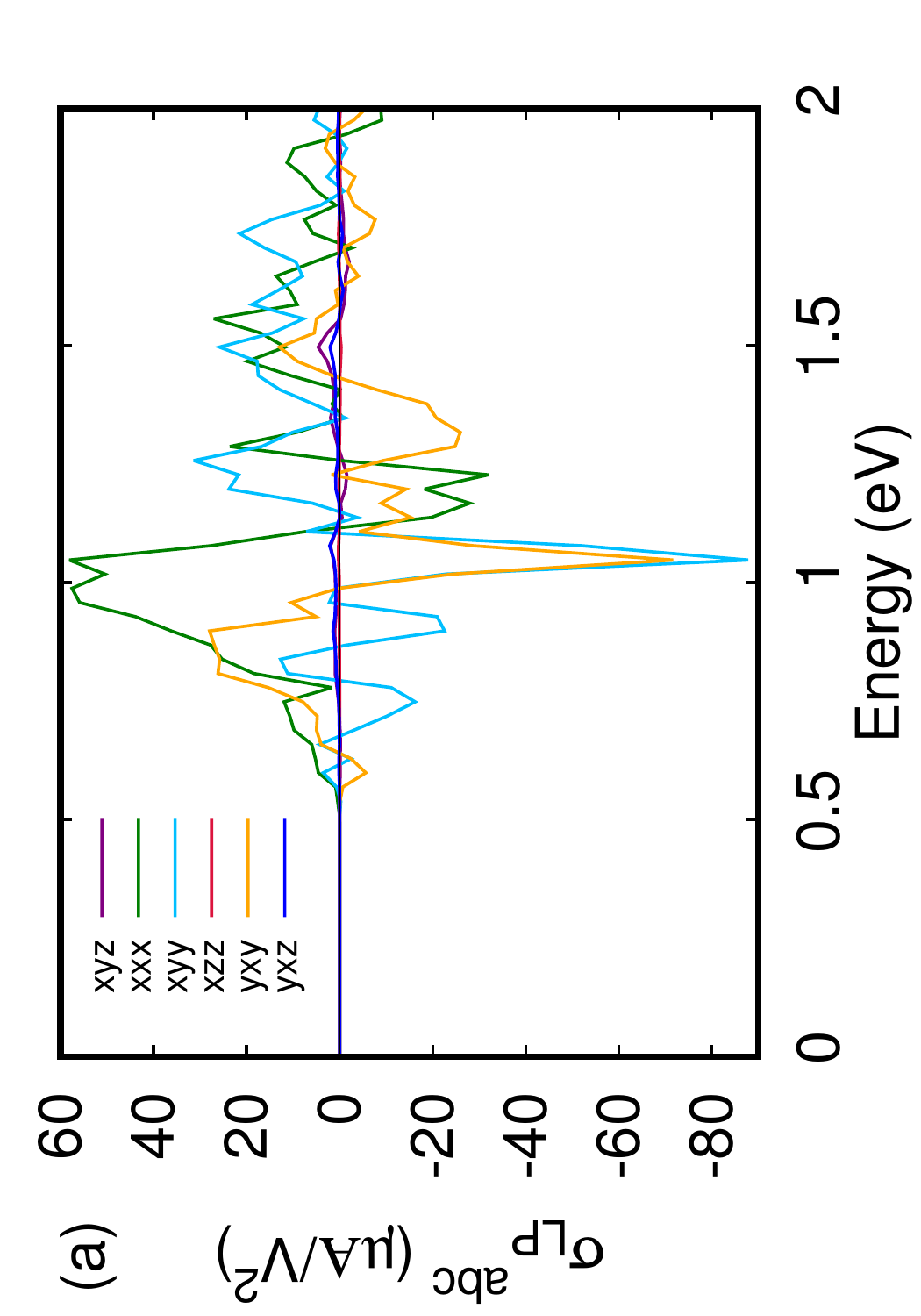}
    \includegraphics[width=5.5cm,angle=270]{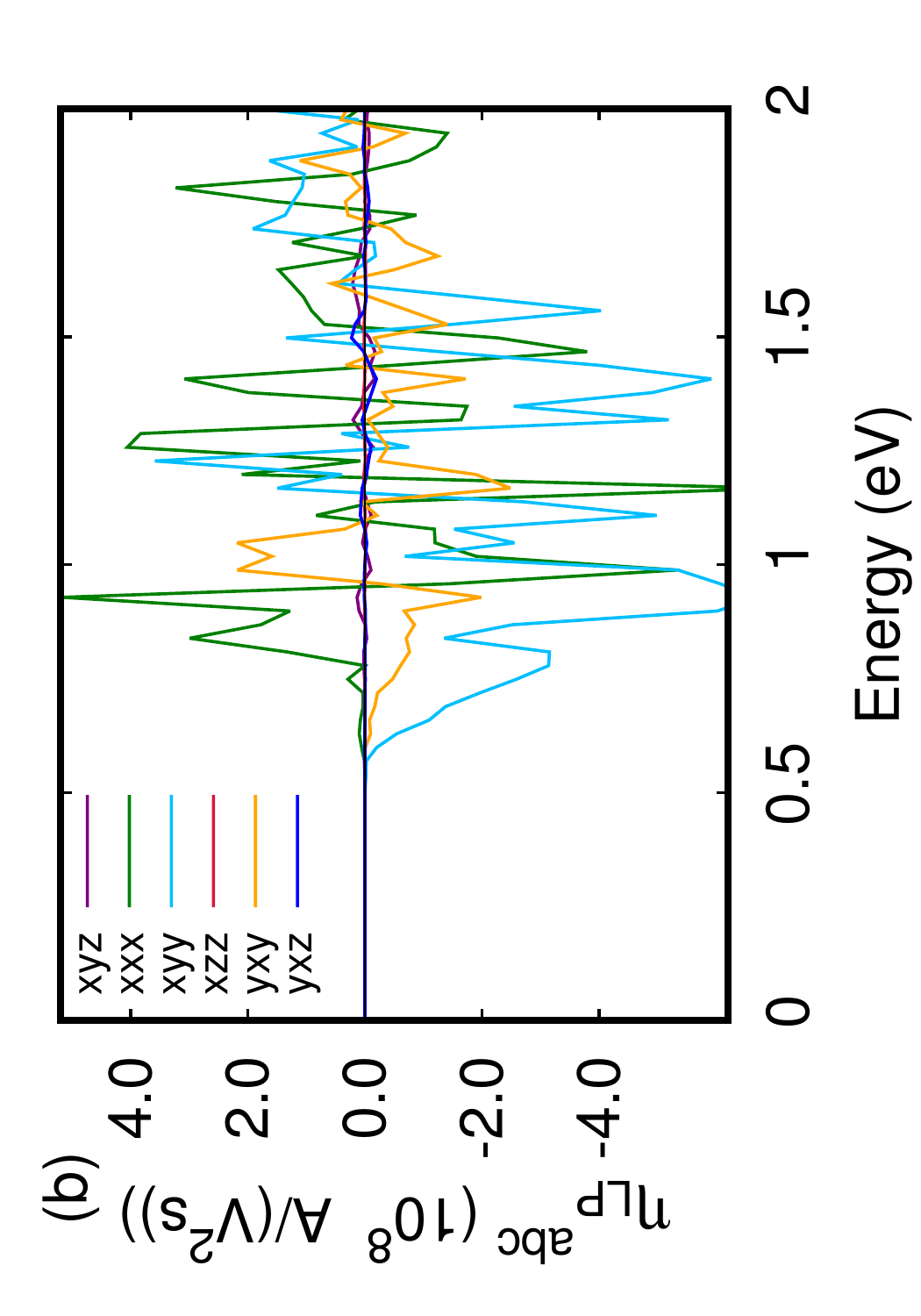}
    \includegraphics[width=5.5cm,angle=270]{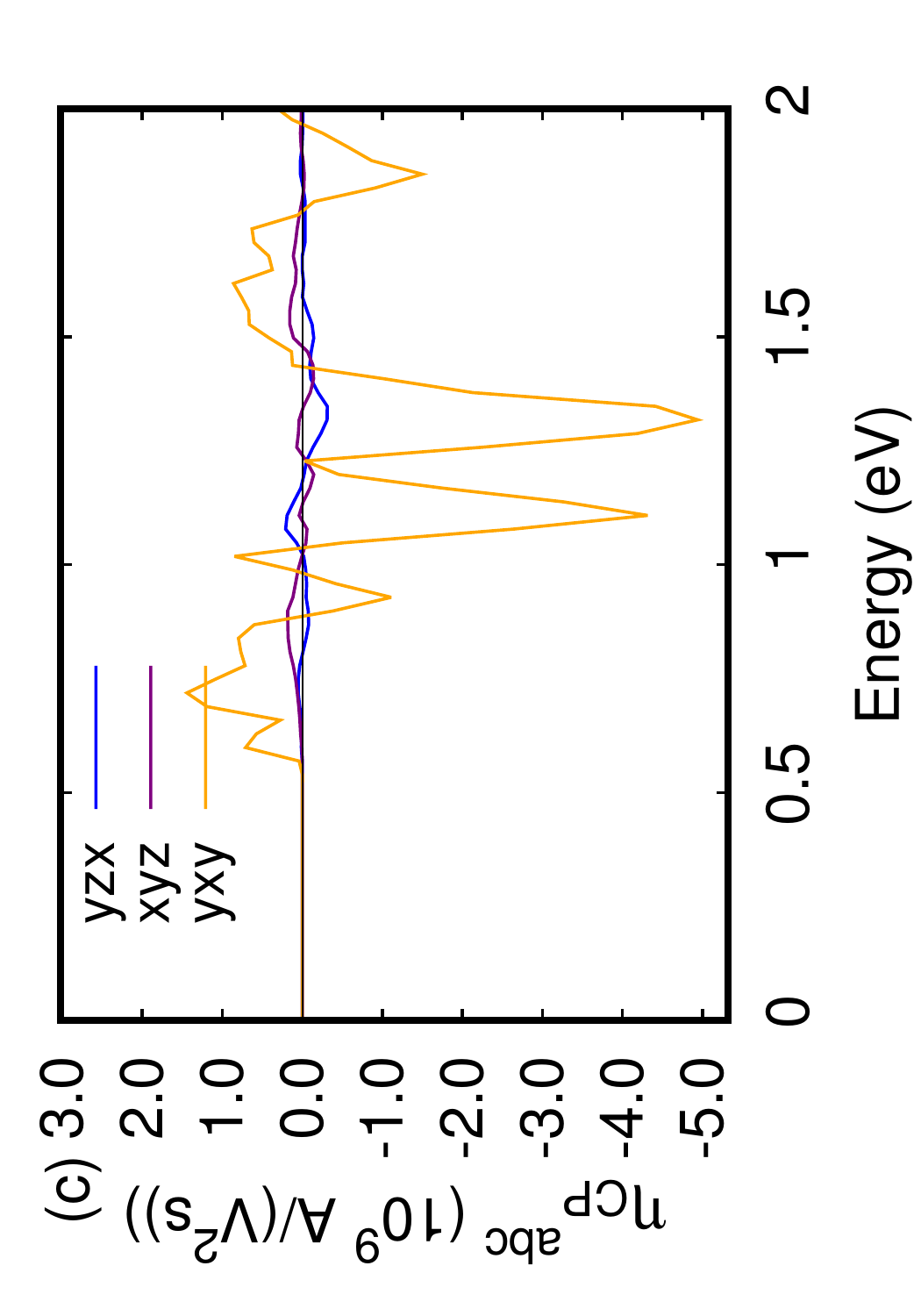}
    \caption{Non-zero components of the charge (a) LP shift photoconductivity, (b) LP injection photoconductivity, and (c) CP injection photoconductivity tensors for the $n=5$ cycloid structure.}
    \label{Chargephotocurrents1x5}
\end{figure}

The spectra of the non-zero components of the LP shift,  LP injection, and CP injection photoconductivity tensors are shown in panels (a), (b), and (c) of Fig.~\ref{Chargephotocurrents1x5}, respectively.

\subsection{LP Charge Shift Photoconductivity}

The components $\sigma^{xxx}_\mathrm{LP}$ and $\sigma^{xyy}_\mathrm{LP}$ dominate the response, with a sizable contribution from $\sigma^{yxy}_\mathrm{LP}$, similarly to the $n=3$ case. 
However, the contributions from the remaining components are further reduced compared to those found for $n=3$ and $n=4$. This suggests that increasing the magnetic period effectively filters out subleading contributions.

\subsection{LP Charge Injection Photoconductivity}

This contribution reflects the nonreciprocal band structure that emerges for odd values of $n$. 
The dominant components are $\eta_\mathrm{LP}^{xxx}$, $\eta_\mathrm{LP}^{xyy}$, and $\eta_\mathrm{LP}^{yxy}$. 
For $n=5$, however, this effect is significantly reduced compared to the $n=3$ case. 
Accordingly, the LP injection photoconductivity reaches peak values approximately one order of magnitude smaller, i.e., $10^{8}$~A/(V$^{2}$s).

\subsection{CP Charge Injection Photoconductivity}

As in the $n=3$ case, the component $\eta_\mathrm{CP}^{yxy}$ dominates the response, reaching peak absolute values of approximately $5 \times 10^{9}$~A/(V$^{2}$s), and is associated with the $p$-wave magnetic state. By contrast, the components $\eta_\mathrm{CP}^{yzy}$ and $\eta_\mathrm{CP}^{xyz}$, which reflect the SOC-induced spin-texture, are suppressed for $n=5$ compared to systems with shorter magnetic periods.

\newpage




\newpage

\section{Charge Photoconductivities for the $n=3$ Proper-Screw Magnetic Structure}\label{sec.screw_conductivity}

Here we analyze the charge photoresponse for the proper-screw spin-spiral structure with period $n=3$, shown in Fig.~\ref{Structures}(a). Since the MSG is the same as for the odd-$n$ cycloidal structure, the system exhibits both chrage shift and injection currents under linearly and circularly polarized light, with the same allowed photoconductivity tensor components.

\begin{figure}[h!]
    \centering
    \includegraphics[width=6.5cm,angle=0]{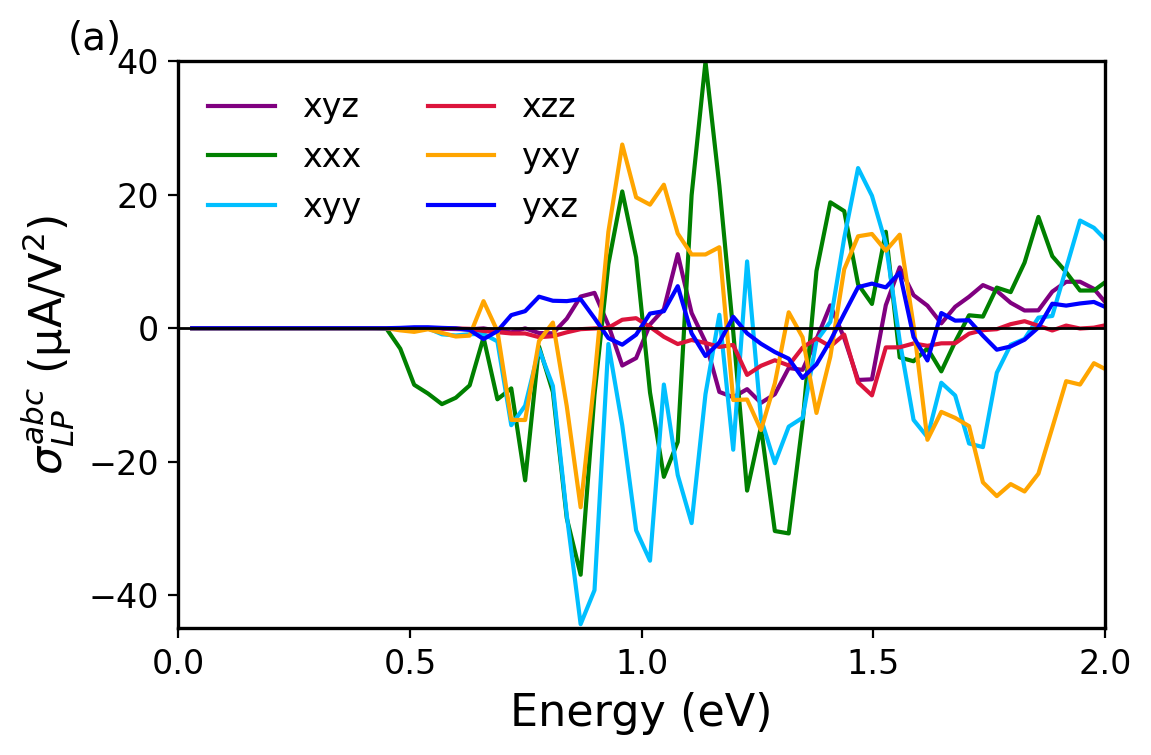}
    \includegraphics[width=6.5cm,angle=0]{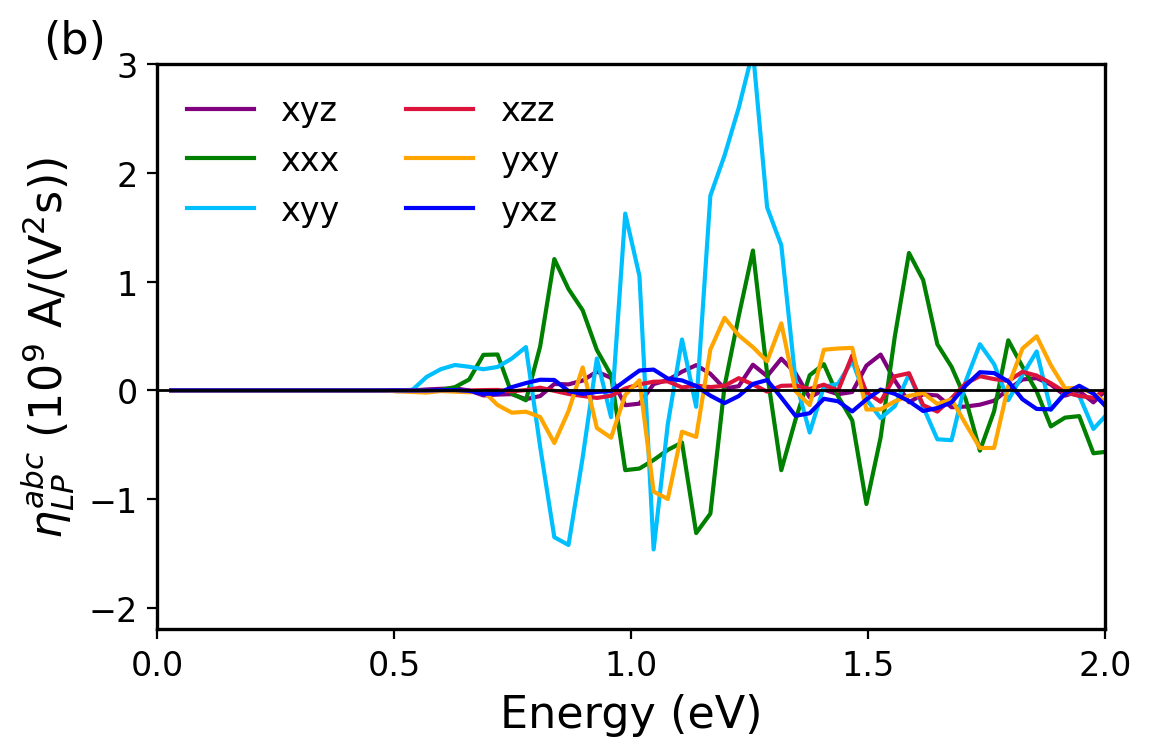}
    \includegraphics[width=6.5cm,angle=0]{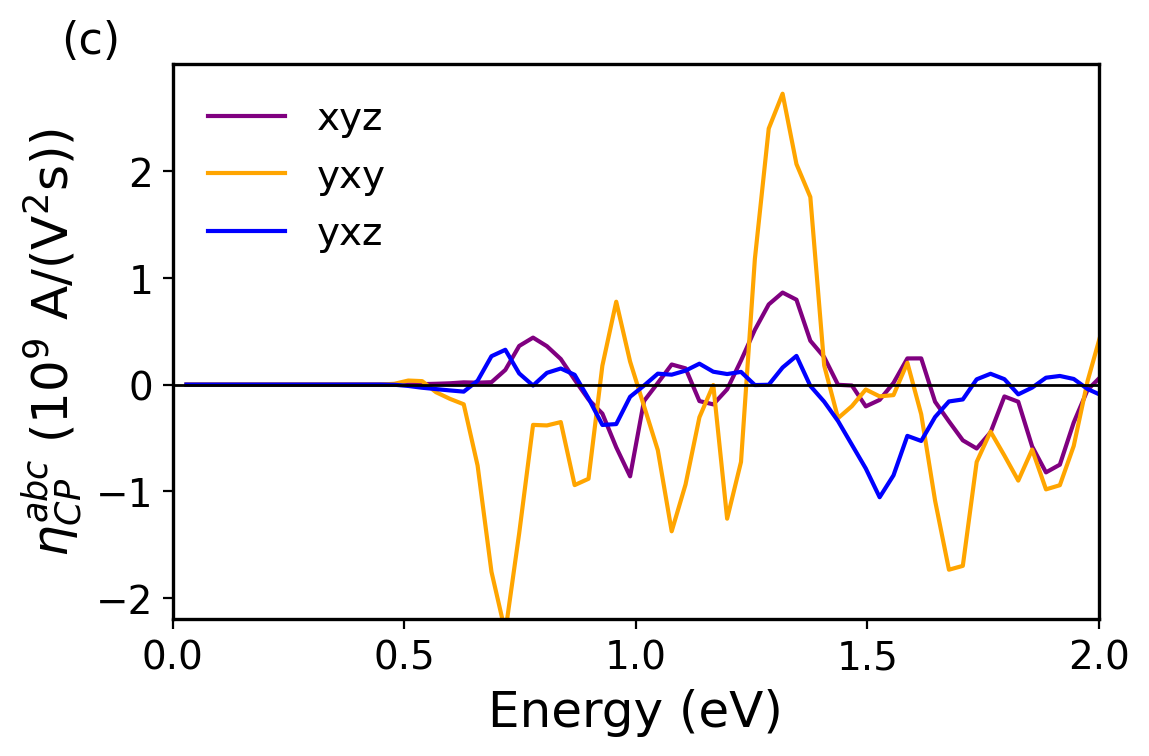}
    \caption{\justifying Non-zero components of the charge (a) LP shift photoconductivity, (b) LP injection photoconductivity, and (c) CP injection photoconductivity tensors for the $n=3$ proper-screw magnetic structure.}
    \label{photocurrents_helical1x3}
\end{figure}

\subsection{LP Charge Shift Photoconductivity}

The shift photoconductivity tensor under LP light exhibits the same six independent nonzero components as in the cyloidal $n=3$ case:
\[
\sigma^{xyz}_\mathrm{LP},\;
\sigma^{xxx}_\mathrm{LP},\;
\sigma^{xyy}_\mathrm{LP},\;
\sigma^{xzz}_\mathrm{LP},\;
\sigma^{yxy}_\mathrm{LP},\;
\sigma^{yxz}_\mathrm{LP},
\]
in agreement with the symmetry analysis in Sec. \ref{sec.symmetry}. Owing to the intrinsic symmetry of the shift tensor under exchange of its last two indices, $\sigma^{abc}_\mathrm{LP} = \sigma^{acb}_\mathrm{LP}$, all remaining nonvanishing components follow from permutation of the polarization indices.

The spectra of the independent components are shown in Fig.~\ref{photocurrents_helical1x3}(a).
$\sigma^{xxx}_\mathrm{LP}$ and $\sigma^{xyy}_\mathrm{LP}$ dominate the response, with a sizable contribution from $\sigma^{yxy}_\mathrm{LP}$, similarly to the cycloidal $n=3$ case. 

\subsection{LP Charge Injection Photoconductivity}

The injection photoconductivity tensor under LP light exhibits the same six independent nonzero components as in the cyloidal $n=3$ case:
\[
\eta^{xyz}_\mathrm{LP},\;
\eta^{xxx}_\mathrm{LP},\;
\eta^{xyy}_\mathrm{LP},\;
\eta^{xzz}_\mathrm{LP},\;
\eta^{yxy}_\mathrm{LP},\;
\eta^{yxz}_\mathrm{LP},
\]
in agreement with the symmetry analysis in Sec. \ref{sec.symmetry}.

This contribution reflects the nonreciprocal band structure that emerges for odd values of $n$. The spectra of the components are shown in Fig.~\ref{photocurrents_helical1x3}(b).  
The dominant components are $\eta_\mathrm{LP}^{xxx}$, $\eta_\mathrm{LP}^{xyy}$, and $\eta_\mathrm{LP}^{yxy}$, as in the $n=3$ cycloidal case, reflecting the fact that the largest band-structure asymmetry occurs along the same directions in the BZ.

\subsection{CP Charge Injection Photoconductivity}

The CP injection photoconductivity exhibits three independent nonzero components:
\[
\eta_\mathrm{CP}^{xyz},\;
\eta_\mathrm{CP}^{yxy},\;
\eta_\mathrm{CP}^{yzx},
\]
in agreement with the symmetry analysis in Sec. \ref{sec.symmetry}.
All other nonvanishing components follow from antisymmetry under exchange of the last two indices, $\eta^{abc} = -\eta^{acb}$. The spectra of the components are shown in Fig.~\ref{photocurrents_helical1x3}(c).  

In contrast to the cycloidal $n=3$ case, in the proper-screw spin-spiral system the component $\eta_\mathrm{CP}^{yzx}$ is associated with $p$-wave magnetism, whereas $\eta_\mathrm{CP}^{yxy}$ and $\eta_\mathrm{CP}^{yzy}$ reflect the SOC-induced spin-texture.

Notably, the $p$-wave-related component $\eta_\mathrm{CP}^{yzx}$ is here slightly smaller than the others. This reflects the fact that the spin-$y$ $p$-wave polarization is reduced in the proper-screw configuration and becomes comparable in magnitude to the SOC-induced spin polarization, as discussed in the band-structure analysis of Sec.~\ref{sec.screw}.

Following Eq.~(2) of the main text, these components can be mapped onto an effective rank-two pseudotensor, with the same nonzero elements $\gamma_{xx}$, $\gamma_{yy}$, and $\gamma_{yz}$ as in the cycloidal case. However, in contrast to that case, here $\gamma_{yy}$—rather than $\gamma_{yz}$—is associated with the $p$-wave spin texture. This component directly reflects the spin–momentum coupling $k_y s^y$, derived from Tab. \ref{tab:spinpol} ($T_yy$ entry). In contrast, $\gamma_{xx}$ and $\gamma_{yz}$ characterize the SOC-induced spin textures.

\section{Spin Photoconductivities for the $n=3$ Cycloidal Magnetic Structure}

The cycloidal magnetic structure with period $n=3$ exhibits spin shift and injection currents under both LP and CP light, as exlained in Sec. \ref{sec.symmetry}.

\begin{figure}[h!]
    \centering
    \includegraphics[width=7.5cm,angle=0]{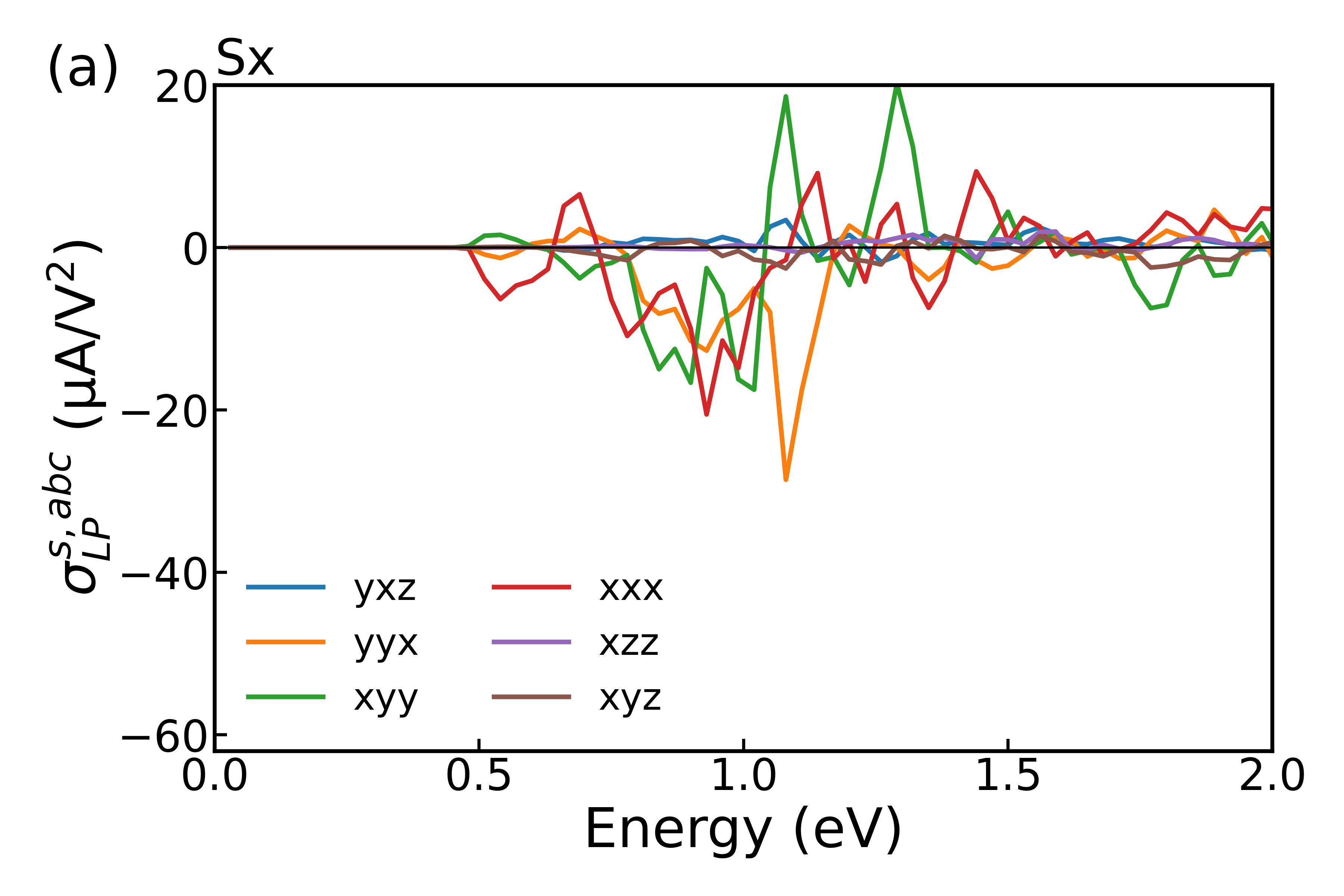}
    \includegraphics[width=7.5cm,angle=0]{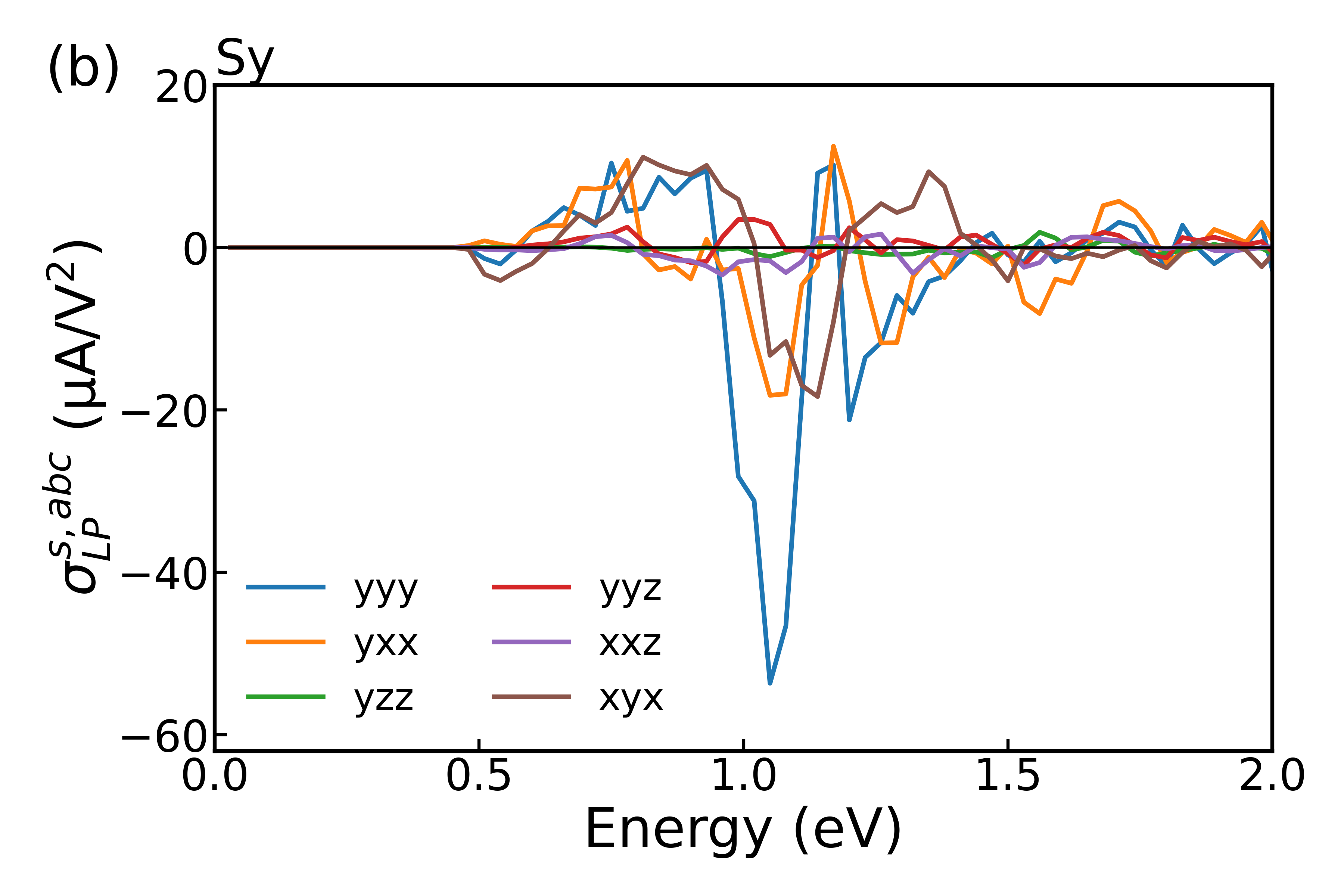}
    \includegraphics[width=7.5cm,angle=0]{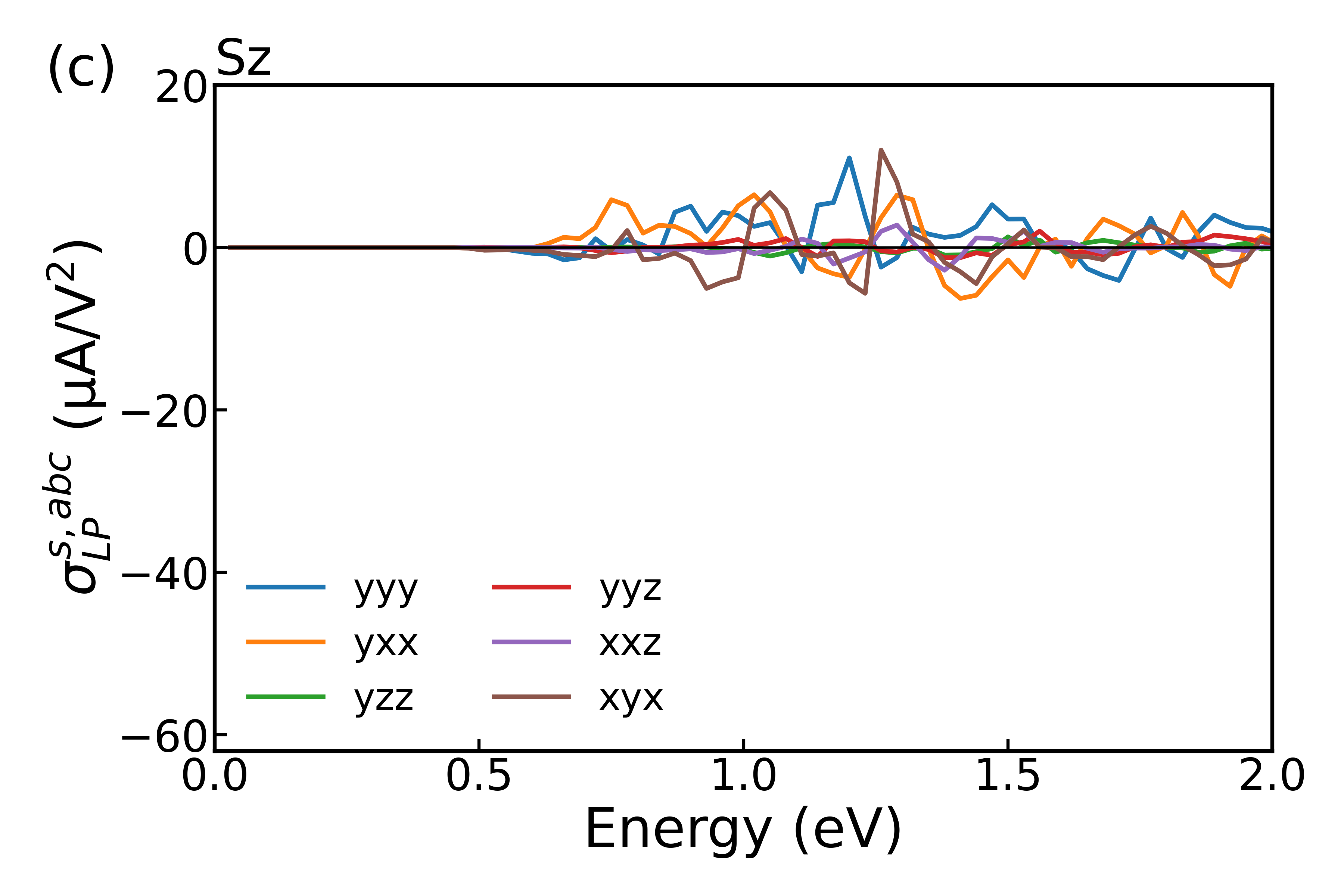}
    \caption{\justifying Non-zero independent components of the LP spin shift photoconductivity tensor for (a) $s^x$, (b) $s^y$ and (c) $s^z$ spin projections and for the $n=3$ cycloid structure.}
    \label{spinlinearshift1x3}
\end{figure}

\subsection{LP Spin Shift Photoconductivity}

The spin shift photoconductivity tensor of the $n=3$ system under LP light exhibits the following independent nonzero components:

\[
\text{spin } s^x:
\quad
\sigma^{x,yxz}_\mathrm{LP},\;
\sigma^{x,yyx}_\mathrm{LP},\;
\sigma^{x,xyy}_\mathrm{LP},\;
\sigma^{x,xxx}_\mathrm{LP},\;
\sigma^{x,xzz}_\mathrm{LP},\;
\sigma^{x,xyz}_\mathrm{LP}.
\]

\[
\text{spin } s^y:
\quad
\sigma^{y,yyy}_\mathrm{LP},\;
\sigma^{y,yxx}_\mathrm{LP},\;
\sigma^{y,yzz}_\mathrm{LP},\;
\sigma^{y,yyz}_\mathrm{LP},\;
\sigma^{y,xxz}_\mathrm{LP},\;
\sigma^{y,xyx}_\mathrm{LP}.
\]

\[
\text{spin } s^z:
\quad
\sigma^{z,yyy}_\mathrm{LP},\;
\sigma^{z,yxx}_\mathrm{LP},\;
\sigma^{z,yzz}_\mathrm{LP},\;
\sigma^{z,yyz}_\mathrm{LP},\;
\sigma^{z,xxz}_\mathrm{LP},\;
\sigma^{z,xyx}_\mathrm{LP}.
\]

All remaining nonvanishing components follow from the intrinsic permutation symmetry of the shift tensor under exchange of the last two indices, $\sigma^{s,abc}_\mathrm{LP} = \sigma^{s,acb}_\mathrm{LP}$.

The spectra are shown in Fig.~\ref{spinlinearshift1x3}. 
The dominant contributions correspond to light polarized in the $xy$ plane. 
For the $s^x$ and $s^y$ spin projections, the peak magnitudes are comparable to those of the LP charge shift current, whereas for the $s^z$ projection the response is slightly reduced.

\begin{figure}[h!]
    \centering
    \includegraphics[width=7.5cm,angle=0]{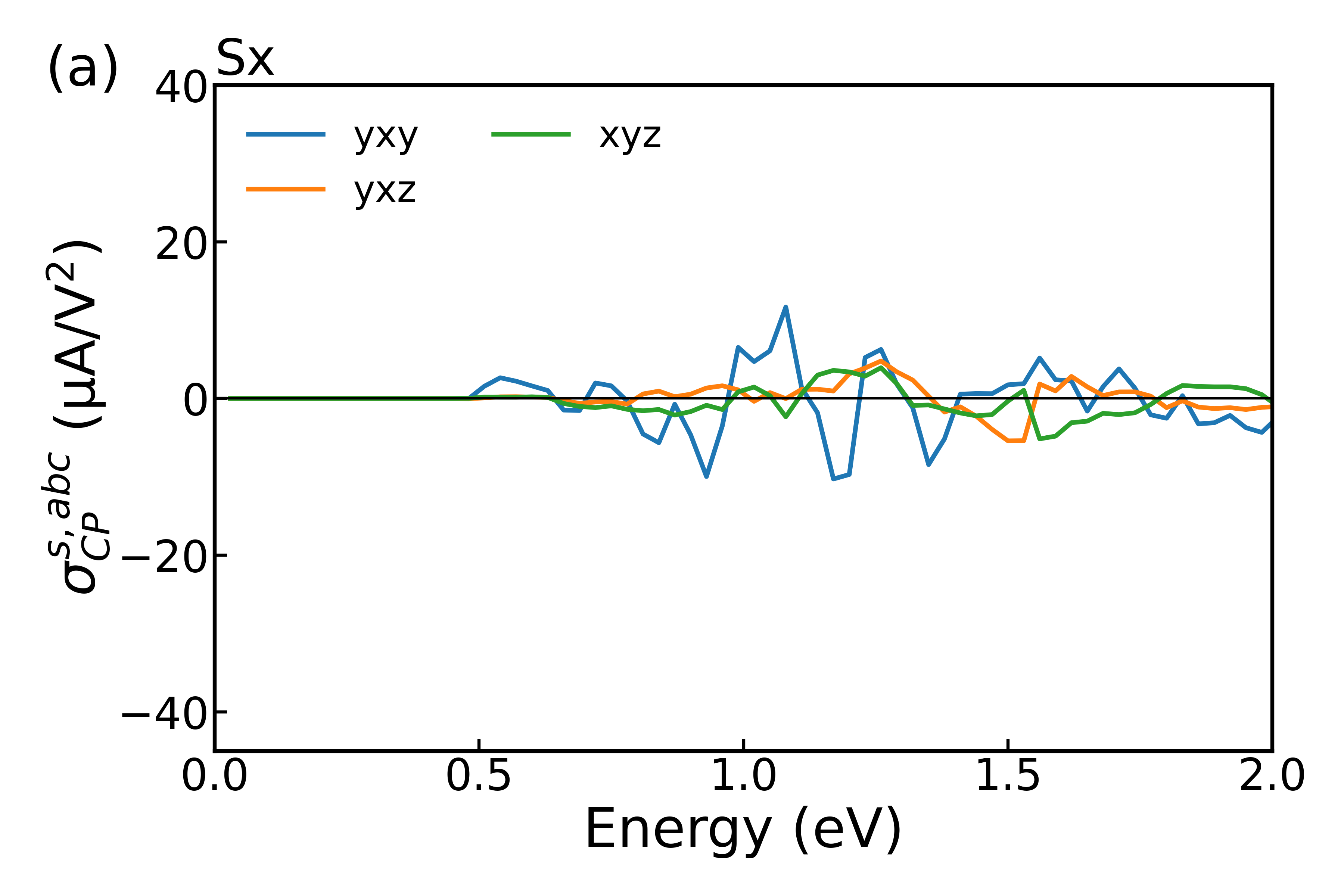}
    \includegraphics[width=7.5cm,angle=0]{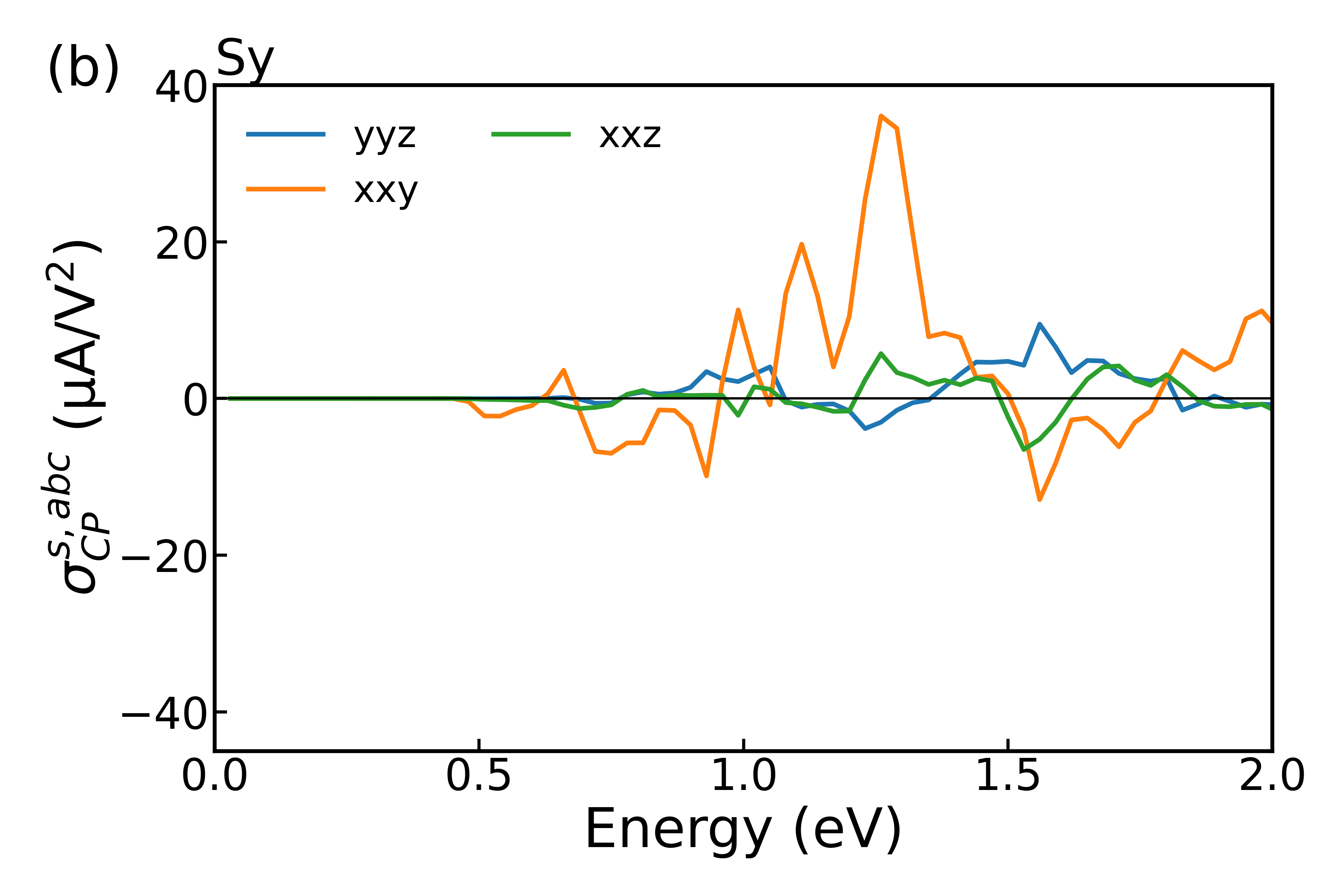}
    \includegraphics[width=7.5cm,angle=0]{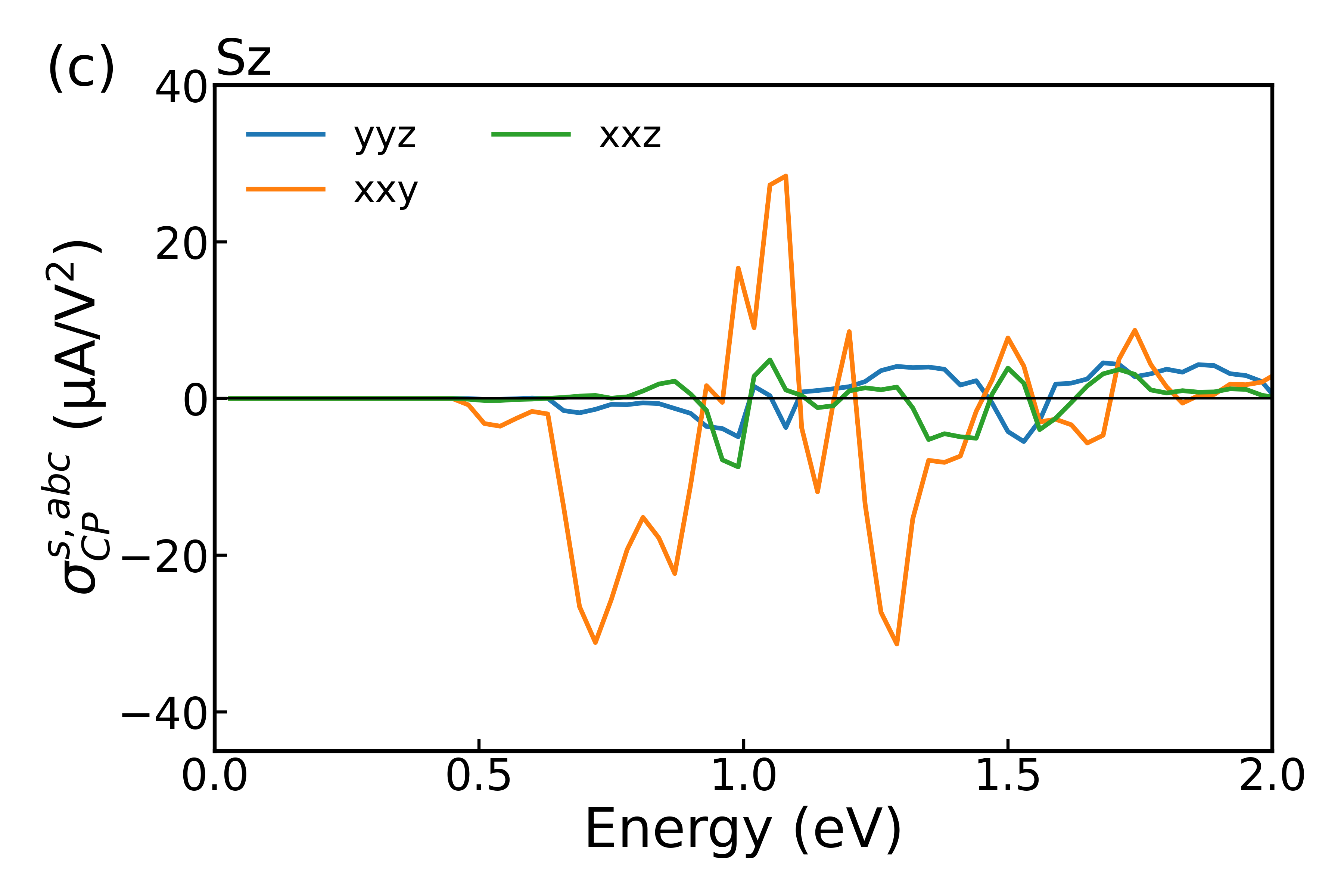}
   \caption{\justifying Non-zero independent components of the CP spin shift photoconductivity tensor for (a) $s^x$, (b) $s^y$ and (c) $s^z$ spin projections and for the $n=3$ cycloid structure.}
    \label{spincircularshift1x3}
\end{figure}

\subsection{CP Spin Shift Photoconductivity}

The spin shift photoconductivity tensor of the $n=3$ system under CP light exhibits the following independent nonzero components:

\[
\text{spin } s^x:
\quad
\sigma^{x,yxy}_\mathrm{CP},\;
\sigma^{x,yxz}_\mathrm{CP},\;
\sigma^{x,xyz}_\mathrm{CP}.
\]

\[
\text{spin } s^y:
\quad
\sigma^{y,yyz}_\mathrm{CP},\;
\sigma^{y,xxy}_\mathrm{CP},\;
\sigma^{y,xxz}_\mathrm{CP}.
\]

\[
\text{spin } s^z:
\quad
\sigma^{z,yyz}_\mathrm{CP},\;
\sigma^{z,xxy}_\mathrm{CP},\;
\sigma^{z,xxz}_\mathrm{CP},
\]
in agreement with the symmetry analysis in Sec. \ref{sec.symmetry}. All other nonvanishing components follow from antisymmetry under exchange of the last two indices, $\sigma^{s,abc}_\mathrm{CP}
=
-\,\sigma^{s,acb}_\mathrm{CP}$.

The spectra are displayed in Fig.~\ref{spincircularshift1x3}. 
As in the LP case, the dominant contributions arise from light polarized in the $xy$ plane, notably 
$\sigma^{x,yxy}_\mathrm{CP}$, 
$\sigma^{y,xxy}_\mathrm{CP}$, and 
$\sigma^{z,xxy}_\mathrm{CP}$.

For the $s^x$ and $s^y$ projections, the spin current flows perpendicular to the spin polarization direction, i.e., they flow along $y$ and $x$, respectively. 
For the $s^z$ projection, the current flows predominantly along $x$, i.e., parallel to the electric polarization axis. 
Remarkably, the peak values slightly exceed those of the charge CP shift current shown in Fig.~\ref{Circshift1x3}.

\begin{figure}[h!]
    \centering
    \includegraphics[width=7.5cm,angle=0]{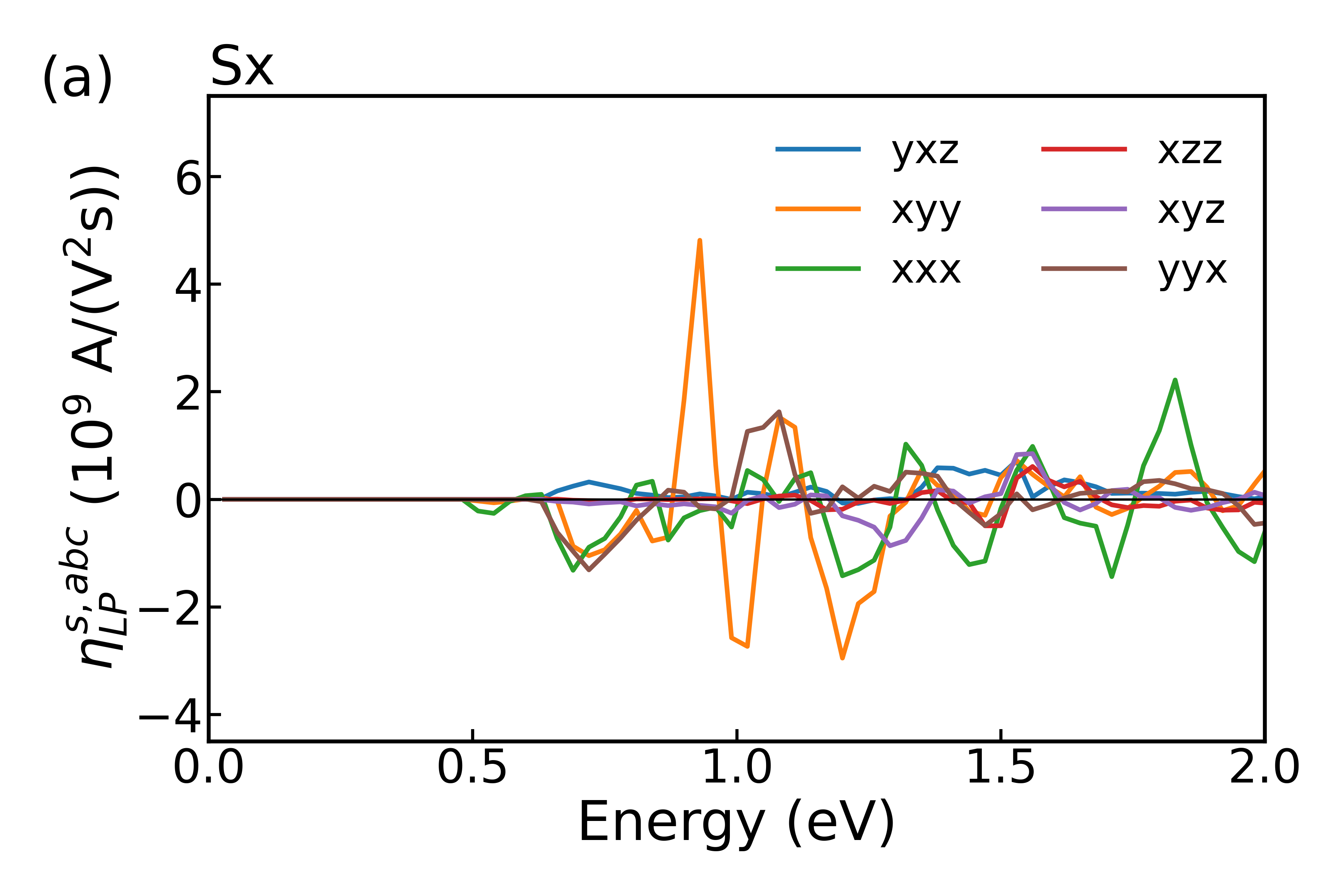}
    \includegraphics[width=7.5cm,angle=0]{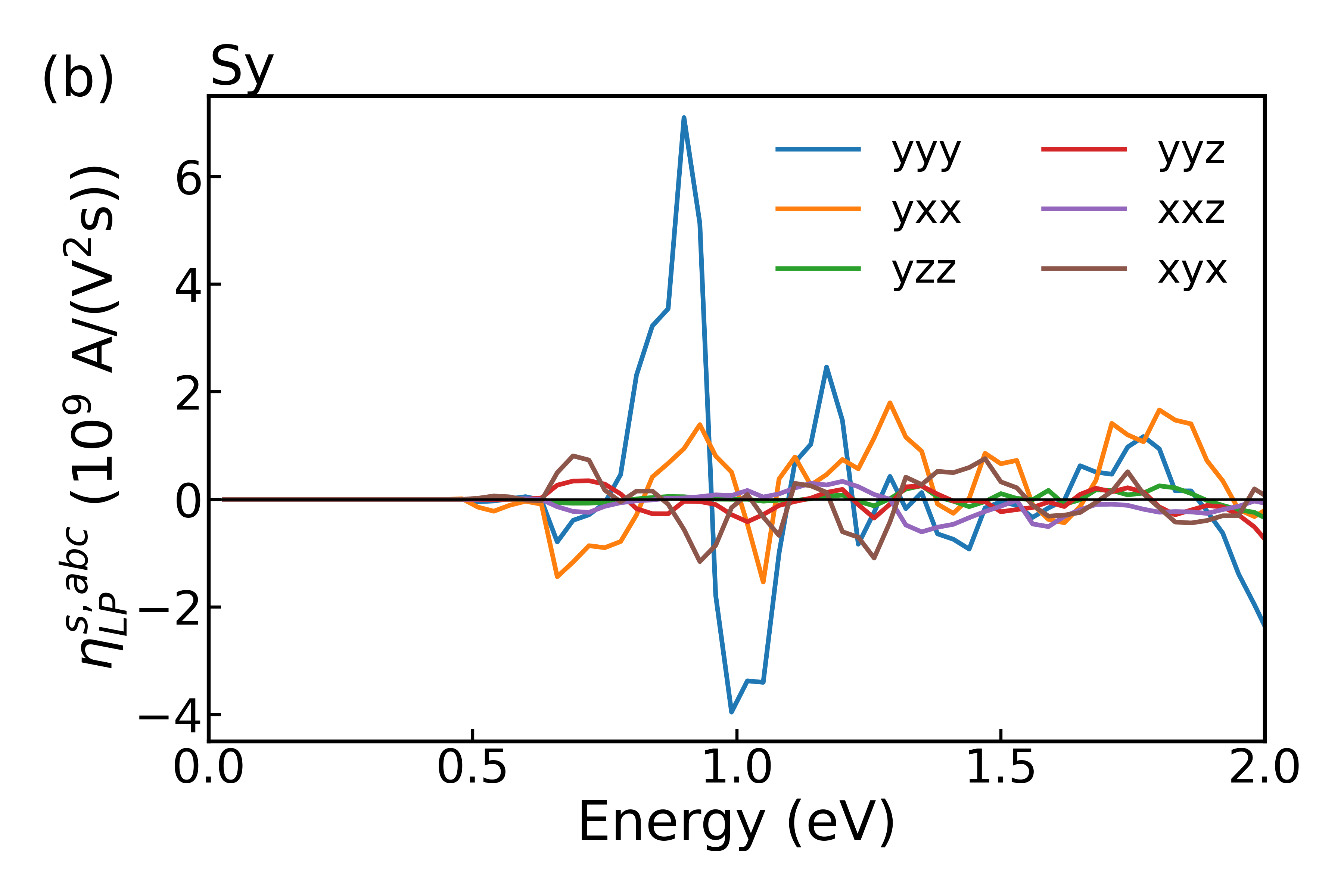}
    \includegraphics[width=7.5cm,angle=0]{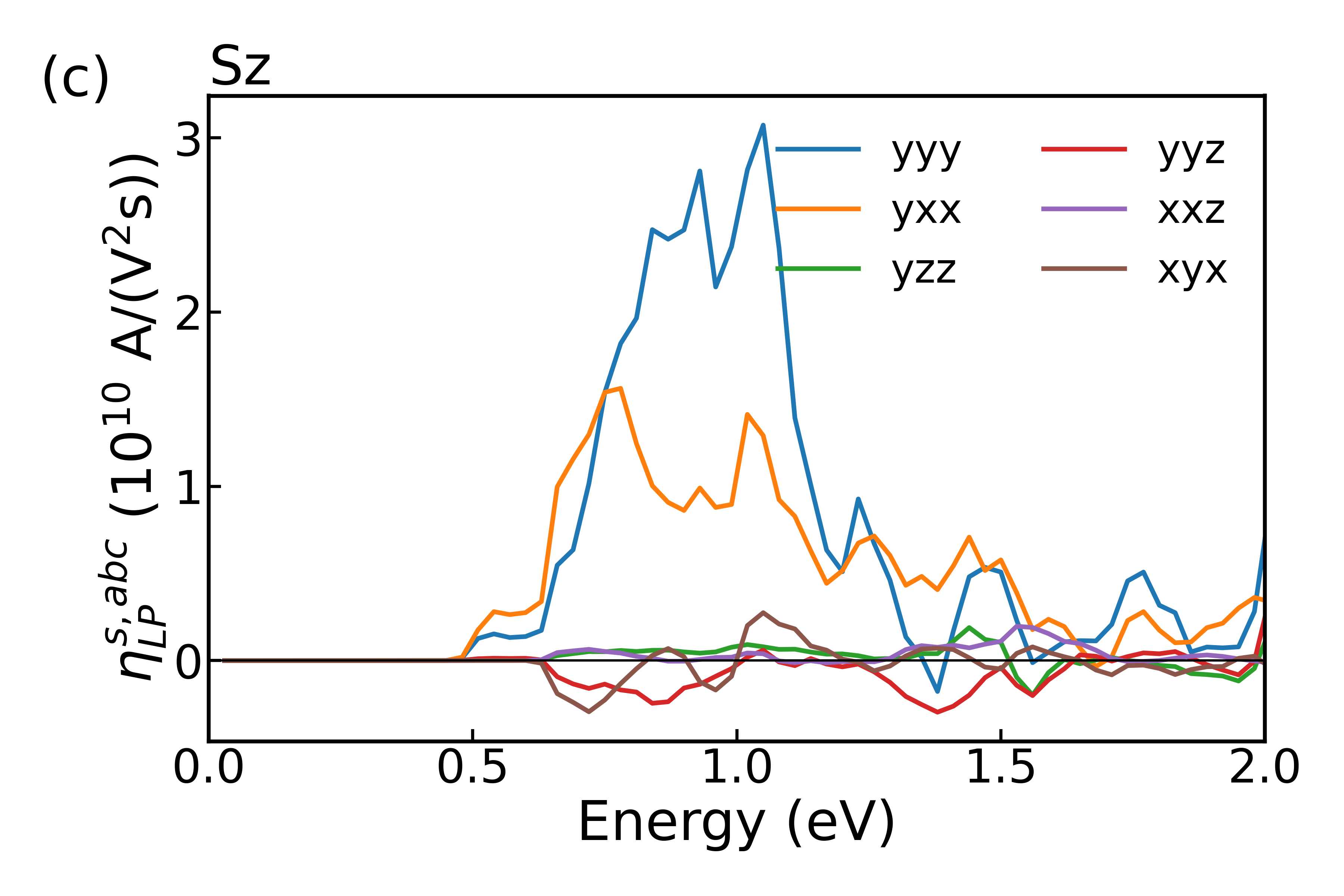}
    \caption{\justifying Non-zero independent components of the LP spin injection photoconductivity tensor for (a) $s^x$, (b) $s^y$ and (c) $s^z$ spin projections and for the $n=3$ cycloid structure.}
    \label{spinlinearinjection1x3}
\end{figure}

\subsection{LP Spin Injection Photoconductivity}

The spin injection photoconductivity tensor of the $n=3$ system under LP light exhibits the following independent nonzero components:

\[
\text{spin } s^x:
\quad
\eta^{x,yxz}_\mathrm{LP},\;
\eta^{x,xyy}_\mathrm{LP},\;
\eta^{x,xxx}_\mathrm{LP},\;
\eta^{x,xzz}_\mathrm{LP},\;
\eta^{x,yyx}_\mathrm{LP},\;
\eta^{x,xyz}_\mathrm{LP}.
\]

\[
\text{spin } s^y:
\quad
\eta^{y,yyy}_\mathrm{LP},\;
\eta^{y,yxx}_\mathrm{LP},\;
\eta^{y,yzz}_\mathrm{LP},\;
\eta^{y,yyz}_\mathrm{LP},\;
\eta^{y,xxz}_\mathrm{LP},\;
\eta^{y,xyx}_\mathrm{LP}.
\]

\[
\text{spin } s^z:
\quad
\eta^{z,yyy}_\mathrm{LP},\;
\eta^{z,yxx}_\mathrm{LP},\;
\eta^{z,yzz}_\mathrm{LP},\;
\eta^{z,yyz}_\mathrm{LP},\;
\eta^{z,xxz}_\mathrm{LP},\;
\eta^{z,xyx}_\mathrm{LP},
\]
in agreement with the symmetry analysis in Sec. \ref{sec.symmetry}.
All other nonvanishing components follow from antisymmetry under exchange of the last two indices,
$\eta^{s,abc}_\mathrm{LP}
=
\,\eta^{s,acb}_\mathrm{LP}
$.

The spectra are displayed in Fig.~\ref{spinlinearinjection1x3}.
For the $s^x$ and $s^y$ projections, the components reaching the largest peak values are 
$\eta^{x,xyy}_\mathrm{LP}$ and $\eta^{y,yyy}_\mathrm{LP}$, respectively, implying that the spin current flows parallel to the spin polarization direction.

However, as already discussed in the main text, the overall largest photoconductivity occurs for the $s^z$ projection with current flowing along $y$. In particular, $\eta^{z,yyy}_\mathrm{LP}$ and $\eta^{z,yxx}_\mathrm{LP}$ reach peak values approximately one order of magnitude larger than the other components ($\sim 10^{10}$~A/(V$^{2}$s) versus $\sim 10^{9}$~A/(V$^{2}$s)). 
This enhancement originates from the underlying nonrelativistic $p$-wave magnetic spin texture.

\begin{figure}[h!]
    \centering
    \includegraphics[width=7.5cm,angle=0]{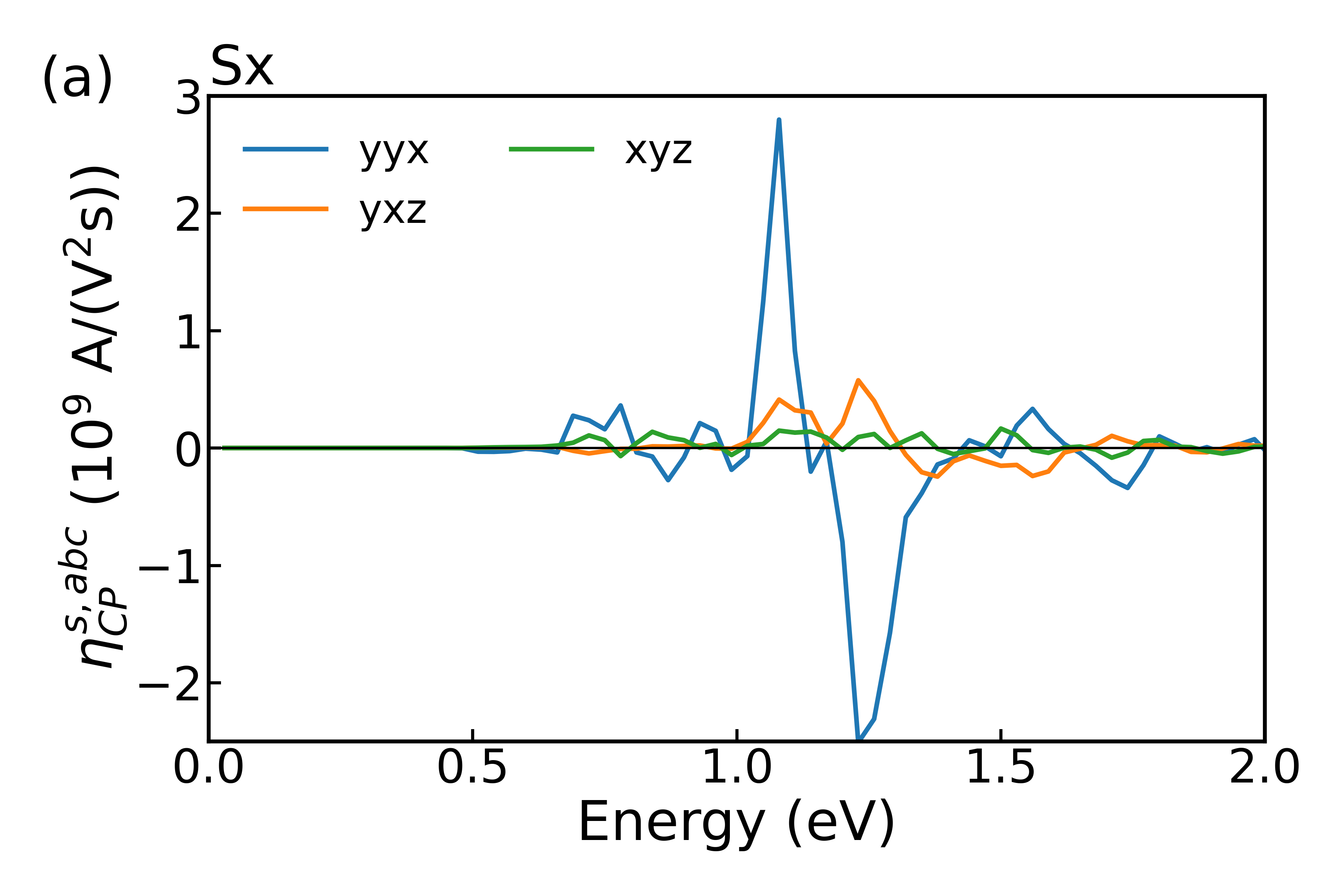}
    \includegraphics[width=7.5cm,angle=0]{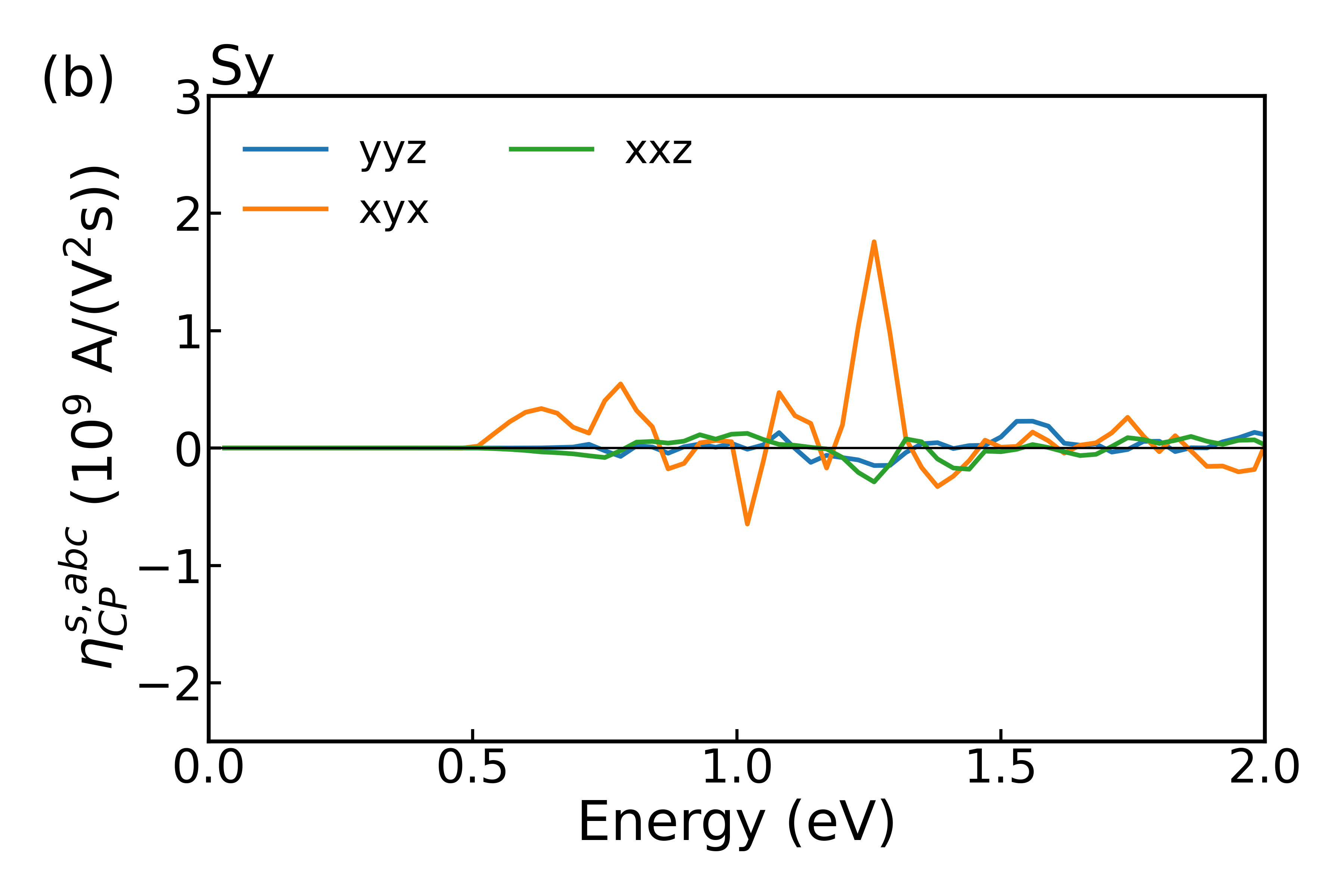}
    \includegraphics[width=7.5cm,angle=0]{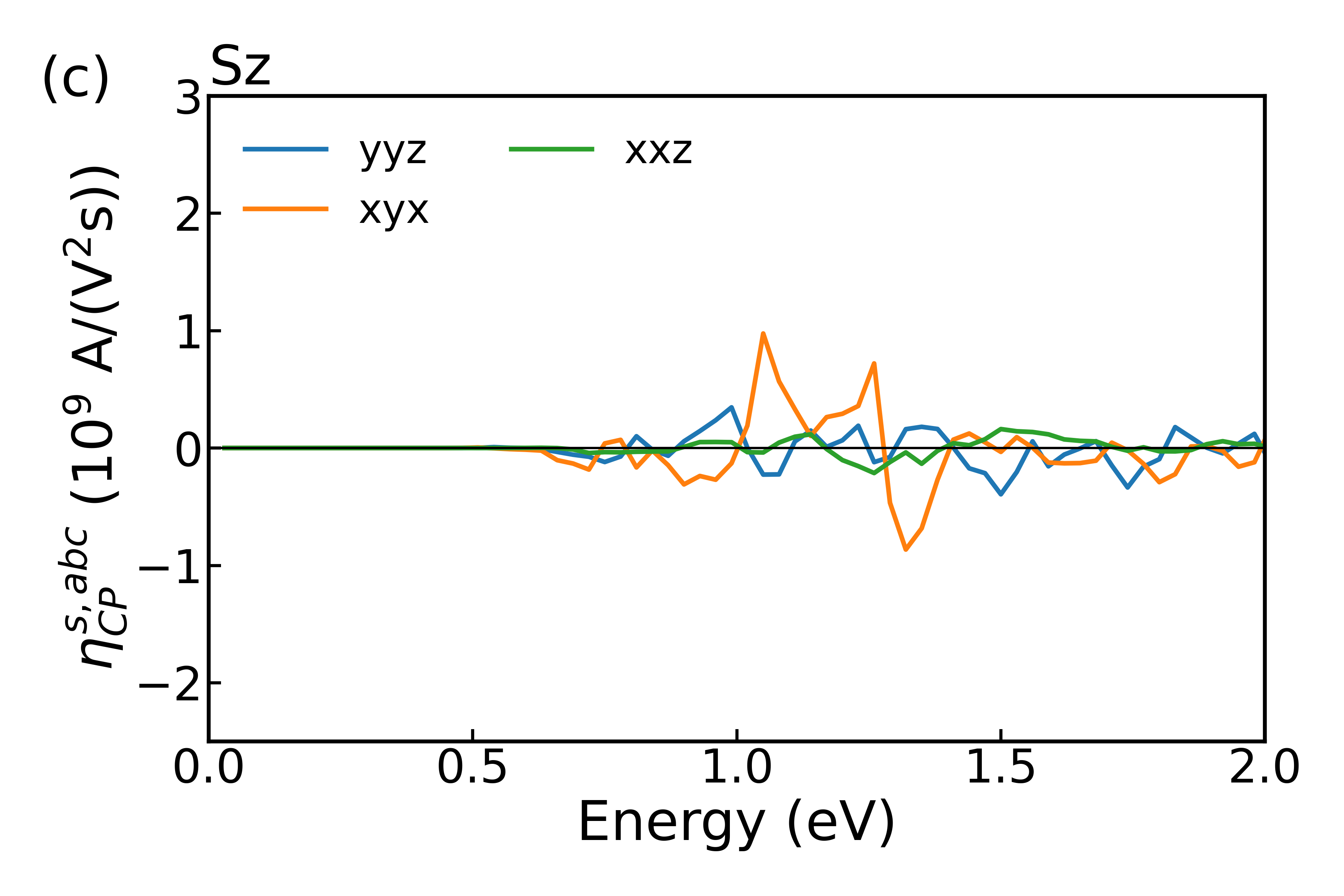}
    \caption{\justifying Non-zero independent components of the CP spin injection photoconductivity tensor for (a) $s^x$, (b) $s^y$ and (c) $s^z$ spin projections and for the $n=3$ cycloid structure.}
    \label{spincircularinjection1x3}
\end{figure}

\subsection{CP Spin Injection Photoconductivity}

The spin injection photoconductivity tensor of the $n=3$ system under CP light exhibits the following independent nonzero components:

\[
\text{spin } s^x:
\quad
\eta^{x,yyx}_\mathrm{CP},\;
\eta^{x,yxz}_\mathrm{CP},\;
\eta^{x,xyz}_\mathrm{CP}.
\]

\[
\text{spin } s^y:
\quad
\eta^{y,yyz}_\mathrm{CP},\;
\eta^{y,xyx}_\mathrm{CP},\;
\eta^{y,xxz}_\mathrm{CP}.
\]

\[
\text{spin } s^z:
\quad
\eta^{z,yyz}_\mathrm{CP},\;
\eta^{z,xyx}_\mathrm{CP},\;
\eta^{z,xxz}_\mathrm{CP}.
\]

All other nonvanishing components follow from antisymmetry under exchange of the last two indices,
$\eta^{s,abc}_\mathrm{CP}=
-\,\eta^{s,acb}_\mathrm{CP}$.

The spectra are displayed in Fig.~\ref{spincircularinjection1x3}. 
The dominant contributions correspond to light polarized in the $xy$ plane, namely 
$\eta^{x,yyx}_\mathrm{CP}$, 
$\eta^{y,xyx}_\mathrm{CP}$, and 
$\eta^{z,xyx}_\mathrm{CP}$ 
for the $S^x$, $S^y$, and $S^z$ spin projections, respectively. 
The peak values are of order $\sim 10^{9}$~A/(V$^{2}$s), comparable for all spin projections, therefore indicating the absence of an enhancement mechanism associated with the nonrelativistic $p$-wave magnetic spin texture.

\newpage

\section{Spin Photoconductivities for the $n=4$ Cycloidal Magnetic Structure}

This section complements the analysis of the spin photoconductivity for the cycloidal magnetic structure with period $n=4$ presented in the main text by providing additional tensor components that were not included in Fig.~4 of the main text.

The magnetic space group allows only spin shift currents under CP light and spin injection currents under LP light, as discussed in Sec. \ref{sec.symmetry}.

\begin{figure}[h!]
    \centering
    \includegraphics[width=7.5cm]{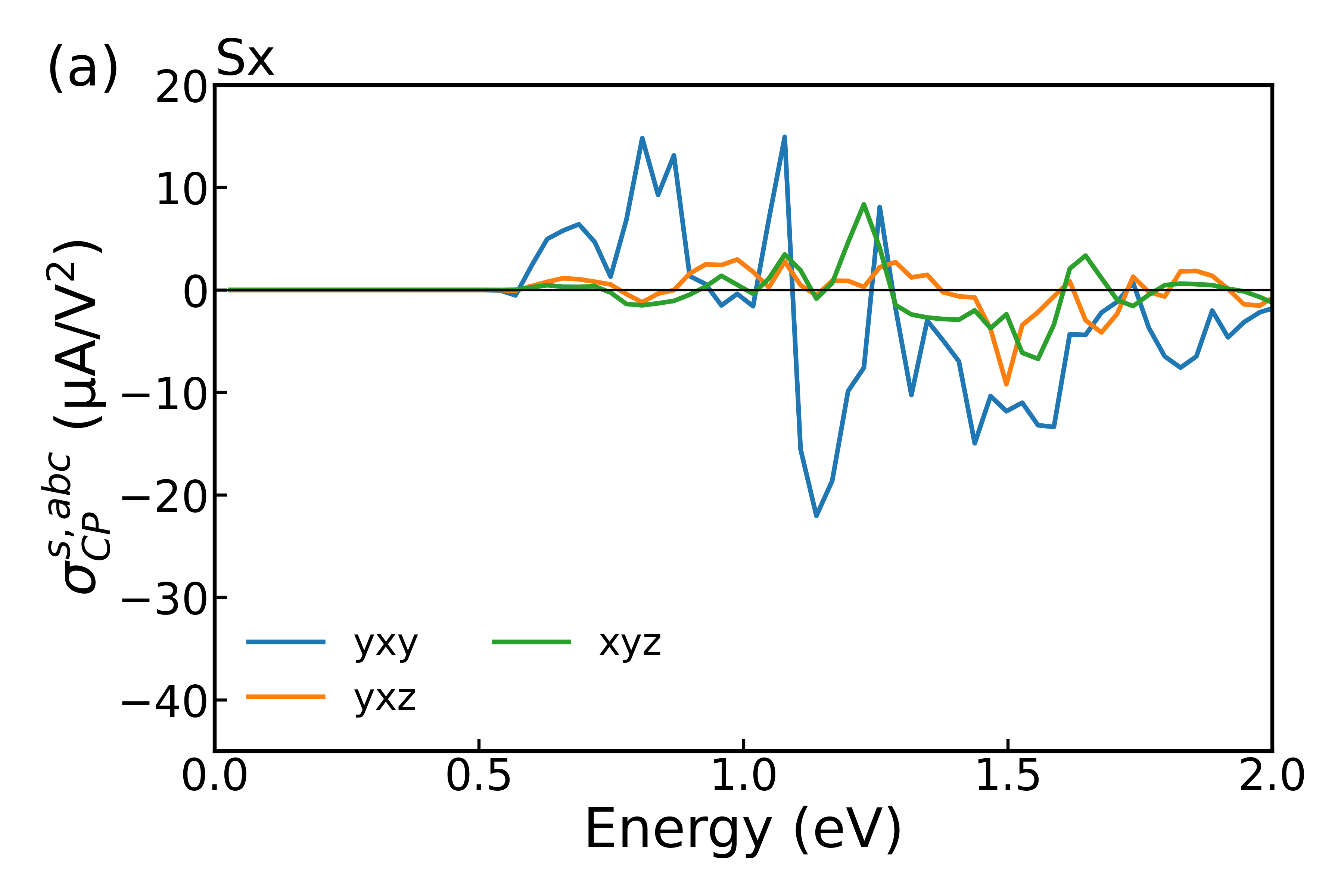}
    \includegraphics[width=7.5cm]{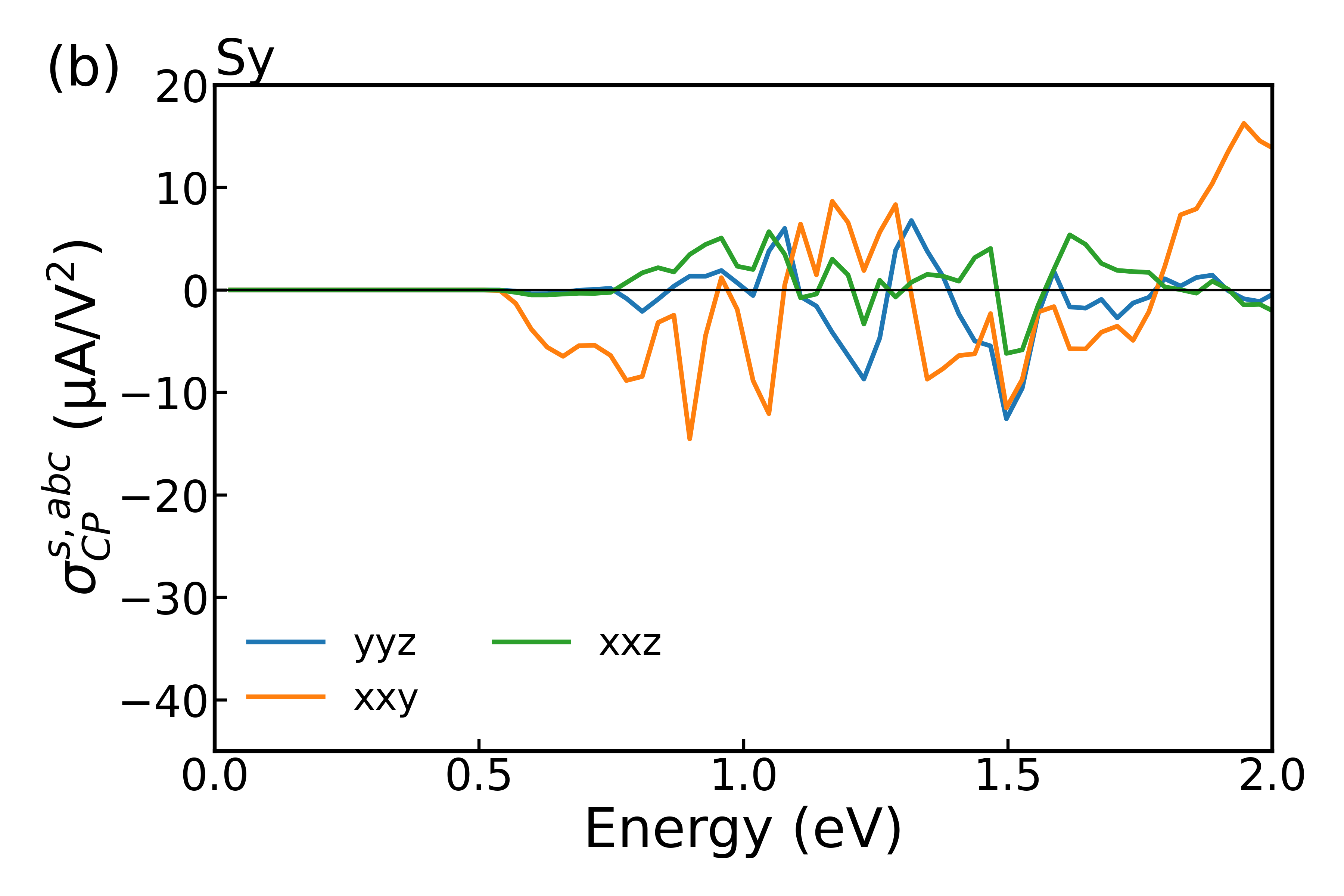}
    \includegraphics[width=7.5cm]{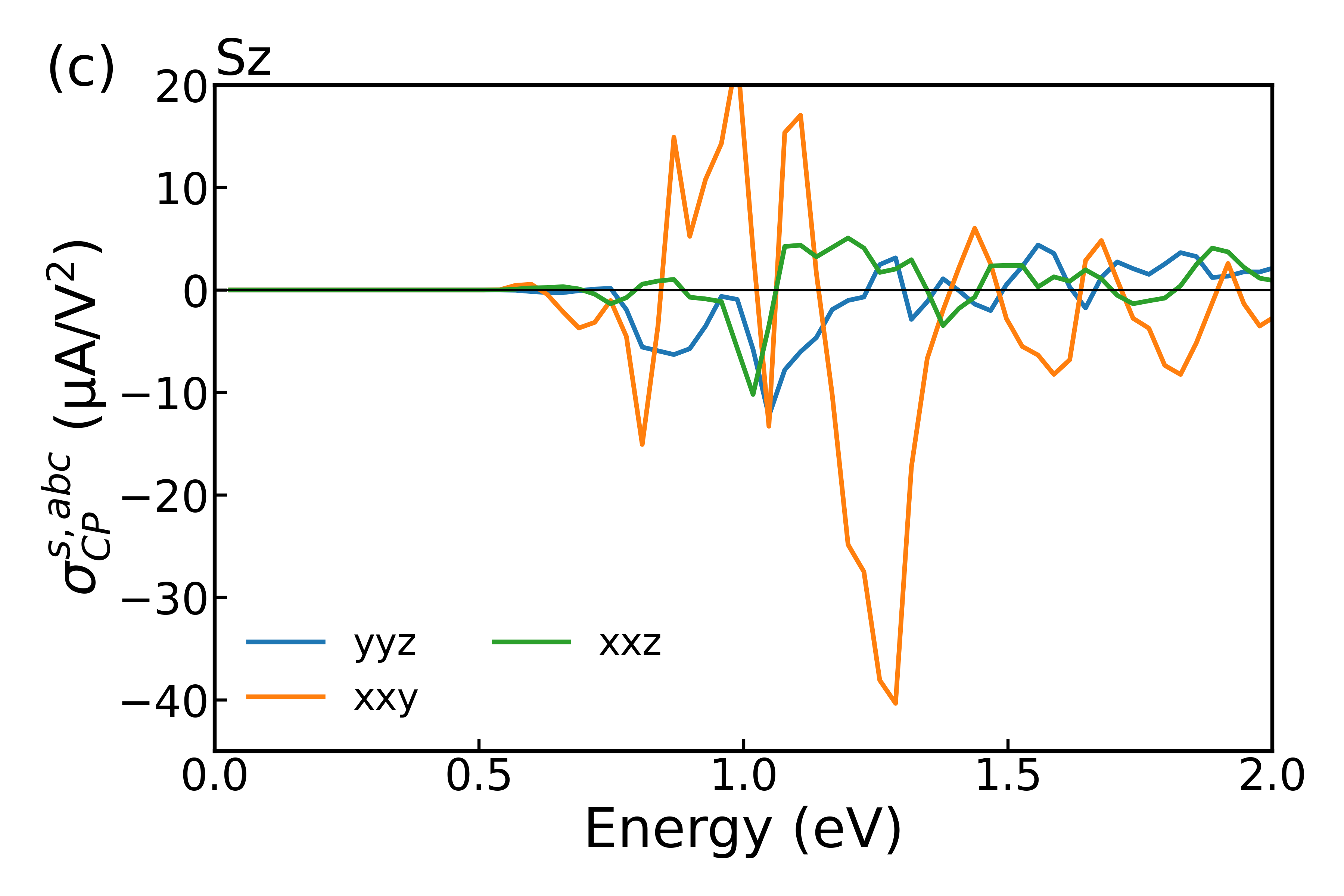}
    \caption{Non-zero independent components of the CP spin shift photoconductivity tensor for (a) $s^x$, (b) $s^y$, and (c) $s^z$ spin projections in the $n=4$ cycloid structure.}
    \label{spincircularshift1x4}
\end{figure}

\subsection{CP Spin Shift Photoconductivity}

The spin shift photoconductivity tensor of the $n=4$ system under CP light exhibits the same independent nonzero components as in the $n=3$ case:

\[
\text{spin } s^x:
\quad
\sigma^{x,yxy}_\mathrm{CP},\;
\sigma^{x,yxz}_\mathrm{CP},\;
\sigma^{x,xyz}_\mathrm{CP}.
\]

\[
\text{spin } s^y:
\quad
\sigma^{y,yyz}_\mathrm{CP},\;
\sigma^{y,xxy}_\mathrm{CP},\;
\sigma^{y,xxz}_\mathrm{CP}.
\]

\[
\text{spin } s^z:
\quad
\sigma^{z,yyz}_\mathrm{CP},\;
\sigma^{z,xxy}_\mathrm{CP},\;
\sigma^{z,xxz}_\mathrm{CP}.
\]

The spectra are displayed in Fig.~\ref{spincircularshift1x4}. 
As in the $n=3$ case, the dominant contributions arise from light polarized in the $xy$ plane, notably 
$\sigma^{x,yxy}_\mathrm{CP}$, 
$\sigma^{y,xxy}_\mathrm{CP}$, and 
$\sigma^{z,xxy}_\mathrm{CP}$. In particular, the latter component, corresponding to the $s^z$ spin projection, reaches the largest peak values, as discussed in the main text. It represents a spin-$z$ current flowing along the $x$ direction.

\begin{figure}[h!]
    \centering
    \includegraphics[width=7.5cm,angle=0]{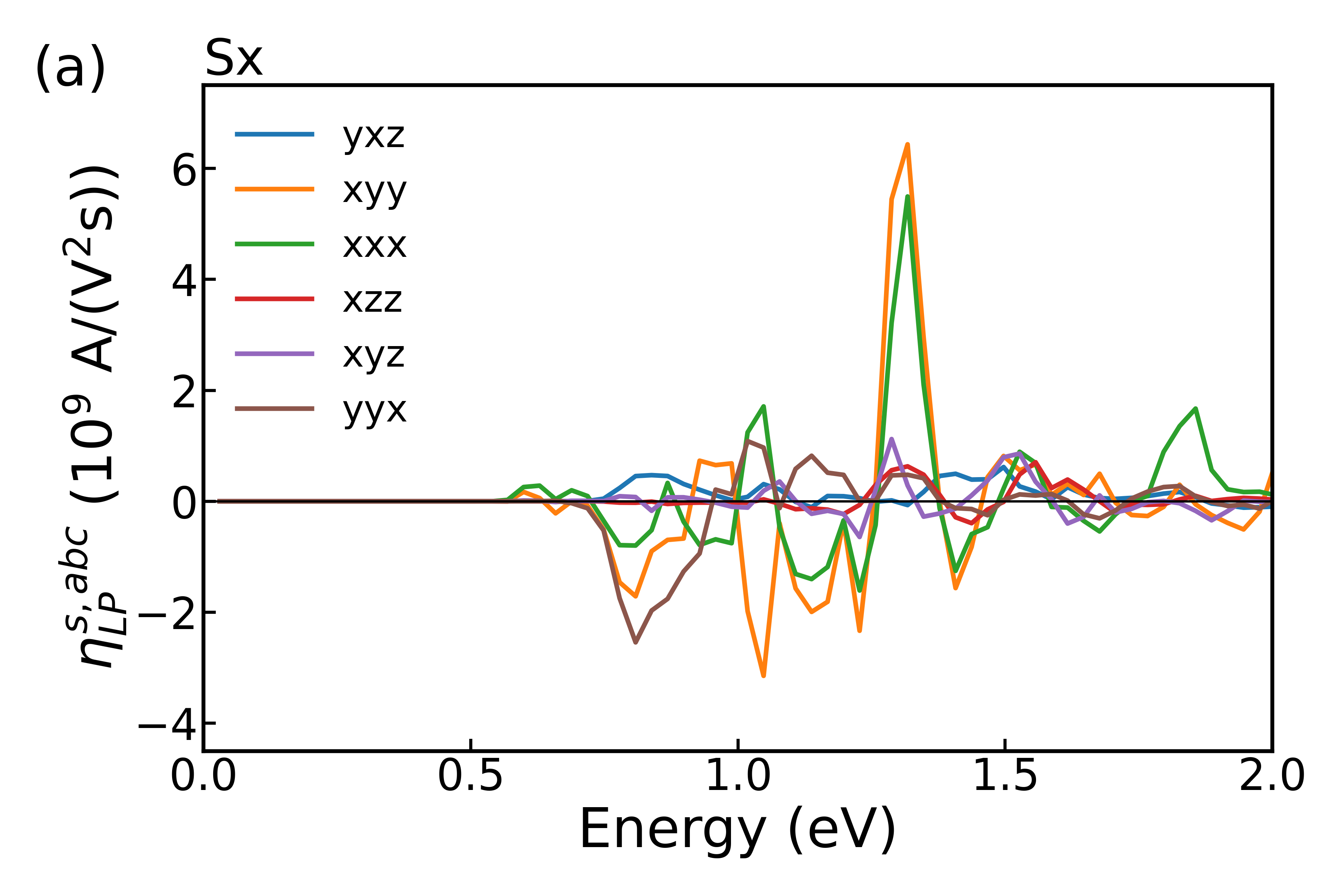}
    \includegraphics[width=7.5cm,angle=0]{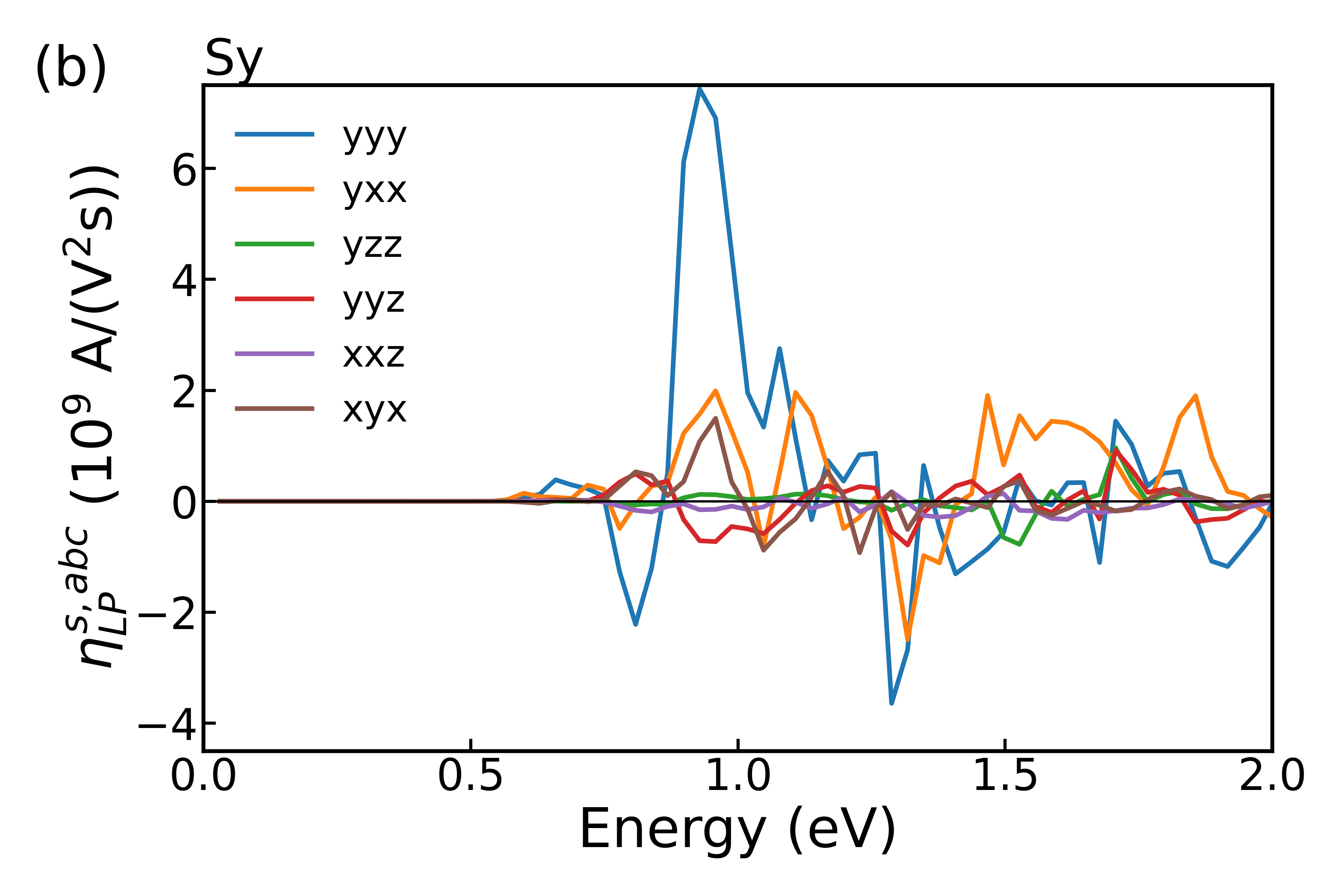}
    \includegraphics[width=7.5cm,angle=0]{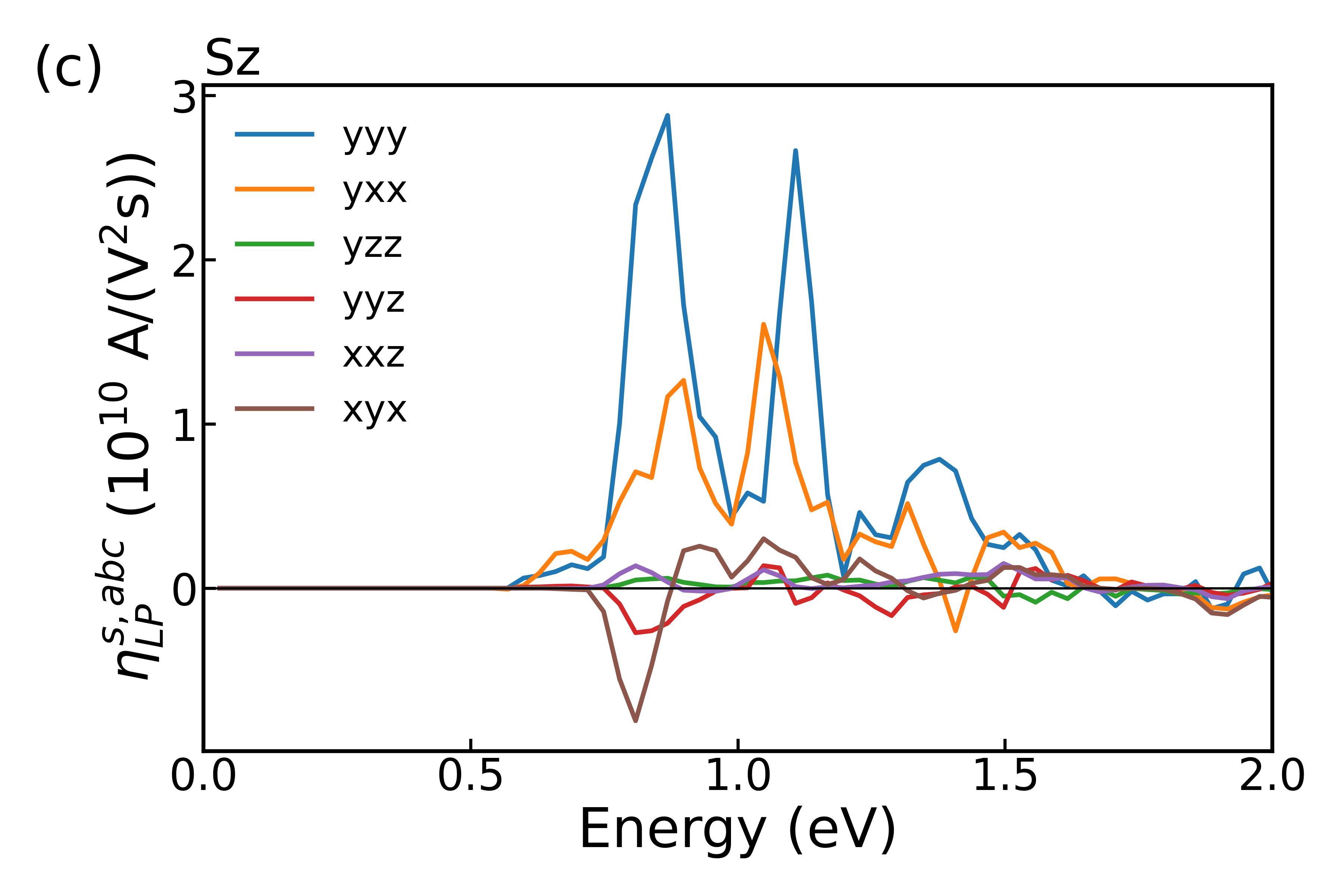}
    \caption{\justifying Non-zero independent components of the LP spin injection photoconductivity tensor for (a) $s^x$, (b) $s^y$ and (c) $s^z$ spin projections and for the $n=4$ cycloid structure.}
    \label{spinlinearinjection1x4}
\end{figure}

\subsection{LP Spin Injection Photoconductivity}
The spin injection photoconductivity tensor of the $n=4$ system under LP light exhibits the same independent nonzero components as in the $n=3$ case:

\[
\text{spin } s^x:
\quad
\eta^{x,yxz}_\mathrm{LP},\;
\eta^{x,xyy}_\mathrm{LP},\;
\eta^{x,xxx}_\mathrm{LP},\;
\eta^{x,xzz}_\mathrm{LP},\;
\eta^{x,yyx}_\mathrm{LP},\;
\eta^{x,xyz}_\mathrm{LP}.
\]

\[
\text{spin } s^y:
\quad
\eta^{y,yyy}_\mathrm{LP},\;
\eta^{y,yxx}_\mathrm{LP},\;
\eta^{y,yzz}_\mathrm{LP},\;
\eta^{y,yyz}_\mathrm{LP},\;
\eta^{y,xxz}_\mathrm{LP},\;
\eta^{y,xyx}_\mathrm{LP}.
\]

\[
\text{spin } s^z:
\quad
\eta^{z,yyy}_\mathrm{LP},\;
\eta^{z,yxx}_\mathrm{LP},\;
\eta^{z,yzz}_\mathrm{LP},\;
\eta^{z,yyz}_\mathrm{LP},\;
\eta^{z,xxz}_\mathrm{LP},\;
\eta^{z,xyx}_\mathrm{LP}.
\]

The spectra are displayed in Fig.~\ref{spinlinearinjection1x4}. 
For the $s^x$ and $s^y$ projections, the components reaching the largest peak values are 
$\eta^{x,xyy}_\mathrm{LP}$ and $\eta^{y,yyy}_\mathrm{LP}$, respectively, implying that the spin current flows parallel to the spin polarization direction.

As discussed in the main text, the largest photoconductivity occurs for the $s^z$ projection with current flowing along $y$. In particular, $\eta^{z,yyy}_\mathrm{LP}$ and $\eta^{z,yxx}_\mathrm{LP}$ reach peak values of the order of $\sim 10^{10}$~A/(V$^{2}$s), approximately one order of magnitude larger than the dominant components for the $s^x$ and $s^y$ projections (which are of order $\sim 10^{9}$~A/(V$^{2}$s)). 
This enhancement originates from the underlying nonrelativistic $p$-wave magnetic spin texture, as discussed in the manuscript.

\medskip

\bibliography{biblio}

@article{Lihm22,
  title = {Comprehensive theory of second-order spin photocurrents},
  author = {Lihm, Jae-Mo and Park, Cheol-Hwan},
  journal = {Phys. Rev. B},
  volume = {105},
  issue = {4},
  pages = {045201},
  numpages = {19},
  year = {2022},
  month = {Jan},
  publisher = {American Physical Society},
  doi = {10.1103/PhysRevB.105.045201},
  url = {https://link.aps.org/doi/10.1103/PhysRevB.105.045201}
}

@article{Amini24,
author = {Amini, Mohammad and Fumega, Adolfo O. and González-Herrero, Héctor and Vaňo, Viliam and Kezilebieke, Shawulienu and Lado, Jose L. and Liljeroth, Peter},
title = {Atomic-Scale Visualization of Multiferroicity in Monolayer NiI2},
journal = {Advanced Materials},
volume = {36},
number = {18},
pages = {2311342},
doi = {https://doi.org/10.1002/adma.202311342},
url = {https://advanced.onlinelibrary.wiley.com/doi/abs/10.1002/adma.202311342},
year = {2024}
}

@article{Tseng25,
author = {Tseng, Yi and Occhialini, Connor A. and Song, Qian and Barone, Paolo and Patel, Sahaj and Shankar, Meghna and Acevedo-Esteves, Raul and Li, Jiarui and Nelson, Christie and Picozzi, Silvia and Sutarto, Ronny and Comin, Riccardo},
title = {Shear-Mediated Stabilization of Spin Spiral Order in Multiferroic NiI2},
journal = {Advanced Materials},
volume = {37},
number = {9},
pages = {2417434},
doi = {https://doi.org/10.1002/adma.202417434},
url = {https://advanced.onlinelibrary.wiley.com/doi/abs/10.1002/adma.202417434},
eprint = {https://advanced.onlinelibrary.wiley.com/doi/pdf/10.1002/adma.202417434},
year = {2025}
}

@article{Ju21,
author = {Ju, Hwiin  and  Lee, Youjin and Kim, Kwang-Tak and Choi, In Hyeok and Roh, Chang Jae and Son, Suhan and Park, Pyeongjae and Kim, Jae Ha and Jung, Taek Sun and Kim, Jae Hoon and Kim, Kee Hoon and Park, Je-Geun and Lee, Jong Seok},
title = {Possible Persistence of Multiferroic Order down to Bilayer Limit of van der Waals Material NiI2},
journal = {Nano Lett.},
volume = {21},
pages = {5126-5132},
year = {2021},
doi = {10.1021/acs.nanolett.1c01095},
URL = {https://doi.org/10.1021/acs.nanolett.1c01095
}}

@article{Song25,
doi = {
DO  - },
url = {https://doi.org/10.1038/s41586-025-09034-7},
year = {2025},
volume = {642},
number = {8066},
pages = {64-70},
author = {Quian Song
and Srdjan Stavrić and Paolo Barone and Andrea Droghetti and DAniil S. Antonenko and Jorn W. F. Venderbos and Connor A. Occhialini and Batyr Ilyas and Emre Ergecen and Nuh Fedik and Sang-Wook Cheong and Rafael M. Fernandes and Silvia Picozzi and Riccardo Comin},
title = {Electrical switching of a p-wave magnet },
journal = {Nature}
}

@article{Cuono25,
  title = {Bulk photovoltaic effect in ferroelectric and antiferroelectric phases of antimony sulphoiodide investigated by means of $\mathit{ab}\text{\ensuremath{-}}\mathit{initio}$ simulations},
  author = {Cuono, Giuseppe and Bandyopadhyay, Subhadeep and Droghetti, Andrea and Picozzi, Silvia},
  journal = {Phys. Rev. Mater.},
  volume = {9},
  issue = {6},
  pages = {064412},
  numpages = {11},
  year = {2025},
  month = {Jun},
  publisher = {American Physical Society},
  doi = {10.1103/crdx-prhl},
  url = {https://link.aps.org/doi/10.1103/crdx-prhl}
}

@article{Stavric25,
  title = {Giant Nonreciprocal Band Structure Effect in a Multiferroic Material},
  author = {Stavri\ifmmode \acute{c}\else \'{c}\fi{}, Srdjan and Cuono, Giuseppe and Yang, Baishun and Puente-Uriona, \'Alvaro R. and Iba\~nez-Azpiroz, Julen and Barone, Paolo and Droghetti, Andrea and Picozzi, Silvia},
  journal = {Phys. Rev. Lett.},
  volume = {135},
  issue = {20},
  pages = {206401},
  numpages = {8},
  year = {2025},
  month = {Nov},
  publisher = {American Physical Society},
  doi = {10.1103/15ws-ftbf},
  url = {https://link.aps.org/doi/10.1103/15ws-ftbf}
}

@article{Song22,
doi = {10.1038/s41586-021-04337-x},
url = {https://doi.org/10.1038/s41586-021-04337-x},
year = {2022},
volume = {602},
number = {7898},
pages = {601-605},
author = {Song, Qian and Occhialini, Connor A. and Ergeçen, Emre and Ilyas, Batyr and Amoroso, Danila and Barone, Paolo and Kapeghian, Jesse and Watanabe, Kenji and Taniguchi, Takashi and Botana, Antia S. and Picozzi, Silvia and Gedik, Nuh and Comin, Riccardo},
title = {Evidence for a single-layer van der Waals multiferroic},
journal = {Nature}
}

@article{Perdew96,
  title = {Generalized Gradient Approximation Made Simple},
  author = {Perdew, John P. and Burke, Kieron and Ernzerhof, Matthias},
  journal = {Phys. Rev. Lett.},
  volume = {77},
  issue = {18},
  pages = {3865--3868},
  numpages = {0},
  year = {1996},
  month = {Oct},
  publisher = {American Physical Society},
  doi = {10.1103/PhysRevLett.77.3865},
  url = {https://link.aps.org/doi/10.1103/PhysRevLett.77.3865}
}

@article{Kresse93,
  title = {Ab initio molecular dynamics for liquid metals},
  author = {Kresse, G. and Hafner, J.},
  journal = {Phys. Rev. B},
  volume = {47},
  issue = {1},
  pages = {558--561},
  numpages = {0},
  year = {1993},
  month = {Jan},
  publisher = {American Physical Society},
  doi = {10.1103/PhysRevB.47.558},
  url = {https://link.aps.org/doi/10.1103/PhysRevB.47.558}
}

@article{Kresse96,
title = {Efficiency of ab-initio total energy calculations for metals and semiconductors using a plane-wave basis set},
journal = {Computational Materials Science},
volume = {6},
number = {1},
pages = {15-50},
year = {1996},
issn = {0927-0256},
doi = {https://doi.org/10.1016/0927-0256(96)00008-0},
url = {https://www.sciencedirect.com/science/article/pii/0927025696000080},
author = {G. Kresse and J. Furthmüller},
}

@article{Kresse96b,
  title = {Efficient iterative schemes for ab initio total-energy calculations using a plane-wave basis set},
  author = {Kresse, G. and Furthm\"uller, J.},
  journal = {Phys. Rev. B},
  volume = {54},
  issue = {16},
  pages = {11169--11186},
  numpages = {0},
  year = {1996},
  month = {Oct},
  publisher = {American Physical Society},
  doi = {10.1103/PhysRevB.54.11169},
  url = {https://link.aps.org/doi/10.1103/PhysRevB.54.11169}
}

@article{Mostofi08,
title = {wannier90: A tool for obtaining maximally-localised Wannier functions},
journal = {Computer Physics Communications},
volume = {178},
number = {9},
pages = {685-699},
year = {2008},
issn = {0010-4655},
doi = {https://doi.org/10.1016/j.cpc.2007.11.016},
url = {https://www.sciencedirect.com/science/article/pii/S0010465507004936},
author = {Arash A. Mostofi and Jonathan R. Yates and Young-Su Lee and Ivo Souza and David Vanderbilt and Nicola Marzari}
}

@article{Azpiroz18,
  title = {Ab initio calculation of the shift photocurrent by Wannier interpolation},
  author = {Iba\~nez-Azpiroz, Julen and Tsirkin, Stepan S. and Souza, Ivo},
  journal = {Phys. Rev. B},
  volume = {97},
  issue = {24},
  pages = {245143},
  numpages = {13},
  year = {2018},
  month = {Jun},
  publisher = {American Physical Society},
  doi = {10.1103/PhysRevB.97.245143},
  url = {https://link.aps.org/doi/10.1103/PhysRevB.97.245143}
}

@article{Dai23,
    author = {Dai, Zhenbang and Rappe, Andrew M.},
    title = "{Recent progress in the theory of bulk photovoltaic effect}",
    journal = {Chemical Physics Reviews},
    volume = {4},
    number = {1},
    pages = {011303},
    year = {2023},
    month = {01},
    issn = {2688-4070},
    doi = {10.1063/5.0101513},
    url = {https://doi.org/10.1063/5.0101513}
}

@article{Sipe00,
  title = {Second-order optical response in semiconductors},
  author = {Sipe, J. E. and Shkrebtii, A. I.},
  journal = {Phys. Rev. B},
  volume = {61},
  issue = {8},
  pages = {5337--5352},
  numpages = {0},
  year = {2000},
  month = {Feb},
  publisher = {American Physical Society},
  doi = {10.1103/PhysRevB.61.5337},
  url = {https://link.aps.org/doi/10.1103/PhysRevB.61.5337}
}

@article{Young12,
  title = {First Principles Calculation of the Shift Current Photovoltaic Effect in Ferroelectrics},
  author = {Young, Steve M. and Rappe, Andrew M.},
  journal = {Phys. Rev. Lett.},
  volume = {109},
  issue = {11},
  pages = {116601},
  numpages = {5},
  year = {2012},
  month = {Sep},
  publisher = {American Physical Society},
  doi = {10.1103/PhysRevLett.109.116601},
  url = {https://link.aps.org/doi/10.1103/PhysRevLett.109.116601}
}

@Article{Butler15,
author ="Butler, Keith T. and Frost, Jarvist M. and Walsh, Aron",
title  ="Ferroelectric materials for solar energy conversion: photoferroics revisited",
journal  ="Energy Environ. Sci.",
year  ="2015",
volume  ="8",
issue  ="3",
pages  ="838-848",
publisher  ="The Royal Society of Chemistry",
doi  ="10.1039/C4EE03523B",
url  ="http://dx.doi.org/10.1039/C4EE03523B"}

@article{Tan16,
author = {Tan, Liang Z and Zheng, Fan and Young, Steve M and Wang, Fenggong and Liu, Shi and Rappe, Andrew M},
title = {Shift current bulk photovoltaic effect in polar materials—hybrid and oxide perovskites and beyond},
journal = {npj Computational Materials},
volume = {2},
issue = {1},
year = {2016},
doi = {10.1038/npjcompumats.2016.26},
URL = {https://doi.org/10.1038/npjcompumats.2016.26}}

@article{Young12b,
  title = {First-Principles Calculation of the Bulk Photovoltaic Effect in Bismuth Ferrite},
  author = {Young, Steve M. and Zheng, Fan and Rappe, Andrew M.},
  journal = {Phys. Rev. Lett.},
  volume = {109},
  issue = {23},
  pages = {236601},
  numpages = {5},
  year = {2012},
  month = {Dec},
  publisher = {American Physical Society},
  doi = {10.1103/PhysRevLett.109.236601},
  url = {https://link.aps.org/doi/10.1103/PhysRevLett.109.236601}
}

@article{Tiwari22,
doi = {10.1088/1361-648X/ac8b50},
url = {https://dx.doi.org/10.1088/1361-648X/ac8b50},
year = {2022},
month = {aug},
publisher = {IOP Publishing},
volume = {34},
number = {43},
pages = {435404},
author = {Rajender Prasad Tiwari},
title = {Enhanced shift current bulk photovoltaic effect in ferroelectric Rashba semiconductor $\alpha$-GeTe: ab initio study from three- to two-dimensional van der Waals layered structures},
journal = {Journal of Physics: Condensed Matter}
}

@article{Rangel17,
  title = {Large Bulk Photovoltaic Effect and Spontaneous Polarization of Single-Layer Monochalcogenides},
  author = {Rangel, Tonatiuh and Fregoso, Benjamin M. and Mendoza, Bernardo S. and Morimoto, Takahiro and Moore, Joel E. and Neaton, Jeffrey B.},
  journal = {Phys. Rev. Lett.},
  volume = {119},
  issue = {6},
  pages = {067402},
  numpages = {6},
  year = {2017},
  month = {Aug},
  publisher = {American Physical Society},
  doi = {10.1103/PhysRevLett.119.067402},
  url = {https://link.aps.org/doi/10.1103/PhysRevLett.119.067402}
}

@article{Puente23,
  title = {Ab initio study of the nonlinear optical properties and dc photocurrent of the Weyl semimetal ${\mathrm{TaIrTe}}_{4}$},
  author = {Puente-Uriona, \'Alvaro R. and Tsirkin, Stepan S. and Souza, Ivo and Iba\~nez-Azpiroz, Julen},
  journal = {Phys. Rev. B},
  volume = {107},
  issue = {20},
  pages = {205204},
  numpages = {9},
  year = {2023},
  month = {May},
  publisher = {American Physical Society},
  doi = {10.1103/PhysRevB.107.205204},
  url = {https://link.aps.org/doi/10.1103/PhysRevB.107.205204}
}

@article{Ganichev14,
author = {Ganichev, Sergey D. and Golub, Leonid E.},
title = {Interplay of Rashba/Dresselhaus spin splittings probed by photogalvanic spectroscopy –A review},
journal = {physica status solidi (b)},
volume = {251},
number = {9},
pages = {1801-1823},
doi = {https://doi.org/10.1002/pssb.201350261},
url = {https://onlinelibrary.wiley.com/doi/abs/10.1002/pssb.201350261},
year = {2014}
}

@article{deJuan2017,
author={de Juan, Fernando
and Grushin, Adolfo G.
and Morimoto, Takahiro
and Moore, Joel E.},
title={Quantized circular photogalvanic effect in Weyl semimetals},
journal={Nature Communications},
year={2017},
month={Jul},
day={06},
volume={8},
number={1},
pages={15995},
issn={2041-1723},
doi={10.1038/ncomms15995},
url={https://doi.org/10.1038/ncomms15995}
}

@Article{Yuan2014,
author={Yuan, Hongtao
and Wang, Xinqiang
and Lian, Biao
and Zhang, Haijun
and Fang, Xianfa
and Shen, Bo
and Xu, Gang
and Xu, Yong
and Zhang, Shou-Cheng
and Hwang, Harold Y.
and Cui, Yi},
title={Generation and electric control of spin--valley-coupled circular photogalvanic current in WSe2},
journal={Nature Nanotechnology},
year={2014},
month={Oct},
day={01},
volume={9},
number={10},
pages={851-857},
issn={1748-3395},
doi={10.1038/nnano.2014.183},
url={https://doi.org/10.1038/nnano.2014.183}
}

@article{Pizzi2020,
doi = {10.1088/1361-648X/ab51ff},
url = {https://dx.doi.org/10.1088/1361-648X/ab51ff},
year = {2020},
month = {jan},
publisher = {IOP Publishing},
volume = {32},
number = {16},
pages = {165902},
author = {Giovanni Pizzi and Valerio Vitale and Ryotaro Arita and Stefan Blügel and Frank Freimuth and Guillaume Géranton and Marco Gibertini and Dominik Gresch and Charles Johnson and Takashi Koretsune and Julen Ibañez-Azpiroz and Hyungjun Lee and Jae-Mo Lihm and Daniel Marchand and Antimo Marrazzo and Yuriy Mokrousov and Jamal I Mustafa and Yoshiro Nohara and Yusuke Nomura and Lorenzo Paulatto and Samuel Poncé and Thomas Ponweiser and Junfeng Qiao and Florian Thöle and Stepan S Tsirkin and Małgorzata Wierzbowska and Nicola Marzari and David Vanderbilt and Ivo Souza and Arash A Mostofi and Jonathan R Yates},
title = {Wannier90 as a community code: new features and applications},
journal = {Journal of Physics: Condensed Matter}
}

@article{Marzari97,
  title = {Maximally localized generalized Wannier functions for composite energy bands},
  author = {Marzari, Nicola and Vanderbilt, David},
  journal = {Phys. Rev. B},
  volume = {56},
  issue = {20},
  pages = {12847--12865},
  numpages = {0},
  year = {1997},
  month = {Nov},
  publisher = {American Physical Society},
  doi = {10.1103/PhysRevB.56.12847},
  url = {https://link.aps.org/doi/10.1103/PhysRevB.56.12847}
}

@Article{Zhang2019,
author={Zhang, Yang
and Holder, Tobias
and Ishizuka, Hiroaki
and de Juan, Fernando
and Nagaosa, Naoto
and Felser, Claudia
and Yan, Binghai},
title={Switchable magnetic bulk photovoltaic effect in the two-dimensional magnet CrI3},
journal={Nature Communications},
year={2019},
month={Aug},
day={22},
volume={10},
number={1},
pages={3783},
issn={2041-1723},
doi={10.1038/s41467-019-11832-3},
url={https://doi.org/10.1038/s41467-019-11832-3}
}

@article{Cheong2007,
author={Cheong, Sang-Wook
and Mostovoy, Maxim},
title={Multiferroics: a magnetic twist for ferroelectricity},
journal={Nature Materials},
year={2007},
month={Jan},
day={01},
volume={6},
number={1},
pages={13-20},
issn={1476-4660},
doi={10.1038/nmat1804},
url={https://doi.org/10.1038/nmat1804}
}

@article{Katsura2005,
  title = {Spin Current and Magnetoelectric Effect in Noncollinear Magnets},
  author = {Katsura, Hosho and Nagaosa, Naoto and Balatsky, Alexander V.},
  journal = {Phys. Rev. Lett.},
  volume = {95},
  issue = {5},
  pages = {057205},
  numpages = {4},
  year = {2005},
  month = {Jul},
  publisher = {American Physical Society},
  doi = {10.1103/PhysRevLett.95.057205},
  url = {https://link.aps.org/doi/10.1103/PhysRevLett.95.057205}
}

@article{Mostovoy2006,
  title = {Ferroelectricity in Spiral Magnets},
  author = {Mostovoy, Maxim},
  journal = {Phys. Rev. Lett.},
  volume = {96},
  issue = {6},
  pages = {067601},
  numpages = {4},
  year = {2006},
  month = {Feb},
  publisher = {American Physical Society},
  doi = {10.1103/PhysRevLett.96.067601},
  url = {https://link.aps.org/doi/10.1103/PhysRevLett.96.067601}
}

@article{Hur2004,
author={Hur, N.
and Park, S.
and Sharma, P. A.
and Ahn, J. S.
and Guha, S.
and Cheong, S.-W.},
title={Electric polarization reversal and memory in a multiferroic material induced by magnetic fields},
journal={Nature},
year={2004},
month={May},
day={01},
volume={429},
number={6990},
pages={392-395},
issn={1476-4687},
doi={10.1038/nature02572},
url={https://doi.org/10.1038/nature02572}
}

@article{Lawes2005,
  title = {Magnetically Driven Ferroelectric Order in ${\mathrm{Ni}}_{3}{\mathrm{V}}_{2}{\mathrm{O}}_{8}$},
  author = {Lawes, G. and Harris, A. B. and Kimura, T. and Rogado, N. and Cava, R. J. and Aharony, A. and Entin-Wohlman, O. and Yildirim, T. and Kenzelmann, M. and Broholm, C. and Ramirez, A. P.},
  journal = {Phys. Rev. Lett.},
  volume = {95},
  issue = {8},
  pages = {087205},
  numpages = {4},
  year = {2005},
  month = {Aug},
  publisher = {American Physical Society},
  doi = {10.1103/PhysRevLett.95.087205},
  url = {https://link.aps.org/doi/10.1103/PhysRevLett.95.087205}
}

@article{Kenzelmann2005,
  title = {Magnetic Inversion Symmetry Breaking and Ferroelectricity in ${\mathrm{TbMnO}}_{3}$},
  author = {Kenzelmann, M. and Harris, A. B. and Jonas, S. and Broholm, C. and Schefer, J. and Kim, S. B. and Zhang, C. L. and Cheong, S.-W. and Vajk, O. P. and Lynn, J. W.},
  journal = {Phys. Rev. Lett.},
  volume = {95},
  issue = {8},
  pages = {087206},
  numpages = {4},
  year = {2005},
  month = {Aug},
  publisher = {American Physical Society},
  doi = {10.1103/PhysRevLett.95.087206},
  url = {https://link.aps.org/doi/10.1103/PhysRevLett.95.087206}
}

@article{Taniguchi2006,
  title = {Ferroelectric Polarization Flop in a Frustrated Magnet ${\mathrm{MnWO}}_{4}$ Induced by a Magnetic Field},
  author = {Taniguchi, K. and Abe, N. and Takenobu, T. and Iwasa, Y. and Arima, T.},
  journal = {Phys. Rev. Lett.},
  volume = {97},
  issue = {9},
  pages = {097203},
  numpages = {4},
  year = {2006},
  month = {Aug},
  publisher = {American Physical Society},
  doi = {10.1103/PhysRevLett.97.097203},
  url = {https://link.aps.org/doi/10.1103/PhysRevLett.97.097203}
}

@misc{hellenes2024,
      title={P-wave magnets}, 
      author={Anna Birk Hellenes and Tomáš Jungwirth and Rodrigo Jaeschke-Ubiergo and Atasi Chakraborty and Jairo Sinova and Libor Šmejkal},
      year={2024},
      eprint={2309.01607},
      archivePrefix={arXiv},
      primaryClass={cond-mat.mes-hall},
      url={https://arxiv.org/abs/2309.01607}, 
}

@article{yamada2025,
author={Yamada, Rinsuke
and Birch, Max T.
and Baral, Priya R.
and Okumura, Shun
and Nakano, Ryota
and Gao, Shang
and Ezawa, Motohiko
and Nomoto, Takuya
and Masell, Jan
and Ishihara, Yuki
and Kolincio, Kamil K.
and Belopolski, Ilya
and Sagayama, Hajime
and Nakao, Hironori
and Ohishi, Kazuki
and Ohhara, Takashi
and Kiyanagi, Ryoji
and Nakajima, Taro
and Tokura, Yoshinori
and Arima, Taka-hisa
and Motome, Yukitoshi
and Hirschmann, Moritz M.
and Hirschberger, Max},
title={A metallic p-wave magnet with commensurate spin helix},
journal={Nature},
year={2025},
month={Oct},
day={01},
volume={646},
number={8086},
pages={837-842},
issn={1476-4687},
doi={10.1038/s41586-025-09633-4},
url={https://doi.org/10.1038/s41586-025-09633-4}
}

@article{Belinicher1980,
doi = {10.1070/PU1980v023n03ABEH004703},
url = {https://doi.org/10.1070/PU1980v023n03ABEH004703},
year = {1980},
month = {mar},
publisher = {},
volume = {23},
number = {3},
pages = {199},
author = {V I Belinicher and B I Sturman},
title = {The photogalvanic effect in media lacking a center of
symmetry},
journal = {Soviet Physics Uspekhi}
}

@article{PismaZhETF.27.640,
    title = {New photogalvanic effect in gyrotropic crystals},
    author = {Ivchenko, E. L. and Pikus, G. E.},
    journal = {ZhETF Pisma Redaktsiiu},
    volume = {27},
    issue = {11},
    pages = {640},
    year = {1978},
    doi = {},
    url = {http://jetpletters.ru/ps/0/article_23792.shtml},
}

@article{Jiang2025,
author = {Jiang, Xiao and Jeong, Uiseok and Sato, Shunsuke and Shin, Dongbin and Yabana, Kazuhiro and Yan, Binghai and Park, Noejung},
title = {Nonlinear Photocurrent as a Hallmark of Altermagnet},
journal = {ACS Nano},
volume = {19},
number = {26},
pages = {23620-23628},
year = {2025},
doi = {10.1021/acsnano.5c01421},
note ={PMID: 40569840},
URL = {https://doi.org/10.1021/acsnano.5c01421}
}

@article{Bhat2005,
  title = {Pure Spin Current from One-Photon Absorption of Linearly Polarized Light in Noncentrosymmetric Semiconductors},
  author = {Bhat, R. D. R and Nastos, F. and Najmaie, Ali and Sipe, J. E.},
  journal = {Phys. Rev. Lett.},
  volume = {94},
  issue = {9},
  pages = {096603},
  numpages = {4},
  year = {2005},
  month = {Mar},
  publisher = {American Physical Society},
  doi = {10.1103/PhysRevLett.94.096603},
  url = {https://link.aps.org/doi/10.1103/PhysRevLett.94.096603}
}

@article{Young2013,
  title = {Prediction of a Linear Spin Bulk Photovoltaic Effect in Antiferromagnets},
  author = {Young, Steve M. and Zheng, Fan and Rappe, Andrew M.},
  journal = {Phys. Rev. Lett.},
  volume = {110},
  issue = {5},
  pages = {057201},
  numpages = {4},
  year = {2013},
  month = {Jan},
  publisher = {American Physical Society},
  doi = {10.1103/PhysRevLett.110.057201},
  url = {https://link.aps.org/doi/10.1103/PhysRevLett.110.057201}
}

@article{Lihm2022,
  title = {Comprehensive theory of second-order spin photocurrents},
  author = {Lihm, Jae-Mo and Park, Cheol-Hwan},
  journal = {Phys. Rev. B},
  volume = {105},
  issue = {4},
  pages = {045201},
  numpages = {19},
  year = {2022},
  month = {Jan},
  publisher = {American Physical Society},
  doi = {10.1103/PhysRevB.105.045201},
  url = {https://link.aps.org/doi/10.1103/PhysRevB.105.045201}
}

@article{Asnin1978,
    title = {Observation of a photo-emf that depends on the sign of the circular polarization of the light},
    author = {Asnin, V. M. and Bakun, A. A. and Danishevskii, A. M. and Ivchenko, E. L. and Pikus, G. E. and Rogachev, A. A.},
    journal = {ZhETF Pisma Redaktsiiu},
    volume = {28},
    issue = {2},
    pages = {80},
    year = {1978},
    doi = {},
    url = {http://jetpletters.ru/ps/0/article_23830.shtml},
}

@article{Asnin1979,
title = {“Circular” photogalvanic effect in optically active crystals},
journal = {Solid State Communications},
volume = {30},
number = {9},
pages = {565-570},
year = {1979},
issn = {0038-1098},
doi = {https://doi.org/10.1016/0038-1098(79)91137-2},
url = {https://www.sciencedirect.com/science/article/pii/0038109879911372},
author = {V.M. Asnin and A.A. Bakun and A.M. Danishevskii and E.L. Ivchenko and G.E. Pikus and A.A. Rogachev}
}

@misc{ganichev2003spinphotocurrentsquantumwells1,
      title={Spin Photocurrents in Quantum Wells review part I, (part II: cond-mat/one of the next numbers)}, 
      author={S. D. Ganichev and W. Prettl},
      year={2003},
      eprint={cond-mat/0304266},
      archivePrefix={arXiv},
      primaryClass={cond-mat},
      url={https://arxiv.org/abs/cond-mat/0304266}, 
}

@misc{ganichev2003spinphotocurrentsquantumwells2,
      title={Spin Photocurrents in Quantum Wells review part II, (part I: cond-mat/0304266)}, 
      author={S. D. Ganichev and W. Prettl},
      year={2003},
      eprint={cond-mat/0304268},
      archivePrefix={arXiv},
      primaryClass={cond-mat},
      url={https://arxiv.org/abs/cond-mat/0304268}, 
}

@Article{Xu2021,
author={Xu, Haowei
and Wang, Hua
and Zhou, Jian
and Li, Ju},
title={Pure spin photocurrent in non-centrosymmetric crystals: bulk spin photovoltaic effect},
journal={Nature Communications},
year={2021},
month={Jul},
day={15},
volume={12},
number={1},
pages={4330},
issn={2041-1723},
doi={10.1038/s41467-021-24541-7},
url={https://doi.org/10.1038/s41467-021-24541-7}
}

@article{Marzari1997,
  title = {Maximally localized Wannier functions: Theory and applications},
  author = {Marzari, Nicola and Mostofi, Arash A. and Yates, Jonathan R. and Souza, Ivo and Vanderbilt, David},
  journal = {Rev. Mod. Phys.},
  volume = {84},
  issue = {4},
  pages = {1419--1475},
  numpages = {0},
  year = {2012},
  month = {Oct},
  publisher = {American Physical Society},
  doi = {10.1103/RevModPhys.84.1419},
  url = {https://link.aps.org/doi/10.1103/RevModPhys.84.1419}
}

@article{Mostofi2008,
title = {wannier90: A tool for obtaining maximally-localised Wannier functions},
journal = {Computer Physics Communications},
volume = {178},
number = {9},
pages = {685-699},
year = {2008},
issn = {0010-4655},
doi = {https://doi.org/10.1016/j.cpc.2007.11.016},
url = {https://www.sciencedirect.com/science/article/pii/S0010465507004936},
author = {Arash A. Mostofi and Jonathan R. Yates and Young-Su Lee and Ivo Souza and David Vanderbilt and Nicola Marzari}
}

@article{Sivianes25,
  title = {Optical Signatures of Spin Symmetries in Unconventional Magnets},
  author = {Sivianes, Javier and Santos, Flaviano Jos\'e dos and Iba\~nez-Azpiroz, Julen},
  journal = {Phys. Rev. Lett.},
  volume = {134},
  issue = {19},
  pages = {196907},
  numpages = {7},
  year = {2025},
  month = {May},
  publisher = {American Physical Society},
  doi = {10.1103/PhysRevLett.134.196907},
  url = {https://link.aps.org/doi/10.1103/PhysRevLett.134.196907}
}

@Article{McIver2012,
author={McIver, J. W.
and Hsieh, D.
and Steinberg, H.
and Jarillo-Herrero, P.
and Gedik, N.},
title={Control over topological insulator photocurrents with light polarization},
journal={Nature Nanotechnology},
year={2012},
month={Feb},
day={01},
volume={7},
number={2},
pages={96-100},
issn={1748-3395},
doi={10.1038/nnano.2011.214},
url={https://doi.org/10.1038/nnano.2011.214}
}

@article{Lechner2011,
  title = {Spin and orbital mechanisms of the magnetogyrotropic photogalvanic effects in GaAs/Al${}_{x}$Ga${}_{1\ensuremath{-}x}$As quantum well structures},
  author = {Lechner, V. and Golub, L. E. and Lomakina, F. and Bel'kov, V. V. and Olbrich, P. and Stachel, S. and Caspers, I. and Griesbeck, M. and Kugler, M. and Hirmer, M. J. and Korn, T. and Sch\"uller, C. and Schuh, D. and Wegscheider, W. and Ganichev, S. D.},
  journal = {Phys. Rev. B},
  volume = {83},
  issue = {15},
  pages = {155313},
  numpages = {9},
  year = {2011},
  month = {Apr},
  publisher = {American Physical Society},
  doi = {10.1103/PhysRevB.83.155313},
  url = {https://link.aps.org/doi/10.1103/PhysRevB.83.155313}
}

@article{Taniuchi25,
author = {Taniuchi, Ibuki and Akiyama, Ryota and Hobara, Rei and Hasegawa, Shuji},
title = {Surface Circular Photogalvanic Effect in Tl–Pb Monolayer Alloys on Si(111) with Giant Rashba Splitting},
journal = {ACS Nano},
volume = {19},
number = {3},
pages = {3147-3154},
year = {2025},
doi = {10.1021/acsnano.4c08742},
note ={PMID: 39792011},
URL = {https://doi.org/10.1021/acsnano.4c08742}
}

@Article{Quereda2018,
author={Quereda, Jorge
and Ghiasi, Talieh S.
and You, Jhih-Shih
and van den Brink, Jeroen
and van Wees, Bart J.
and van der Wal, Caspar H.},
title={Symmetry regimes for circular photocurrents in monolayer MoSe2},
journal={Nature Communications},
year={2018},
month={Aug},
day={21},
volume={9},
number={1},
pages={3346},
issn={2041-1723},
doi={10.1038/s41467-018-05734-z},
url={https://doi.org/10.1038/s41467-018-05734-z}
}

@article{Yu20,
author = {Yu, Jinling and Xia, Lijia and Zhu, Kejing and Pan, Qinggao and Zeng, Xiaolin and Chen, Yonghai and Liu, Yu and Yin, Chunming and Cheng, Shuying and Lai, Yunfeng and He, Ke and Xue, Qikun},
title = {Control of Circular Photogalvanic Effect of Surface States in the Topological Insulator Bi2Te3 via Spin Injection},
journal = {ACS Applied Materials \& Interfaces},
volume = {12},
number = {15},
pages = {18091-18100},
year = {2020},
doi = {10.1021/acsami.9b23389},
note ={PMID: 32212669},
URL = {https://doi.org/10.1021/acsami.9b23389}
}

@article{Hirose2018,
    author = {Hirose, Hana and Ito, Naoto and Kawaguchi, Masashi and Lau, Yong-Chang and Hayashi, Masamitsu},
    title = {Circular photogalvanic effect in Cu/Bi bilayers},
    journal = {Applied Physics Letters},
    volume = {113},
    number = {22},
    pages = {222404},
    year = {2018},
    month = {11},
    issn = {0003-6951},
    doi = {10.1063/1.5047418},
    url = {https://doi.org/10.1063/1.5047418}
}

@article{Ganichev_2003,
doi = {10.1088/0953-8984/15/20/204},
url = {https://doi.org/10.1088/0953-8984/15/20/204},
year = {2003},
month = {may},
publisher = {},
volume = {15},
number = {20},
pages = {R935},
author = {S D Ganichev and W Prettl},
title = {Spin photocurrents in quantum wells},
journal = {Journal of Physics: Condensed Matter}
}

@article{Wang22,
  title = {Circular Photogalvanic Effect in Oxide Two-Dimensional Electron Gases},
  author = {Wang, Shuanhu and Zhang, Hui and Zhang, Jine and Li, Shuqin and Luo, Dianbing and Wang, Jianyuan and Jin, Kexin and Sun, Jirong},
  journal = {Phys. Rev. Lett.},
  volume = {128},
  issue = {18},
  pages = {187401},
  numpages = {6},
  year = {2022},
  month = {May},
  publisher = {American Physical Society},
  doi = {10.1103/PhysRevLett.128.187401},
  url = {https://link.aps.org/doi/10.1103/PhysRevLett.128.187401}
}

@Article{Liu2020,
author={Liu, Xiaojie
and Chanana, Ashish
and Huynh, Uyen
and Xue, Fei
and Haney, Paul
and Blair, Steve
and Jiang, Xiaomei
and Vardeny, Z. V.},
title={Circular photogalvanic spectroscopy of Rashba splitting in 2D hybrid organic--inorganic perovskite multiple quantum wells},
journal={Nature Communications},
year={2020},
month={Jan},
day={16},
volume={11},
number={1},
pages={323},
issn={2041-1723},
doi={10.1038/s41467-019-14073-6},
url={https://doi.org/10.1038/s41467-019-14073-6}
}

@article{Ganichev2000,
    author = {Ganichev, S. D. and Ketterl, H. and Prettl, W. and Ivchenko, E. L. and Vorobjev, L. E.},
    title = {Circular photogalvanic effect induced by monopolar spin orientation in p-GaAs/AlGaAs multiple-quantum wells},
    journal = {Applied Physics Letters},
    volume = {77},
    number = {20},
    pages = {3146-3148},
    year = {2000},
    month = {11},
    issn = {0003-6951},
    doi = {10.1063/1.1326488},
    url = {https://doi.org/10.1063/1.1326488}
}

@article{Niesner18,
author = {Daniel Niesner  and Martin Hauck  and Shreetu Shrestha  and Ievgen Levchuk  and Gebhard J. Matt  and Andres Osvet  and Miroslaw Batentschuk  and Christoph Brabec  and Heiko B. Weber  and Thomas Fauster },
title = {Structural fluctuations cause spin-split states in tetragonal (CH<sub>3</sub>NH<sub>3</sub>)PbI<sub>3</sub> as evidenced by the circular photogalvanic effect},
journal = {Proceedings of the National Academy of Sciences},
volume = {115},
number = {38},
pages = {9509-9514},
year = {2018},
doi = {10.1073/pnas.1805422115},
URL = {https://www.pnas.org/doi/abs/10.1073/pnas.1805422115}
}

@article{Ganichev2003,
  title = {Resonant inversion of the circular photogalvanic effect in n-doped quantum wells},
  author = {Ganichev, S. D. and Bel'kov, V. V. and Schneider, Petra and Ivchenko, E. L. and Tarasenko, S. A. and Wegscheider, W. and Weiss, D. and Schuh, D. and Beregulin, E. V. and Prettl, W.},
  journal = {Phys. Rev. B},
  volume = {68},
  issue = {3},
  pages = {035319},
  numpages = {6},
  year = {2003},
  month = {Jul},
  publisher = {American Physical Society},
  doi = {10.1103/PhysRevB.68.035319},
  url = {https://link.aps.org/doi/10.1103/PhysRevB.68.035319}
}

@article{BELKOV2003,
title = {Circular photogalvanic effect at inter-band excitation in semiconductor quantum wells},
journal = {Solid State Communications},
volume = {128},
number = {8},
pages = {283-286},
year = {2003},
issn = {0038-1098},
doi = {https://doi.org/10.1016/j.ssc.2003.08.022},
url = {https://www.sciencedirect.com/science/article/pii/S0038109803007555},
author = {V.V. Bel'kov and S.D. Ganichev and Petra Schneider and C. Back and M. Oestreich and J. Rudolph and D. Hägele and L.E. Golub and W. Wegscheider and W. Prettl}
}

@article{Weber2005,
    author = {Weber, W. and Ganichev, S. D. and Danilov, S. N. and Weiss, D. and Prettl, W. and Kvon, Z. D. and Bel’kov, V. V. and Golub, L. E. and Cho, Hyun-Ick and Lee, Jung-Hee},
    title = {Demonstration of Rashba spin splitting in GaN-based heterostructures},
    journal = {Applied Physics Letters},
    volume = {87},
    number = {26},
    pages = {262106},
    year = {2005},
    month = {12},
    issn = {0003-6951},
    doi = {10.1063/1.2158024},
    url = {https://doi.org/10.1063/1.2158024}
}

@article{Giglberger2007,
  title = {Rashba and Dresselhaus spin splittings in semiconductor quantum wells measured by spin photocurrents},
  author = {Giglberger, S. and Golub, L. E. and Bel'kov, V. V. and Danilov, S. N. and Schuh, D. and Gerl, C. and Rohlfing, F. and Stahl, J. and Wegscheider, W. and Weiss, D. and Prettl, W. and Ganichev, S. D.},
  journal = {Phys. Rev. B},
  volume = {75},
  issue = {3},
  pages = {035327},
  numpages = {8},
  year = {2007},
  month = {Jan},
  publisher = {American Physical Society},
  doi = {10.1103/PhysRevB.75.035327},
  url = {https://link.aps.org/doi/10.1103/PhysRevB.75.035327}
}

@article{GOLUB2003,
title = {Spin splitting induced circular photocurrent in quantum wells},
journal = {Physica E: Low-dimensional Systems and Nanostructures},
volume = {17},
pages = {342-344},
year = {2003},
note = {Proceedings of the International Conference on Superlattices, Nano-structures and Nano-devices ICSNN 2002 o-structures and Nano-devices ICSNN 2002},
issn = {1386-9477},
doi = {https://doi.org/10.1016/S1386-9477(02)00828-7},
url = {https://www.sciencedirect.com/science/article/pii/S1386947702008287},
author = {L.E Golub}
}

@article{Sun2021,
author = {X. Sun  and G. Adamo  and M. Eginligil  and H. N. S. Krishnamoorthy  and N. I. Zheludev  and C. Soci },
title = {Topological insulator metamaterial with giant circular photogalvanic effect},
journal = {Science Advances},
volume = {7},
number = {14},
pages = {eabe5748},
year = {2021},
doi = {10.1126/sciadv.abe5748},
URL = {https://www.science.org/doi/abs/10.1126/sciadv.abe5748}}

@article{Hosur11,
  title = {Circular photogalvanic effect on topological insulator surfaces: Berry-curvature-dependent response},
  author = {Hosur, Pavan},
  journal = {Phys. Rev. B},
  volume = {83},
  issue = {3},
  pages = {035309},
  numpages = {7},
  year = {2011},
  month = {Jan},
  publisher = {American Physical Society},
  doi = {10.1103/PhysRevB.83.035309},
  url = {https://link.aps.org/doi/10.1103/PhysRevB.83.035309}
}

@Article{Ji2019,
author={Ji, Zhurun
and Liu, Gerui
and Addison, Zachariah
and Liu, Wenjing
and Yu, Peng
and Gao, Heng
and Liu, Zheng
and Rappe, Andrew M.
and Kane, Charles L.
and Mele, Eugene J.
and Agarwal, Ritesh},
title={Spatially dispersive circular photogalvanic effect in a Weyl semimetal},
journal={Nature Materials},
year={2019},
month={Sep},
day={01},
volume={18},
number={9},
pages={955-962},
issn={1476-4660},
doi={10.1038/s41563-019-0421-5},
url={https://doi.org/10.1038/s41563-019-0421-5}
}

@article{Sivianes25_2,
  title = {Surface-State Engineering for Generation of Nonlinear Charge and Spin Photocurrents},
  author = {Sivianes, Javier and Garcia-Goiricelaya, Peio and Hernang\'omez-P\'erez, Daniel and Iba\~nez-Azpiroz, Julen},
  journal = {Phys. Rev. Lett.},
  volume = {135},
  issue = {25},
  pages = {256201},
  numpages = {9},
  year = {2025},
  month = {Dec},
  publisher = {American Physical Society},
  doi = {10.1103/h8rp-rtn8},
  url = {https://link.aps.org/doi/10.1103/h8rp-rtn8}
}

@article{Kim1017,
  title = {Shift charge and spin photocurrents in Dirac surface states of topological insulator},
  author = {Kim, Kun Woo and Morimoto, Takahiro and Nagaosa, Naoto},
  journal = {Phys. Rev. B},
  volume = {95},
  issue = {3},
  pages = {035134},
  numpages = {6},
  year = {2017},
  month = {Jan},
  publisher = {American Physical Society},
  doi = {10.1103/PhysRevB.95.035134},
  url = {https://link.aps.org/doi/10.1103/PhysRevB.95.035134}
}

@article{KUINDERSMA1981,
title = {Magnetic and structural investigations on NiI2 and CoI2},
journal = {Physica B+C},
volume = {111},
number = {2},
pages = {231-248},
year = {1981},
issn = {0378-4363},
doi = {https://doi.org/10.1016/0378-4363(81)90100-5},
url = {https://www.sciencedirect.com/science/article/pii/0378436381901005},
author = {S.R. Kuindersma and J.P. Sanchez and C. Haas}
}

@article{Friedt76,
    author = {Friedt, J. M. and Sanchez, J. P. and Shenoy, G. K.},
    title = {Electronic and magnetic properties of metal diiodides MI2 (M=V, Cr, Mn, Fe, Co, Ni, and Cd) from 129I Mössbauer spectroscopy},
    journal = {The Journal of Chemical Physics},
    volume = {65},
    number = {12},
    pages = {5093-5102},
    year = {1976},
    month = {12},
    issn = {0021-9606},
    doi = {10.1063/1.433072},
    url = {https://doi.org/10.1063/1.433072}
}

@article{BILLEREY1977,
title = {Neutron diffraction study and specific heat of antiferromagnetic NiI2},
journal = {Physics Letters A},
volume = {61},
number = {2},
pages = {138-140},
year = {1977},
issn = {0375-9601},
doi = {https://doi.org/10.1016/0375-9601(77)90863-5},
url = {https://www.sciencedirect.com/science/article/pii/0375960177908635},
author = {D. Billerey and C. Terrier and N. Ciret and J. Kleinclauss}
}

@article{Ibanez2018,
  title = {Ab initio calculation of the shift photocurrent by Wannier interpolation},
  author = {Iba\~nez-Azpiroz, Julen and Tsirkin, Stepan S. and Souza, Ivo},
  journal = {Phys. Rev. B},
  volume = {97},
  issue = {24},
  pages = {245143},
  numpages = {13},
  year = {2018},
  month = {Jun},
  publisher = {American Physical Society},
  doi = {10.1103/PhysRevB.97.245143},
  url = {https://link.aps.org/doi/10.1103/PhysRevB.97.245143}
}

@article{Nastos2012,
  title = {Optical rectification and current injection in unbiased semiconductors},
  author = {Nastos, F. and Sipe, J. E.},
  journal = {Phys. Rev. B},
  volume = {82},
  issue = {23},
  pages = {235204},
  numpages = {13},
  year = {2010},
  month = {Dec},
  publisher = {American Physical Society},
  doi = {10.1103/PhysRevB.82.235204},
  url = {https://link.aps.org/doi/10.1103/PhysRevB.82.235204}
}

@article{Aversa1995,
  title = {Nonlinear optical susceptibilities of semiconductors: Results with a length-gauge analysis},
  author = {Aversa, Claudio and Sipe, J. E.},
  journal = {Phys. Rev. B},
  volume = {52},
  issue = {20},
  pages = {14636--14645},
  numpages = {0},
  year = {1995},
  month = {Nov},
  publisher = {American Physical Society},
  doi = {10.1103/PhysRevB.52.14636},
  url = {https://link.aps.org/doi/10.1103/PhysRevB.52.14636}
}

@Article{Wang2020,
author={Wang, Hua
and Qian, Xiaofeng},
title={Electrically and magnetically switchable nonlinear photocurrent in РТ-symmetric magnetic topological quantum materials},
journal={npj Computational Materials},
year={2020},
month={Dec},
day={18},
volume={6},
number={1},
pages={199},
issn={2057-3960},
doi={10.1038/s41524-020-00462-9},
url={https://doi.org/10.1038/s41524-020-00462-9}
}

@article{Tsirkin2018,
  title = {Gyrotropic effects in trigonal tellurium studied from first principles},
  author = {Tsirkin, Stepan S. and Puente, Pablo Aguado and Souza, Ivo},
  journal = {Phys. Rev. B},
  volume = {97},
  issue = {3},
  pages = {035158},
  numpages = {15},
  year = {2018},
  month = {Jan},
  publisher = {American Physical Society},
  doi = {10.1103/PhysRevB.97.035158},
  url = {https://link.aps.org/doi/10.1103/PhysRevB.97.035158}
}

@article{Xiang.prl.2011,
  title = {General Theory for the Ferroelectric Polarization Induced by Spin-Spiral Order},
  author = {Xiang, H. J. and Kan, E. J. and Zhang, Y. and Whangbo, M.-H. and Gong, X. G.},
  journal = {Phys. Rev. Lett.},
  volume = {107},
  issue = {15},
  pages = {157202},
  numpages = {5},
  year = {2011},
  month = {Oct},
  publisher = {American Physical Society},
  doi = {10.1103/PhysRevLett.107.157202},
  url = {https://link.aps.org/doi/10.1103/PhysRevLett.107.157202}
}

@article{kaplan.prb.2011,
  title = {Canted-spin-caused electric dipoles: A local symmetry theory},
  author = {Kaplan, T. A. and Mahanti, S. D.},
  journal = {Phys. Rev. B},
  volume = {83},
  issue = {17},
  pages = {174432},
  numpages = {10},
  year = {2011},
  month = {May},
  publisher = {American Physical Society},
  doi = {10.1103/PhysRevB.83.174432},
  url = {https://link.aps.org/doi/10.1103/PhysRevB.83.174432}
}

@Article{edstrom.2025,
author={Edstr{\"o}m, Alexander
and Barone, Paolo
and Picozzi, Silvia
and Stengel, Massimiliano},
title={Magnetoelectricity of topological solitons in 2D magnets},
journal={npj Computational Materials},
year={2025},
month={Sep},
day={29},
volume={11},
number={1},
pages={295},
issn={2057-3960},
doi={10.1038/s41524-025-01795-z},
url={https://doi.org/10.1038/s41524-025-01795-z}
}

@article{Kurumaji.prb.2013,
  title = {Magnetoelectric responses induced by domain rearrangement and spin structural change in triangular-lattice helimagnets NiI${}_{2}$ and CoI${}_{2}$},
  author = {Kurumaji, T. and Seki, S. and Ishiwata, S. and Murakawa, H. and Kaneko, Y. and Tokura, Y.},
  journal = {Phys. Rev. B},
  volume = {87},
  issue = {1},
  pages = {014429},
  numpages = {9},
  year = {2013},
  month = {Jan},
  publisher = {American Physical Society},
  doi = {10.1103/PhysRevB.87.014429},
  url = {https://link.aps.org/doi/10.1103/PhysRevB.87.014429}
}

@article{Etxebarria.stensor.25,
author = "Etxebarria, Jesus and Perez-Mato, J. Manuel and Tasci, Emre S. and Elcoro, Luis",
title = "{Crystal tensor properties of magnetic materials with and without spin{--}orbit coupling. Application of spin point groups as approximate symmetries}",
journal = "Acta Crystallographica Section A",
year = "2025",
volume = "81",
number = "4",
pages = "317--338",
month = "Jul",
doi = {10.1107/S2053273325004127},
url = {https://doi.org/10.1107/S2053273325004127},
}

@article{radaelli.25,
  title = {Color symmetry and altermagneticlike spin textures in noncollinear antiferromagnets},
  author = {Radaelli, Paolo G. and Gurung, Gautam},
  journal = {Phys. Rev. B},
  volume = {112},
  issue = {1},
  pages = {014431},
  numpages = {12},
  year = {2025},
  month = {Jul},
  publisher = {American Physical Society},
  doi = {10.1103/r34k-xjpx},
  url = {https://link.aps.org/doi/10.1103/r34k-xjpx}
}

@article{brinkman66,
    author = {Brinkman, W. F. and Elliott, Roger James},
    title = {Theory of spin-space groups},
    journal = {Proceedings of the Royal Society of London. A. Mathematical and Physical Sciences},
    volume = {294},
    number = {1438},
    pages = {343-358},
    year = {1966},
    month = {10},
    issn = {0080-4630},
    doi = {10.1098/rspa.1966.0211},
    url = {https://doi.org/10.1098/rspa.1966.0211},
}

@article{litvin77,
author = "Litvin, D. B.",
title = "{Spin point groups}",
journal = "Acta Crystallographica Section A",
year = "1977",
volume = "33",
number = "2",
pages = "279--287",
month = "Mar",
doi = {10.1107/S0567739477000709},
url = {https://doi.org/10.1107/S0567739477000709},
}

@article{RevModPhys.87.1213,
  title = {Spin Hall effects},
  author = {Sinova, Jairo and Valenzuela, Sergio O. and Wunderlich, J. and Back, C. H. and Jungwirth, T.},
  journal = {Rev. Mod. Phys.},
  volume = {87},
  issue = {4},
  pages = {1213--1260},
  numpages = {47},
  year = {2015},
  month = {Oct},
  publisher = {American Physical Society},
  doi = {10.1103/RevModPhys.87.1213},
  url = {https://link.aps.org/doi/10.1103/RevModPhys.87.1213}
}

@article{PhysRevB.110.L041104,
  title = {Bulk photovoltaic effects in helimagnets},
  author = {Zhang, Chunmei and Pi, Hanqi and Zhou, Jian},
  journal = {Phys. Rev. B},
  volume = {110},
  issue = {4},
  pages = {L041104},
  numpages = {6},
  year = {2024},
  month = {Jul},
  publisher = {American Physical Society},
  doi = {10.1103/PhysRevB.110.L041104},
  url = {https://link.aps.org/doi/10.1103/PhysRevB.110.L041104}
}

\end{document}